\documentstyle[nato,numreferences,epsfig,amsmath,verbatim,cite]{crckapb}

\makeatletter
\renewcommand\citeform{\thesection.}
\renewcommand{\@biblabel}[1]{[\thesection.#1]}
\makeatother

\renewcommand{\thefigure}{\thesection.\arabic{figure}}
\renewcommand{\thetable}{\thesection.\arabic{table}}

\newcommand{\ndt}{\noindent}

\newcommand{\nue}{\nu_{e}}
\newcommand{\num}{\nu_{\mu}}
\newcommand{\nut}{\nu_{\tau}}

\begin{opening}

\title{ASTROPARTICLE PHYSICS}
\author{G. GIACOMELLI}
\author{M. SIOLI}
\institute{Dipartimento di Fisica dell'Universit\`a di Bologna \\ and INFN,
 Sezione di Bologna\\
           Viale C. Berti Pichat 6/2, I-40127, Bologna, Italy\\
e--mail: Giacomelli@bo.infn.it, Sioli@bo.infn.it}

\end{opening}

\begin{document}

%%%%%%%%%%%%%%%%%%%%%%%%%%%%%%%%%%%%%%
\vspace{-6.35cm}
\hspace{9cm}
\bf{DFUB 13/2002}

\hspace{9cm}
\bf{Bologna, 05/11/02}
\vspace{4.5cm}
%%%%%%%%%%%%%%%%%%%%%%%%%%%%%%%%%%%%%%

\begin{center}
 {\bf Lectures at the 6th Constantine School on}

 {\bf "Weak and Strong Interactions Phenomenology"}

 {\bf Mentouri University, Constantine, Algeria , 6-12 April 2002}
\end{center}

\vspace{0.5cm}

{\small Abstract. The main issues in non-accelerator astroparticle physics are 
reviewed and discussed. A short description is given of the experimental methods,
of many experiments and of their experimental results.}

%\tableofcontents
\normalfont

\setcounter{figure}{0}\setcounter{table}{0}\setcounter{equation}{0}
\section{Introduction}
The present Standard Model (SM) of particle physics,
which includes the Glashow-Weinberg-Salam theory of Electroweak
Interactions (EW) and Quantum Chromodynamics (QCD) for the Strong Interaction, 
explains quite well all available experimental results. The theory had
surprising confirmations from the precision measurements performed at LEP \cite{ictp6,oujda}.
There is one important particle still missing in the SM: the Higgs Boson. 
Hints may have been obtained at LEP and lower mass limits established at $\sim$ 114 GeV.

On the other hand few physicists believe that the SM is the ultimate
theory, because it contains too many free parameters, there is a 
proliferation of quarks and leptons and because it seems 
unthinkable that there is no further unification with the strong
interaction and eventually with the gravitational interaction. 
Thus most physicists consider the present Standard Model 
of particle physics as a low energy limit of a more fundamental theory,
which should reveal itself at higher energies. 
It is possible that there are at least two energy thresholds \cite{oujda}.

The first threshold could be associated to supersymmetric particles or to a 
new level of compositeness, a substructure of quarks and leptons. 
This first threshold could be at energies of few TeV and could be revealed 
with the next generation of colliders (LHC, etc).

\begin{figure}[t]
  \begin{center}
  \mbox{ \epsfysize=13.0cm \epsffile{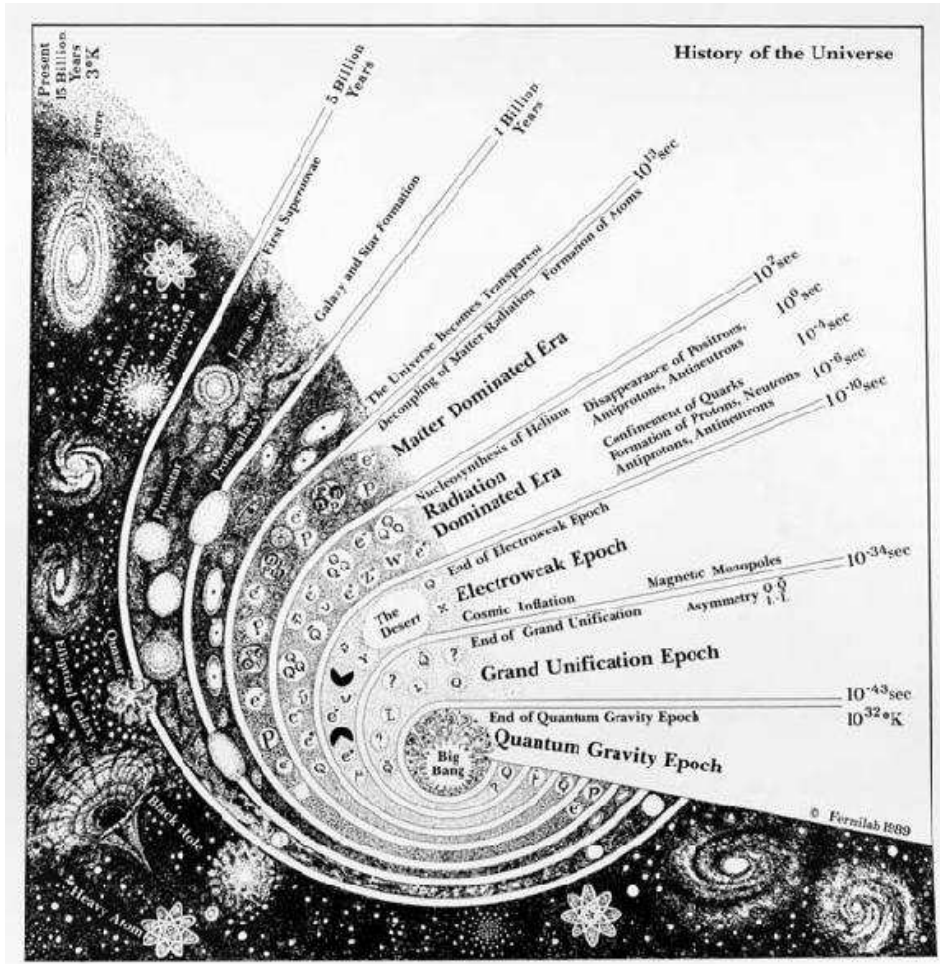} }
\caption{History of the Universe}
\label{f:cosmo}
\end{center}
\end{figure}

The second threshold is associated with the Grand Unification Theory (GUT) of the
EW and Strong Interactions and of quarks with leptons. It would appear at 
extremely high energies, $>10^{14}$ GeV. It is unthinkable 
to reach these energies with any earth-based accelerator; they were instead
available in the first instants of our Universe, at a cosmic time 
$\sim 10^{-35}$~s (Fig. \ref{f:cosmo}). It is in this context that
non-accelerator astroparticle physics plays a very important role.

Astroparticle physics is addressing some of the most fundamental questions:
Does the proton decay? 
Which are the giant accelerators in the cosmos which are capable to
accelerate cosmic ray particles 
to much higher energies that we can possibly obtain in our laboratories?
Are there other fundamental 
particles in the cosmic radiation?  Why do we live in a Universe made only
of matter and not a mix of matter and antimatter? What is the nature of the
dark matter (DM) 
which constitutes more than 90\% of the mass of the Universe? And what
about the dark energy, 
which seems to dominate the energy content of the Universe? Do neutrinos
oscillate? And 
what can we possibly learn with new astronomies, such as neutrino
astronomy, 
dark matter astronomy, gravitational wave astronomy, etc?

There are important connections between astrophysics, particle physics and cosmology, in particular in the Early Universe, which is seen as a gas of very fast particles. As time went by the Universe expanded, the energy per particle decreased, there were phase transitions, the nature of particles changed and one went from unified to non unified interactions, Fig. \ref{f:cosmo}.

Many searches for new particles have been performed at all high-energy accelerators and tight limits have been
obtained for relatively low-mass particles \cite{fossil}.
One may search in the Cosmic Radiation for ``fossil'' particles left over from the Big Bang. Magnetic monopoles, the lightest supersymmetric particles, nuclearites, etc. are examples of such ``fossil'' particles. 
There is now a considerable effort for Fundamental Physics research in Space, see Ref. \cite{space} and the list
of experiments in the Appendix.

GUTs violate the conservation of Baryon Number and of Leptonic Numbers. These violations would be common 
at extremely high energies, while they would lead to very rare phenomena in our ordinary world of low energies. 
Proton decay would be one explicit example and so is neutrinoless double
beta decay \cite{pdg}.

Solar neutrinos and the neutrinos from supernova 1987A opened up the field of Neutrino Astronomy; and one may hope to have, in the future, new astronomies with: High Energy $\nu_{\mu}$, Gravitational Waves and Dark Matter.

Many astroparticle physics experiments (APEs) are performed underground in order to reduce the background coming from cosmic ray muons; but they are also performed at high altitudes, with balloons and in space.

Refined experiments are needed for direct dark matter searches and to search for neutrinoless  double beta decay. Very large experiments are needed to study the highest energy cosmic rays, neutrino physics and astrophysics, to search for proton decay and for magnetic monopoles. The searches for rare phenomena require to reduce the cosmic ray background and thus to go to underground laboratories and to use sophisticated methods for background rejections.

In these lectures notes we shall discuss the present status of many astroparticle physics subjects
and give summaries of non-accelerator APEs; their number is very large and diversified: in the Appendix 
we tried to make an extensive list of the experiments in this wide field.

\setcounter{figure}{0}\setcounter{table}{0}\setcounter{equation}{0}
\section{Astroparticle physics laboratories and detectors}
Astroparticle physics experiments are located underground, on the earth surface, on balloons and in space; 
their list is very long: we refer to the Appendix and to Ref. [1.1, 2.1, 2.2]. %\cite{ictp6,listaexp,mauri}.
Here we shall briefly consider only underground detectors.

The experimental problems of underground detectors are to a large extent related to the energy
range covered by the experiments:\\
i) Low energy phenomena, $E \leq 20$ MeV, for which the main problem is the \textit{radioactivity background};
refined detectors, often of large mass, are needed. For the detection of low energy neutrinos the most important parameters are the detector mass and the energy threshold ($\sim$MeV);\\
ii) Study of $\sim$1 GeV events, like in nucleon decay and in atmospheric neutrino oscillations. 
The main feature of a detector is its \textit{mass} (1--50 kt) and
the capability of identifying neutrino events;\\
iii) Detection of throughgoing particles, high energy muons, monopole candidates, etc. 
The main feature of these detectors is \textit{area}.

Present large underground detectors are: i) water Cherenkov and ii) liquid scintillator detectors; iii) tracking calorimeters.
%Table 2.1 gives the main features of these three types of detectors, with relation to 
%the three kinds of underground experiments defined above.

For nucleon decay searches the competition is between water Cherenkov
detectors and tracking calorimeters. The latter consist of sandwiches of iron plates and ionization/scintillation detectors. The Cherenkov technique allows for larger masses, while tracking calorimeters provide better space resolutions and good identification of electrons, muons and charged kaons.

The largest water Cherenkov detector is Superkamiokande (SK) (50 kton; fiducial mass 22 kton) \cite{skweb}. 
The Cherenkov light, collected by its large phototubes, covered 40\% of its cylindrical surface; 
it will be reduced to about one half when SK will soon resume operation.
The originality of the SNO Cherenkov project \cite{sno1}, designed for the study of solar neutrinos, is the use of heavy water, see Section 5.

Cherenkov detectors are good for determining the direction (versus) of fast charged particles, but have a relatively poor spatial resolution. Large liquid scintillators may identify upward going muons by time-of-flight measurements, but their space resolution is not so good. The combination of scintillators with tracking calorimeters offers both advantages.

The first underground experiments studied solar neutrinos and searched for proton decay and neutrinoless
double $\beta$-decay. At present there is a large number
of underground experiments, large and small, located in mines (Soudan,
Boulby mine, SK, SNO, Canfranc, etc) or in 
underground halls close to a highway tunnel (Mt. Blanc, Frejus, Gran Sasso, etc) \cite{listaexp}.

The largest underground facility is the Gran Sasso Lab. (LNGS) of INFN,
located on the highway Rome--Teramo, 120 km east of Rome. The lab consists of three underground 
tunnels, each about 100 m long; it is at an altitude of 963 m above sea level, is well shielded 
from cosmic rays (by 3700 mwe of rock) and it has a low activity environment. 
The physics aims of the experiments may be classified as follows \cite{ar2001}:\\
\noindent
1) \emph{Detection of particles from external sources.}
                 \par - Solar neutrinos ($E_{\nu_{e}} <$ 14 MeV) ({\scriptsize GALLEX-GNO, BOREXINO, ICARUS})
                 \par - Neutrinos from gravitational stellar collapses ($E_{\nu_{e}} <$30 MeV) 
                 ({\scriptsize LVD, \par ICARUS, MACRO})
                 \par - Study of atmospheric neutrinos and their oscillations, 
                 search for high energy $\nu_{\mu}$ from point sources ($E_{\nu_{\mu}} >$ 1 GeV) ({\scriptsize MACRO})
                 \par - High energy muons, muon bundles and HE CR composition ({\scriptsize LVD, MACRO})
                 \par - Search for magnetic monopoles and other exotica ({\scriptsize MACRO, LVD})
                 \par - Direct searches for dark matter ({\scriptsize DAMA, CRESST, GENIUS, HDMS,} etc)\\
2) \emph{Detection of particles from internal sources.}
                 \par - Search for proton decay, in particular in specific channels ({\scriptsize ICARUS})
                 \par - Search for double beta decay 
                 ({\scriptsize HEIDELBERG-MOSCOW, MIBETA, CUORI- \par CINO, CUORE,} etc)\\
3) \emph{Geophysics experiments.}
\par Here follows a short description of some of the large experiments. They are general purpose 
detectors with a primary item, but capable of giving significant information in many of the 
physics topics listed above.

LVD (Large Volume Detector) in hall A, uses $\sim$ 1000 t of liquid scintillators and it has horizontal 
and vertical layers of limited streamer tubes. Its main purpose is the search for $\bar{\nu_{e}}$ 
from gravitational stellar collapses.

MACRO, with global dimensions of 77 $\times$ 12 $\times$ 9.3 m$^{3}$, was made of
3 horizontal planes of liquid scintillators, 14 horizontal layers of streamer tubes 
and one horizontal and one vertical layer of nuclear track detectors \cite{macro}.
The sides were sealed by one layer of scintillators and 6 layers of streamer tubes.
It studied atmospheric neutrino oscillations, high energy cosmic rays, searched for 
magnetic monopoles, for $\bar{\nu_{e}}$ from stellar gravitational collapses and
HE $\num$'s from astrophysical sources.

ICARUS (Imaging Cosmic And Rare Underground Signals) is a second generation multipurpose
experiment consisting of a liquid argon drift chamber, where one can observe long tracks 
with a space resolution comparable to that of bubble chambers. A 600 t prototype will be 
installed at Gran Sasso next year. The final detector should have a mass of about 3000 t.

GALLEX (GNO) is a radiochemical detector with 30 t of gallium in a GaCl$_{3}$ form
and is used to study solar neutrinos with energies $>$ 233 keV.

BOREXINO is a liquid scintillator detector designed for the study of solar neutrino
events, in particular the $^{7}Be$ neutrinos, see Section 5.

Several small and sophisticated detectors (DAMA, MIBETA, HDMS, CRESST, etc)
are searching for dark matter particles and for neutrinoless double $\beta$
decay, see Sections 6, 7, 8.

An Extensive Air Shower Detector (EASTOP) was located on top of the Gran Sasso mountain.

\setcounter{figure}{0}\setcounter{table}{0}\setcounter{equation}{0}
\section{High energy cosmic rays}
Cosmic rays (CRs) are the only sample of matter from outside our Solar System
that reaches the Earth. High energy (HE) primary CR particles are shielded by the
Earth atmosphere and are indirectly detected on the earth ground via ionization/excitation and 
showers of secondary (pions) and tertiary particles (muons, atmospheric neutrinos, etc). 

Cosmic rays were discovered by V. Hess in 1912 by measuring ionization in a counter
on a balloon. In 1938 Auger detected extensive air showers (EAS) caused by primary 
particles with energies $> 10^{15}$ eV; he detected secondary 
particles in ground detectors spaced several meters apart \cite{gaisser}.

Primary cosmic rays with energies larger than 1 GeV are composed (in number)
of protons (92\%), helium nuclei (6\%), heavier nuclei, lithium to uranium (1\%), 
electrons (1\%) and gamma rays (0.1\%). At $\sim$1 GeV energy the chemical abundance
of the cosmic radiation is similar to the chemical abundance in our galaxy \cite{gaisser}.
This means that the source of heavy cosmic rays cannot be the nucleosysnthesis which
happened 200 $s$ after the Big Bang, when only p and He nuclei were generated.
Cosmic rays must come from stellar sources where iron and other nuclei are produced.

\begin{figure}[t]
\begin{center}
\mbox{ \epsfysize=6.3cm
       \epsffile{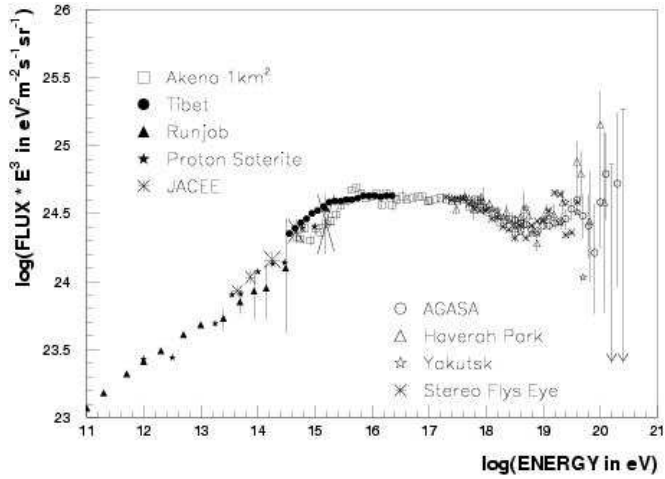}}
\end{center}
\caption{The CR all--particle spectrum observed by different experiments above 10$^{11}$ eV.
The differential flux in units of events per area, time, energy, and solid angle was multiplied by
$E^{3}$ to project out the steeply falling character. The ``knee'' is at $E \simeq 4 \times 10^{15}$ eV
and the ``ankle'' at $E \simeq 5 \times 10^{18}$ eV \protect\cite{sigl}.}
\label{f:allparticle}
\end{figure}

Above 1 GeV the all-particle CR energy spectrum exhibits little structure and is
approximated by a power dependence 
$\ \ d\Phi/dE dS dt \propto E^{-\gamma} \ \ $
with $\gamma \simeq$ 2.7 for $10^{10} < E < 4 \cdot 10^{15}$ eV, 
and $\gamma \simeq$ 3.0 for $4 \cdot 10^{15} < E < 10^{17}$ eV
for the flux per unit area, time and energy; $E \simeq 4 \cdot 10^{15}$ eV is
called the \textit{knee}; at about $E \simeq 5 \cdot 10^{18}$ eV there is the
\textit{ankle}, above which $\gamma \simeq$ 2.8 (see Fig. \ref{f:allparticle}).
Several events have been detected with energies above $10^{20}$ eV
(this corresponds to a particle with about 50 Joules, an energy $\sim 10^{8}$ times larger 
than energies achievable with our accelerators).

The measurement of relatively low energy CR primaries is made directly using detectors
on balloons and satellites. At higher energies we need larger detectors and
very large Extensive Air Shower Arrays on Earth. The Akeno Giant Air Shower Array 
(AGASA) in Japan covers an area of $\sim$100 km$^{2}$ using counters separated by about 1 km \cite{sigl}.

\begin{figure}[t]
  \begin{center}
  \mbox{ \epsfysize=6.0cm
         \epsffile{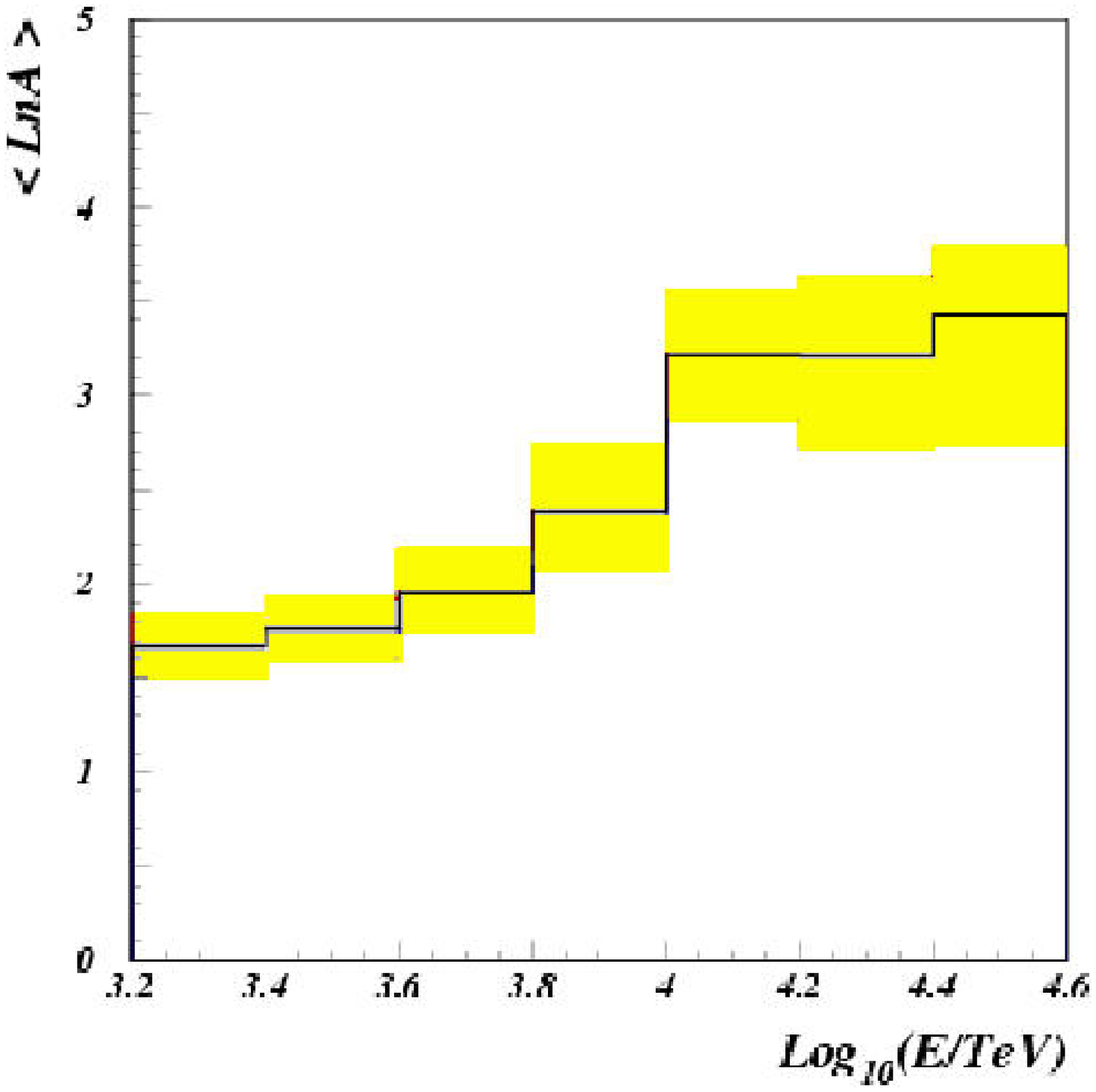} 
         \epsfysize=6.0cm
         \epsffile{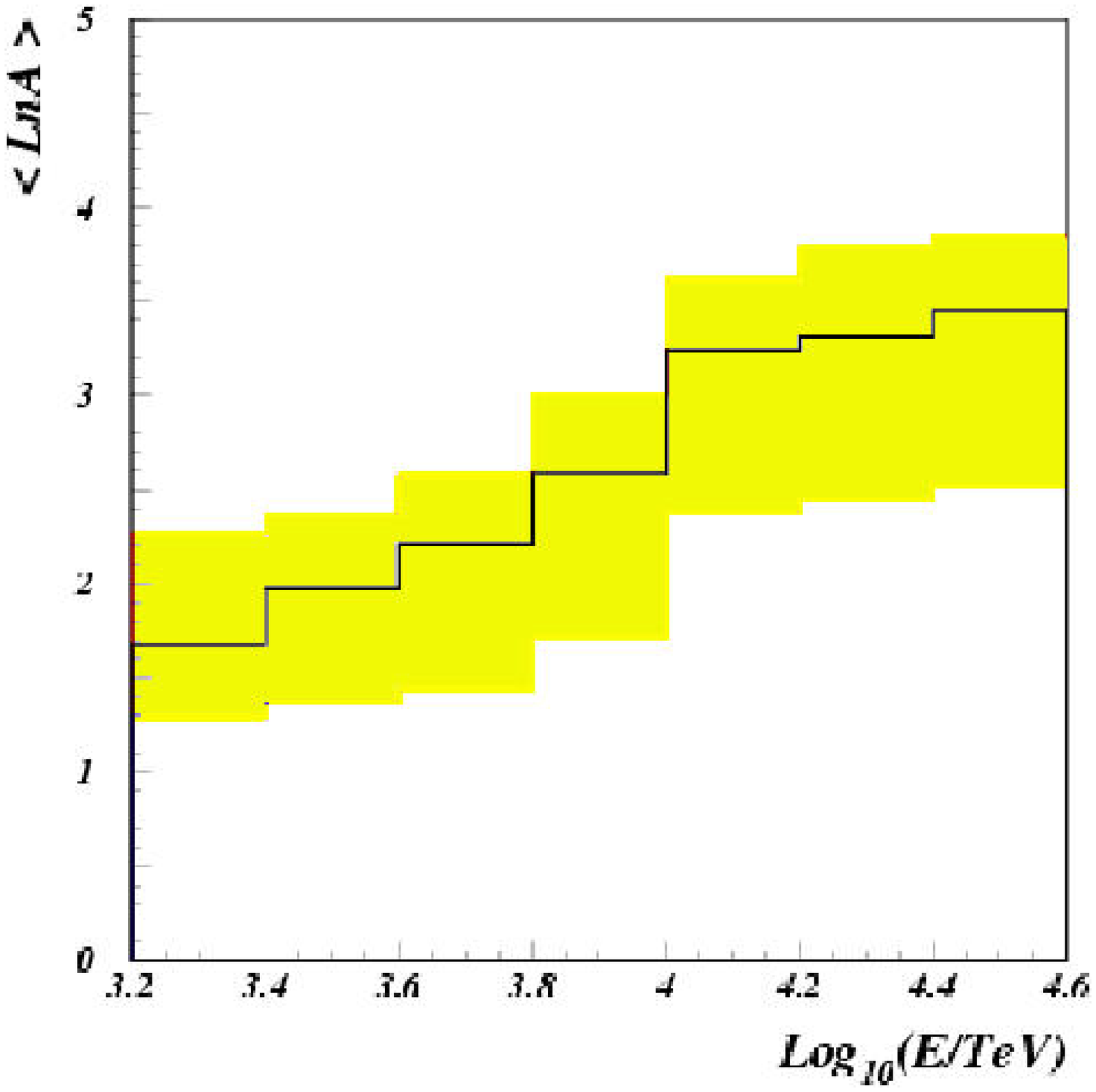} 
         }
{\small \hskip 7.0 truecm (a) \hskip 7.0 truecm (b)}
\caption{The average logarithm of the CR mass number, $\langle ln A \rangle$, 
vs primary energy for: (a) $p/Fe$ and (b) $Light/Heavy$ compositions. Histogram 
lines are obtained from the data, the shaded areas include uncertainties \protect\cite{eastop}.}
\label{f:eastop}
\end{center}
\end{figure}

Measurements of the CR composition around the knee of the energy spectrum can only be performed
indirectly making use of models. The MACRO--EASTOP Collaboration measured the composition
by fitting the underground muon multiplicity distribution ($\approx$ composition) 
for different shower size ($\approx$ energy) bins [2.5, 3.3]. %\cite{eastop}.
Assuming a pure $p-Fe$ model (Fig. \ref{f:eastop}a), or a multi-component model (Fig. \ref{f:eastop}b), 
they found that $\left\langle A \right\rangle$ increases when crossing the knee.

The galactic power in cosmic rays, estimated to be of the order of $10^{33} \div 10^{34}$ W 
(about 1 eV/cm$^{3}$), can be compared with the 
$10^{37}$ W of visible light,
$10^{34}$ W of infrared,
$10^{32} \div 10^{33}$ W of $x$-rays, 
$10^{32}$ W of radiowaves,
$10^{32}$ W of $\gamma$-rays.
Order of magnitude estimates of energy densities in our Galaxy (in
eV/cm$^{3}$) are:
cosmic Rays $\sim$1, 
visible matter $\sim$3,
turbolent gas $\sim$0.3, 
magnetic field $\sim$0.3, 
starlight $\sim$0.3,
CMB (2.7$^{o}$ K) 0.27,
cosmic $\nu$'s $\sim$0.1.
%\begin{table}[b]
%  \begin{center}
%    \begin{tabular}{llll}
%      \hline
%      Cosmic Rays       &  $\sim$ 1    & Turbolent gas   & $\sim$ 0.3   \\
%      Magnetic field    &  $\sim$ 0.3  & Starlight       & $\sim$ 0.3   \\
%      CMB (2.7$^{o}$ K) &         0.27 & Visible matter  &        3     \\
%      Cosmic $\nu$      &  $\sim$ 0.1  &                 &              \\
%      \hline
%    \end{tabular}
%  \end{center}
%  \caption{Order of magnitude estimates of energy densities in our Galaxy (in eV/cm$^{3}$)}
%  \label{t:densities}
%\end{table}

Most astrophysicists favor discrete sources as the originators of cosmic rays. 
Supernovae shells are favoured for medium--high energies (see later).
Hypernovae could be the sources of the highest energy CRs; 
there are other more exotic possibilities: they could be due to magnetic monopoles 
with masses of $\sim$ 10$^{10}$ GeV \cite{mmcr}.

Many air Cherenkov detectors are operating or are planned, see Appendix and Ref. \cite{cecco}.

The largest project presently under construction is the Pierre Auger Giant Observatory \cite{auger}
planned for two sites, one in Argentina and another in Utah, USA. Each site will have a 3000 km$^{2}$
ground array, with 1600 detectors separated by 1.5 km; there will also be 4 fluorescence
detectors. The detection energy threshold will be $\sim 10^{18}$ eV; each detector should yield about
50 events per year with energies above $10^{20}$ eV.

Planned experiments in space would detect EASs from space (OWL and EUSO, see Appendix).
Others may include the near horizontal detection of air showers with ground arrays and detection 
of radio pulses emitted by neutrino--induced electromagnetic showers \cite{jelley}.

What about $\bar{p}$ and antimatter? Conventional models of CR
propagation predict the production of secondary $\bar{p}$ through collisions of high energy CRs
(p, He, etc.) with the interstellar medium. For GeV energies the predicted $\bar{p}/p$
ratio is $\sim 10^{-4}$ and is expected to decrease with increasing energy as $E^{-0.6}$.
For E$<$1 GeV, secondary production is suppressed and the $\bar{p}/p$ ratio should be
smaller than $10^{-5}$. Present data are not conclusive, though it seems that there are no
indications for extra $\bar{p}$'s. Searches have been made for $\bar{He}$ antinuclei: 
the limits are improving. The best
proof would be the presence of heavier antinuclei, like $\overline{Fe}$. Notice that
with optical telescopes one observes collisions of 2 galaxies, 
and not of galaxy--antigalaxy \cite{gaisser}.

\begin{figure}
\begin{center}
\mbox{ \epsfysize=6.0cm
       \epsffile{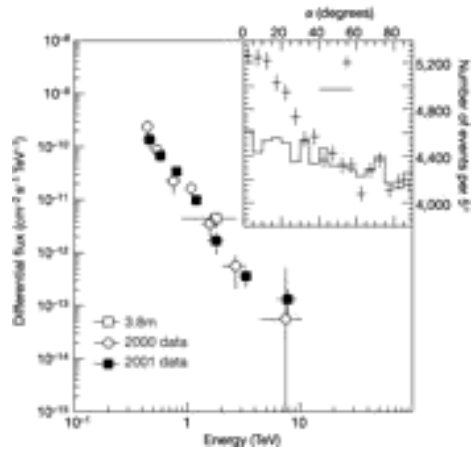}}
\end{center}
\caption{Differential $\gamma$--ray flux vs $E_{\gamma}$ measured by the Cangaroo 
experiment \protect\cite{enomoto}.}
\label{f:cangaroo}
\end{figure}

\begin{figure}[p]
\begin{center}
\mbox{ \epsfysize=6.0cm
       \epsffile{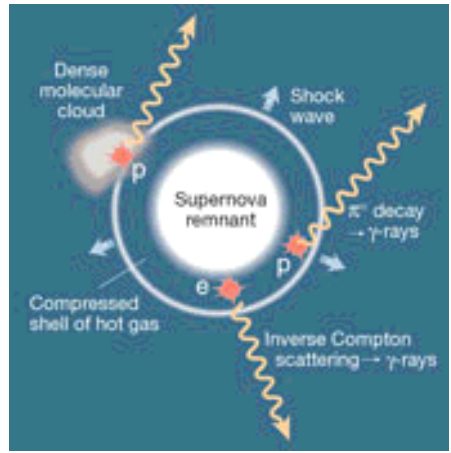}}
\end{center}
\caption{Possible mechanism to produce high energy protons and $\gamma$--rays in a supernova remnant.
Accelerated protons interact with the surrounding gas yielding $\pi^{0}$'s, which 
decay into 2$\gamma$ [3.11]. A second mechanism, where accelerated electrons energize 
cosmic background photons to produce $\gamma$--rays by inverse Compton scattering, 
is not adequate to produce the high flux of measured HE $\gamma$--rays \protect\cite{aharonian}.}
\label{f:supernova}
\end{figure}

The Cangaroo Collaboration operates a 3.8 m and a 10 m air Cherenkov
reflector telescopes in Australia to detect HE $\gamma$--rays.
They reported the differential $\gamma$--ray flux from RX J1713.7-3946 
plotted in Fig. \ref{f:cangaroo} \cite{enomoto}. 
The spectrum extends up to 10 TeV, and it has a power law shape, $E_{\gamma}^{-2.5}$.
RX J1713.7-3946 is a shell-type supernova remnant found in the ROSAT
all-sky survey. One may hypothesize that the $\gamma$--rays come from
the interaction of protons accelerated in the expanding shell and which hit
the material of the shell yielding $\pi^{0}$'s which decay into $\gamma\gamma$.
It should be remembered that the energy needed to maintain the galactic
population of cosmic rays is a few per cent of the
total mechanical energy released by supernovae explosions in our Galaxy.
After the collapse of the core of a massive star a strong shock wave is emitted.
The diffusive shock wave mechanism may trasfer a sizable amount of energy to protons, 
electrons and nuclei in the surrounding interstellar gas, see Fig. \ref{f:supernova}.
The energy spectrum observed is consistent with 
this mechanism and inconsistent with others, like bremsstrahlung radiation from electrons.
$\gamma$--rays of TeV energies have been detected from other two shell type
supernovae remnants \cite{enomoto,aharonian}. One may thus conclude that a considerable part of the
intermediate energy cosmic rays come from supernovae remnants.

\begin{figure}[p] 
\begin{center}
\mbox{ \epsfysize=6.6cm
       \epsffile{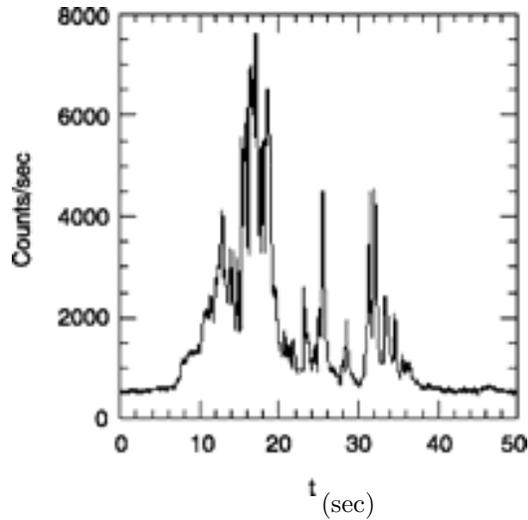}}
%\vspace{-1cm}
\end{center}
\vskip -0.5 truecm
{\small \hskip 7.0 truecm (sec)}
\vskip  0.5 truecm
\caption{Time distribution of a GRB. The photon count in the 0.05--0.5 MeV range 
is plotted versus time (in sec). $\gamma$--rays are not seen before nor after the burst \protect\cite{uno}.}
\label{f:grb}
\end{figure}
%figure 1 e 2 di nature,vol 416, 25 aprile 2002.

\par \textit{$\gamma$--RAY BURSTS (GRBs).}
They were first reported in the early 1970's by the VELA military satellites
monitoring nuclear explosions on the Earth. GRBs are sudden, intense flashes of $\gamma$--rays that
for few seconds appear in a sky which is fairly ``dark'' 
in $\gamma$--rays, see Fig. \ref{f:grb} \cite{uno}.
The BATSE detector on the Compton Gamma Ray Observatory (CGRO) measured about 3000 bursts,
distributed isotropically in the sky. New bursts are now seen at a rate of about one a day \cite{due}.
Recently ``afterglows'' from the direction of some of the GRBs have been detected: the afterglow
includes the emission of X--rays lasting few days \cite{tre},
which makes possible to individuate the later optical and radio part of the afterglow, 
which lasts for months.
This in turn allows the measurement of redshift distances, the identification of the host galaxies and
the confirmation that GRBs come from cosmological distances, billions of light years away.
The energy output of GRBs seems to be $10^{51} \div 10^{54}$ erg s$^{-1}$,
which is larger than that from any other source: it is comparable to burning the mass of
our sun in few seconds. 
%The emitted energy spectrum softens in time: the energy starts concentrated
%in MeV $\gamma$--rays and progressively evolves into an afterglow radiation that peaks in X--rays, then
%ultraviolet, optical, infrared and radio.

GRBs arise in regions where there are large mass star formations; the stars undergo a catastrophic 
energy release toward the end of their evolution. In this case one sees optical lines related 
to elements like iron. A possible model could be the collapse of a massive object into a neutron 
star or black hole, generating jets that create $\gamma$--ray bursts; jets are collimated flows of 
plasma that travel at almost the speed of light. These jets could produce GRBs, ultrahigh energy 
cosmic rays and other high--energy phenomena. If the $\gamma$--rays come from jets then the source 
does not emit isotropically, thus the emitted total energies have to be decreased. Other possible 
sources could involve massive stars whose cores collapse, in the course of merging with a 
companion star (\textit{hypernovae}, \textit{collassars}).
The same shock wave that accelerates electrons should also accelerate protons 
up to $\sim 10^{20}$ eV. The protons could interact with the fireball photons 
yielding high energy charged $\pi^{\pm}, \mu^{\pm}$ and neutrinos. The $\nu$'s could be 
detected by neutrino telescopes. GRBs are also expected to be sources of gravitational waves.

%\begin{figure}
%\begin{center}
%\mbox{ \epsfysize=5.0cm
%       \epsffile{tre_4.eps}}
%%\vspace{-1cm}
%\end{center}
%{\small Figure~3.4: Possible GRB mechanism from a massive stellar progenitor resulting in a relativistic
%jet that undergoes internal shocks producing a burst of $\gamma$--rays and an afterglow which
%leads successively to X--rays, UV, optical, IR and radio.}
%\end{figure}

\setcounter{figure}{0}\setcounter{table}{0}\setcounter{equation}{0}
\section{Neutrino physics (oscillations and masses)}

\begin{figure}[t]
\begin{center}
\epsfclipon
\mbox{\epsfysize=6.0 cm \epsffile{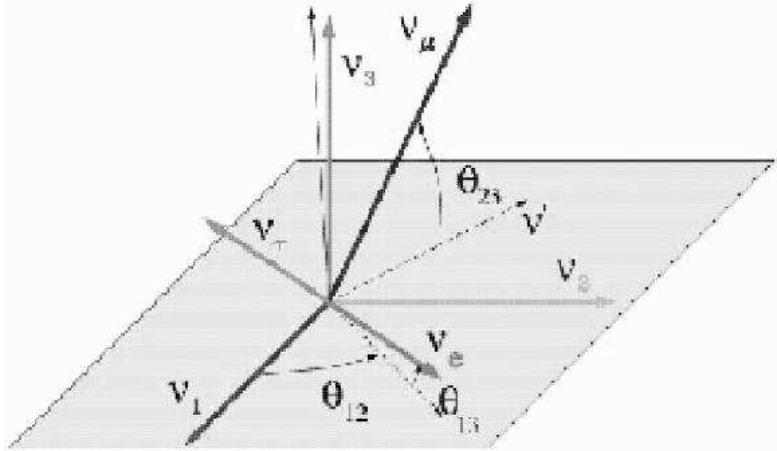}}
\epsfclipoff
\end{center}
\caption{Rotations between mass and flavour eigenstates in the 3 family neutrino
oscillation scheme 
($\theta_{23} \simeq 45^{o}$, $\theta_{12} \simeq 40^{o}$ and $\theta_{13} < 11^{o}$).}
\vspace{-0.5cm}
\label{f:mixing}
\end{figure}

Neutrino physics is extensively discussed in the lectures of P. Lipari \cite{lipa}.
Here we summarize some experimental aspects.
For $\beta \beta$--decay see Sect. 7.

\textit{NEUTRINO OSCILLATIONS.}
If neutrinos have non--zero masses, one has to consider the 3 
\textit{weak flavour eigenstates} $\nue, \num, \nut$
and the 3 \textit{mass eigenstates} $\nu_{1}, \nu_{2}, \nu_{3}$.
The flavour eigenstates $\nu_{l}$ are linear combinations of the
mass eigenstates $\nu_{m}$ via the elements of the unitary mixing
matrix $U_{lm}$:
\begin{equation}
        \nu_{l}=\sum_{m=1}^{3} U_{lm} \; \nu_{m}
\end{equation}
In the conventional parametrization U reads as follows
\footnotesize
\begin{equation}
U\equiv
        \left( \begin{array}{ccc}
        1 &      0  &     0  \\
        0 &  c_{23} & s_{23} \\
        0 & -s_{23} & c_{23}
  \end{array} \right)
        \left( \begin{array}{ccc}
                     c_{13} & 0 &  s_{13}e^{i\delta}  \\
                          0 & 1 &                  0  \\
        -s_{13}e^{-i\delta} & 0 &             c_{13}
  \end{array} \right)
        \left( \begin{array}{ccc}
         c_{12} & s_{12} & 0 \\
        -s_{12} & c_{12} & 0 \\
       0  &      0 & 1
  \end{array} \right)
\end{equation}
\normalsize
with $s_{12}\equiv$ sin $\theta_{12}$, and similarly for the other sines and cosines.
The action of the three rotation matrices is illustrated in Fig. \ref{f:mixing} with approximate
values of the three mixing angles $\theta_{12}$, $\theta_{13}$ and $\theta_{23}$,
as presently known.

\begin{figure}[t]
  \begin{center}
    \epsfclipon
    \mbox{\epsfysize=7.5 cm \epsffile{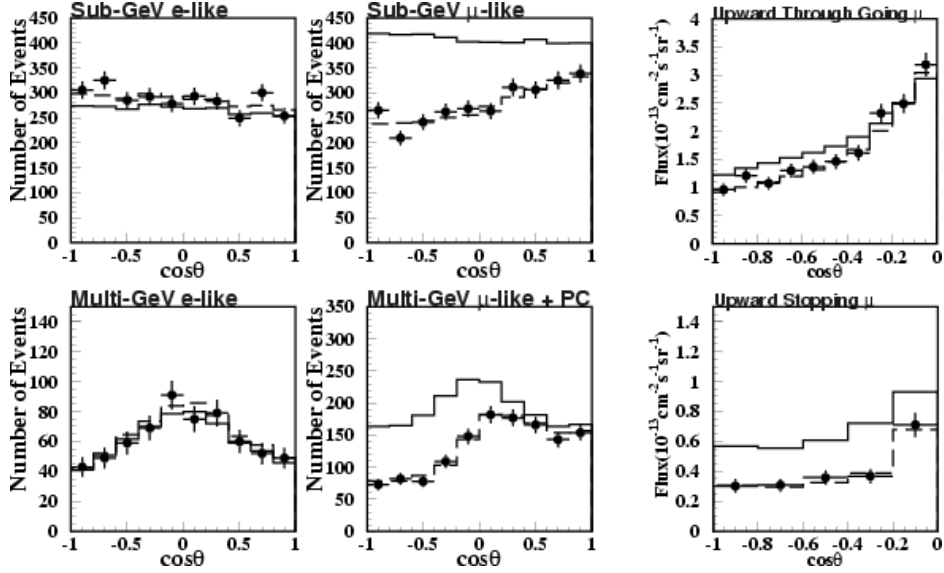}}
    \epsfclipoff
  \end{center}
  \caption{Zenith distributions for SK data (black points) for $e$-like and $\mu$-like sub-GeV and multi-GeV
    events and for up-throughgoing and up-stopping muons. 
    The solid line histograms are the no oscillation MC predictions, the
    dashed lines refer to $\num \leftrightarrow \nut$ oscillations with maximal mixing and 
    $\Delta m^{2} = 2.5 \times 10^{-3}$ eV$^{2}$.}
  \label{f:superk}
\end{figure}

Neutrino oscillations depend on six independent parameters: two mass-squared
differences, $\Delta m^{2}_{12}$ and $\Delta m^{2}_{23}$, three angles 
$\theta_{12}$, $\theta_{13}$, $\theta_{23}$ and the CP-violating phase $\delta$.
In the simple case of two flavour eigenstates ($\num$, $\nut$) which oscillate with two
mass eigenstates ($\nu_2$, $\nu_3$) and $\delta$ = 0 one has:

\begin{equation}
\begin{cases}
   \num =  \nu_{2} \: \text{cos} \: \theta_{23} + \nu_{3} \: \text{sin} \: \theta_{23} \\
   \nut = -\nu_{2} \: \text{sin} \: \theta_{23} + \nu_{3} \: \text{cos} \: \theta_{23}
\end{cases}
\end{equation}

The survival probability of a $\num$ beam is
\footnotesize
\begin{equation}
P(\num\rightarrow\num) 
          = 1 - \text{sin}^{2} \: 2\theta_{23} \: \text{sin}^{2} \left( \frac{E_{2}-E_{1}}{2}t \right)
 \simeq 1 - \text{sin}^{2} \: 2\theta_{23} \: \text{sin}^{2} \left( \frac{1.27 \Delta m_{23}^{2} \cdot L}{E_{\nu}} \right) 
\end{equation}
\normalsize
$\Delta m_{23}^{2}$ = $m_{3}^{2} - m_{2}^{2}$, $L$ is the distance travelled by the 
neutrino from production to detection. The probability for the initial $\num$ to 
oscillate into a $\nut$ is $P(\num\rightarrow\nut) = 1 - P(\num\rightarrow\num)$.

\begin{figure}%[p]
  \begin{center}
    \epsfclipon
    \mbox{\hspace{-1.4cm} \epsfig{file=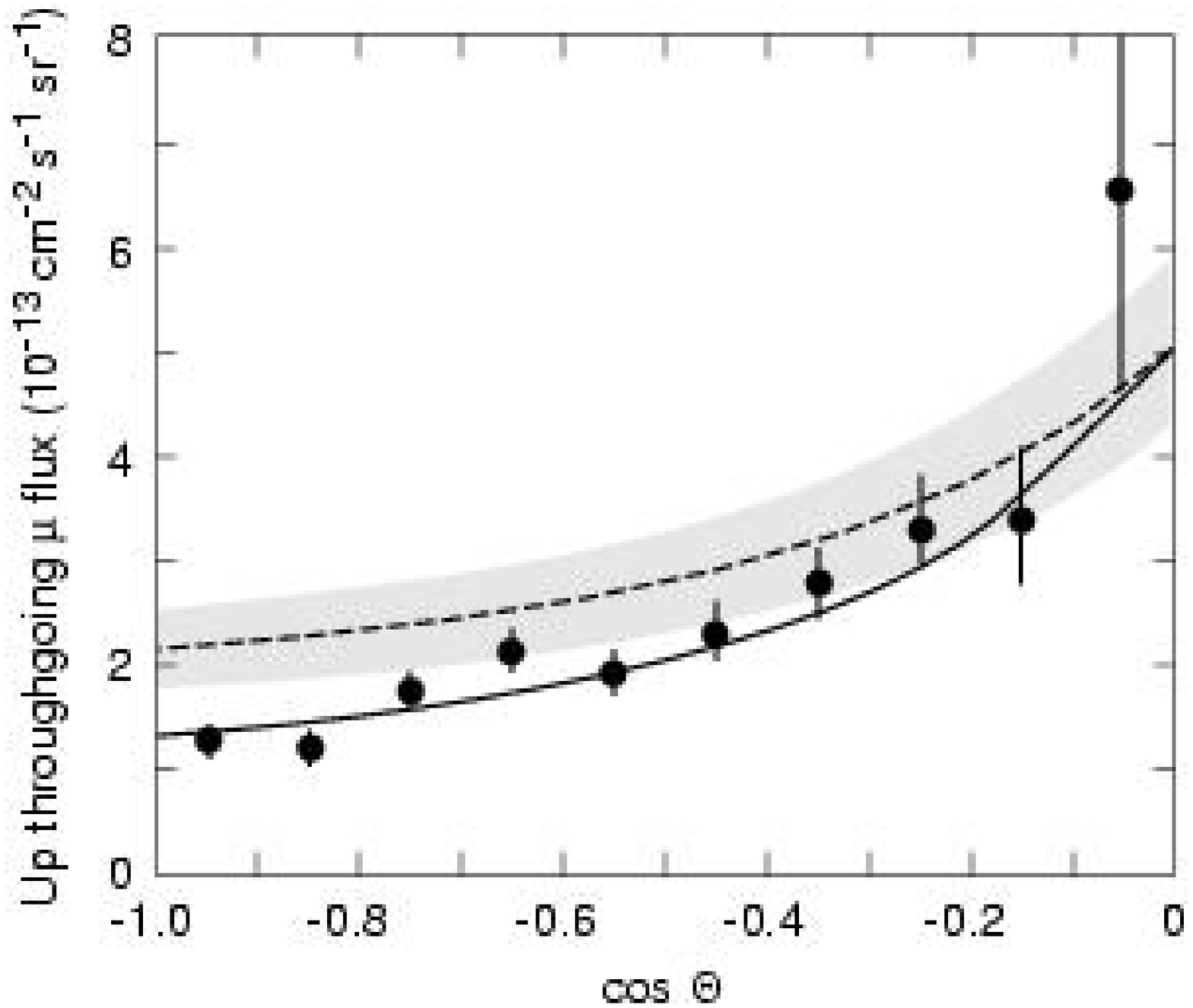, width=4.75cm} 
                          \epsfig{file=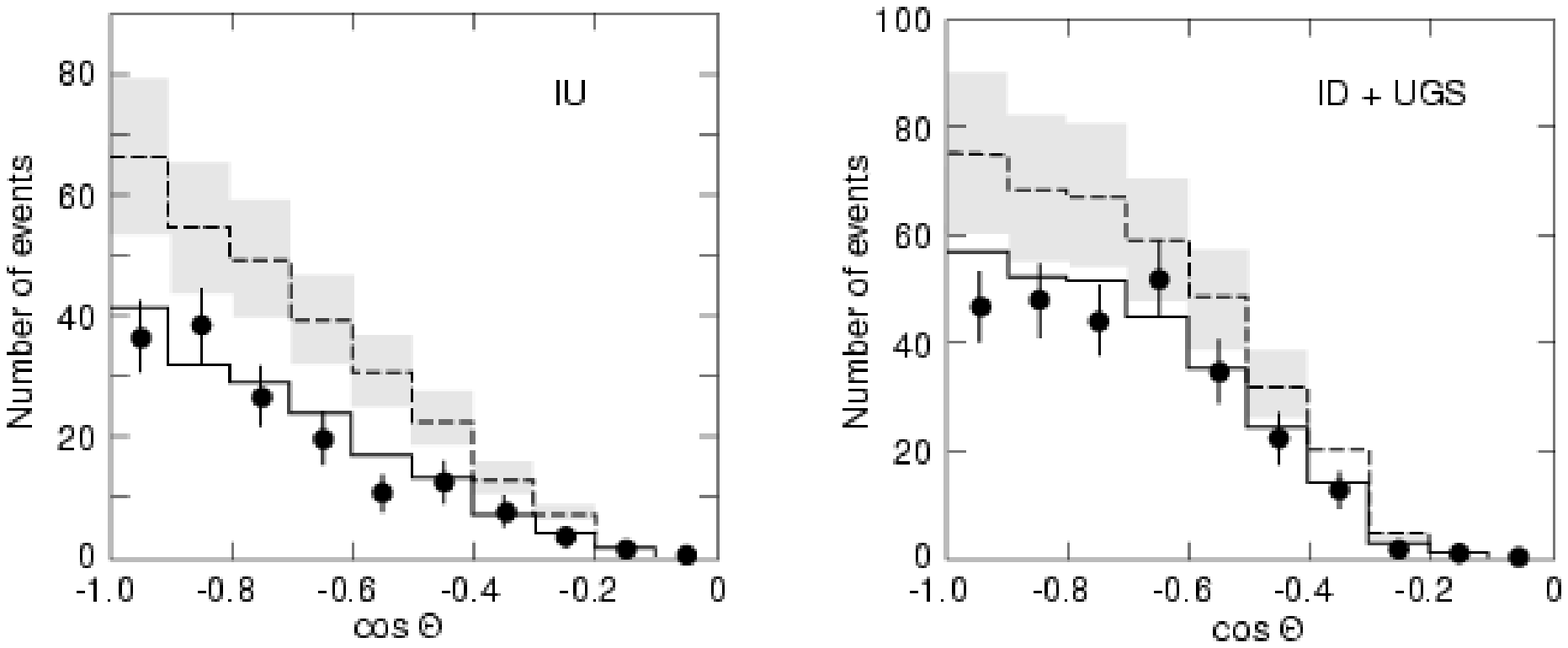, width=9.5cm} }
    \epsfclipoff
  \end{center}
  \caption{Zenith distributions for the MACRO data (black points) for
    (a) upthroughgoing, (b) semicontained and (c) up--stopping muons + down semicontained.
    The dashed lines are the no--oscillation MC predictions (with scale error
    bars); the solid lines refer to $\num \leftrightarrow \nut$ oscillations 
    with maximum mixing and $\Delta m^{2} = 2.5 \times 10^{-3}$ eV$^{2}$.}
  \label{f:macro}
\end{figure}

\textit{ATMOSPHERIC NEUTRINO OSCILLATIONS.}
HE primary CRs, protons and nuclei, interact in the upper atmosphere
producing a large number of pions and kaons, which decay yielding muons and muon
neutrinos; then muons decay yielding $\num$'s and $\nue$'s. The ratios of
their numbers are $N_{\num}/N_{\nue} \simeq 2$ and 
$N_{\nu}/N_{\bar{\nu}} \simeq 1$. 
The neutrinos are produced in a spherical shell at about 10-20 km above ground and
proceed towards the Earth.

Atmospheric neutrinos are well suited for the study of neutrino oscillations, since they have energies
from a fraction of GeV up to more than 100 GeV and they travel distances L from few tens of km up
to 13000 km; thus $L/E_{\nu}$ ranges from $\sim$ 1 km/GeV to $\sim 10^{5}$ km/GeV. One may consider that there
are two sources for a single detector: a near one (downgoing neutrinos) and a far one (upgoing
neutrinos). Results have been obtained for $10^{-3} < \Delta m_{23}^{2} < 10^{-2}$ eV$^{2}$.
$\theta_{23}$ and $\Delta m_{23}^{2}$ are determined from the variation of $P(\num\rightarrow\num)$
vs zenith angle $\theta$, or from the variation in $L/E_{\nu}$.

\begin{figure}%[b]
\begin{center}
\epsfclipon
\mbox{\epsfysize=7.0 cm \epsffile{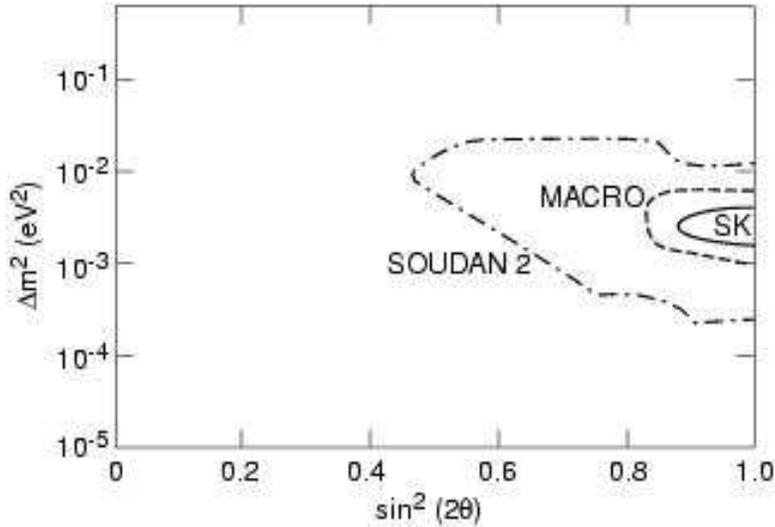}}
\epsfclipoff
\end{center}
\caption{90\% CL contour plots of the allowed regions of SK, MACRO and Soudan 2
(the Soudan 2 and MACRO regions are now somewhat smaller).}
\label{f:contour}
\end{figure}

The zenith angle distributions and the observed number of events in Soudan 2, 
MACRO (Fig. \ref{f:macro}) and SK (Fig. \ref{f:superk}) agree with two flavour 
$\num \rightarrow \nut$ oscillations with $sin^{2} 2\theta_{23} \simeq 1$ and 
$\Delta m^{2}_{23} \simeq 2.5 \times 10^{-3}$ eV$^{2}$ [4.2-4.5].
The 90\% CL contours of the allowed regions of Soudan 2, MACRO and SK overlap, see Fig. \ref{f:contour}.
For $\nue$--induced electrons there are no deviations from the no-oscillation MC predictions.

Matter effects have been studied with HE atmospheric neutrinos by MACRO
\cite{macronu} and SK \cite{superknu}.
Compared to $\num \leftrightarrow \nut$,
$\num \leftrightarrow \nu_{sterile}$ oscillations are disfavoured at 99\% CL for any mixing. 

Present atmospheric neutrino experiments are disappearance experiments; future atmospheric
$\nu$ experiments are under study \cite{lipa}. Also long baseline 
experiments using $\num$ from accelerators are planned. The main goals of the experiments are the
detection of the first oscillation in $L/E_{\nu}$ and the appearance of $\nut$, to really prove the oscillation hypothesis. They should also yield improved values of the oscillation parameters. Other goals include
the detection of a possible small $\num \rightarrow \nue$ contribution. 
Eventually one would like complete information on the 3 $\times$ 3 oscillation matrix. 

\textit{SOLAR NEUTRINO OSCILLATIONS.}
Solar $\nu$'s and their oscillations are discussed in the lectures
of P. Lipari \cite{lipa}. In Section 5 we recall some features of 
solar neutrinos and the recent SNO experimental results.

\begin{figure}%[t]
\begin{center}
\epsfclipon
\mbox{\epsfysize=4.5 cm \epsffile{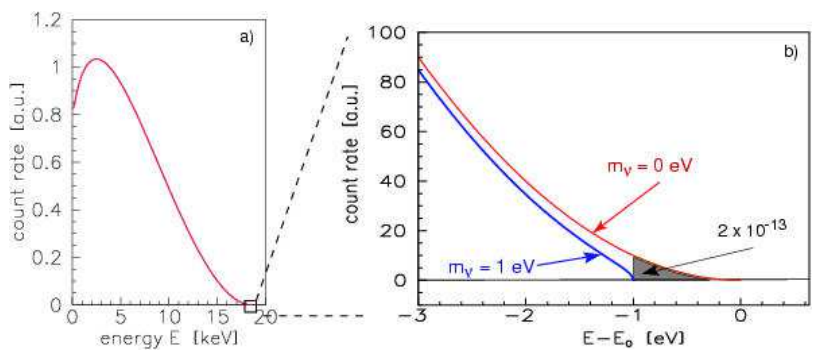}}
\epsfclipoff
\end{center}
\caption{Kurie plot for tritium $\beta$ decay: (a) complete and (b) narrow
region around the endpoint $E_{0}$. The $\beta$ spectrum is shown for $\overline{\nue}$ 
masses of 0 and 1 eV.}
\label{f:endpoint}
\end{figure}

\textit{DIRECT MEASUREMENTS OF NEUTRINO MASSES.}
Many direct measurements have been performed using tritium decay,
$^{3}H \rightarrow ^{3}He^{+} + e^{-} + \overline{\nue}$,
measuring, with magnetic spectrometers of ever increasing precision,
the electron spectrum near its kinematical limit where the number of
events is small, $\sim 10^{-13}$ times the whole sample (Fig. \ref{f:endpoint}). 
The present best limit is $m(\overline{\nue}) < 2.2$ eV (95\% CL) \cite{tritium}.
Limits have been obtained at accelerators using charged muons and charged
tau decays, obtaining $m(\num) < $170 keV, $m(\nut) <$ 15 MeV.
The last limit comes from the combination of results from different
experiments, mainly at LEP (see for example Ref. \cite{taumass}). In these experiments
one considers in particular the $\tau$ decays into many charged pions +
$\nut$: $\ \ \ e^{+}e^{-} \rightarrow \tau^{+}\tau^{-} \rightarrow 
(\pi^{-}\pi^{+}\pi^{-}\pi^{+}\pi^{+}\nut) + 
(\pi^{-}\pi^{+}\pi^{-}\pi^{+}\pi^{-}\nut)$.

\setcounter{figure}{0}\setcounter{table}{0}\setcounter{equation}{3}
\section{Neutrino astronomy (solar, SN, HE $\nu$'s)}
\begin{figure}[t!]
\begin{center}
\epsfclipon
\mbox{\epsfysize=6.6 cm \epsffile{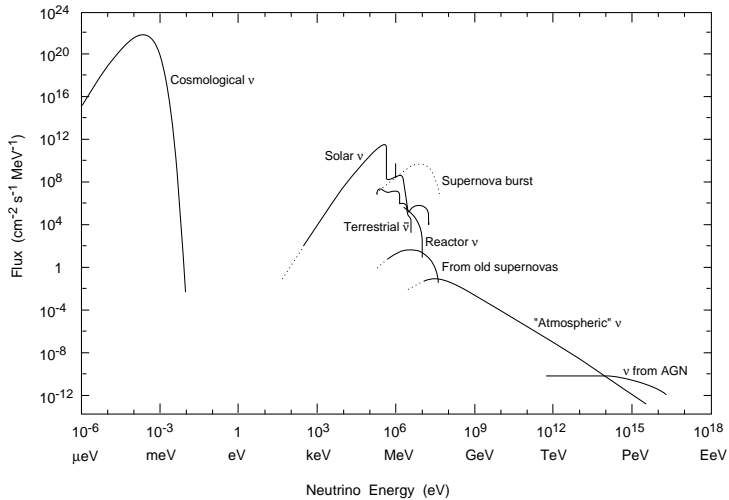}}
\epsfclipoff
\end{center}
\caption{Energy spectrum of neutrinos emitted by different celestial bodies.}
\label{f:nuspectra}
\end{figure}

Underground (underwater) detectors of large area and mass may yield
important information on the astrophysics of neutrinos. Neutrino 
astronomy is a new observational window on the Universe. 
Low energy neutrinos of few MeV come from the interior of stars like 
the Sun; bursts of slightly higher energy neutrinos, with few tens of MeV, 
are emitted in stellar gravitational collapses (supernovae).
High energy neutrinos with hundreds of GeV should come from
non--thermal powerful astrophysical sources. Since neutrinos interact rarely,
the observed $\nu$'s come directly from their sources, without suffering
the many interactions typical of photons. Because of all these sources,
the number of neutrinos in the Universe is increasing. It has to be
remembered that the Universe should be filled with ``fossil'' low energy
neutrinos ($\sim 2 \times 10^{-4}$eV) from the Big Bang; their number
is comparable to that of the Cosmic Microwave Background (CMB) radiation, see Sect. 9;
at present there is no possibility of detecting them. The Earth emits
MeV antineutrinos from radioactive decay. The global $\nu$ spectrum
is shown in Fig. \ref{f:nuspectra}. Neutrinos of (1-100) GeV may also come from the 
interior of celestial bodies, like the Earth or the Sun, where annihilations
of Weakly Interactive Massive Particles (WIMPs) could take place, Sect. 8.

Neutrinos travel at the speed of light
(or almost, if $m_{\nu} \neq 0$) and are electrically neutral; thus they are 
not deflected in magnetic fields and carry
direct information from their celestial source and may be observed day and night.
But because of their weak interaction they are difficult to detect and
one needs large detectors.

\begin{figure}
\begin{center}
\epsfclipon
  \mbox{ \hspace{-0.5cm} \epsfysize=4.5cm
         \epsffile{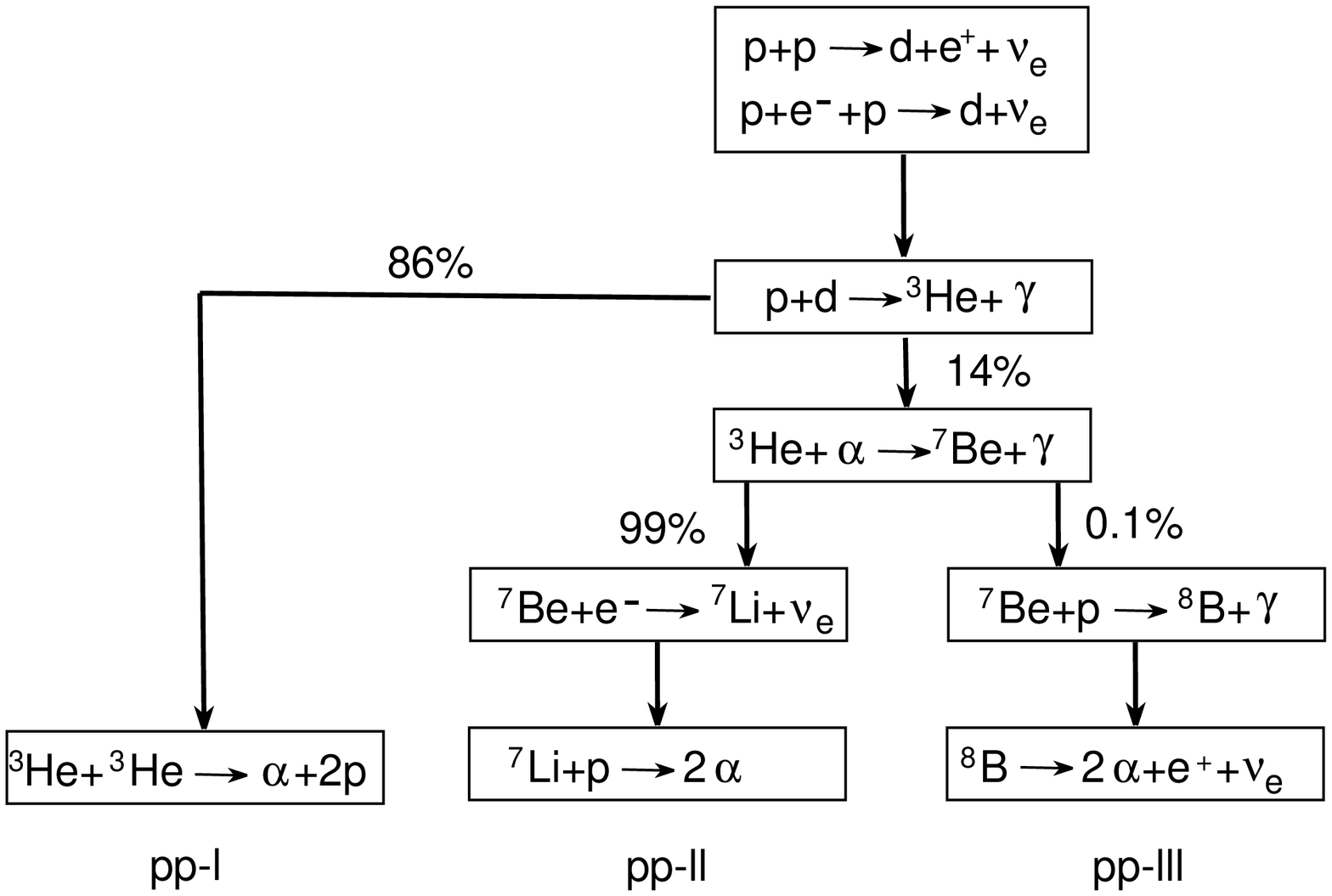} 
         \hspace{0cm} \epsfysize=4.5cm
         \epsffile{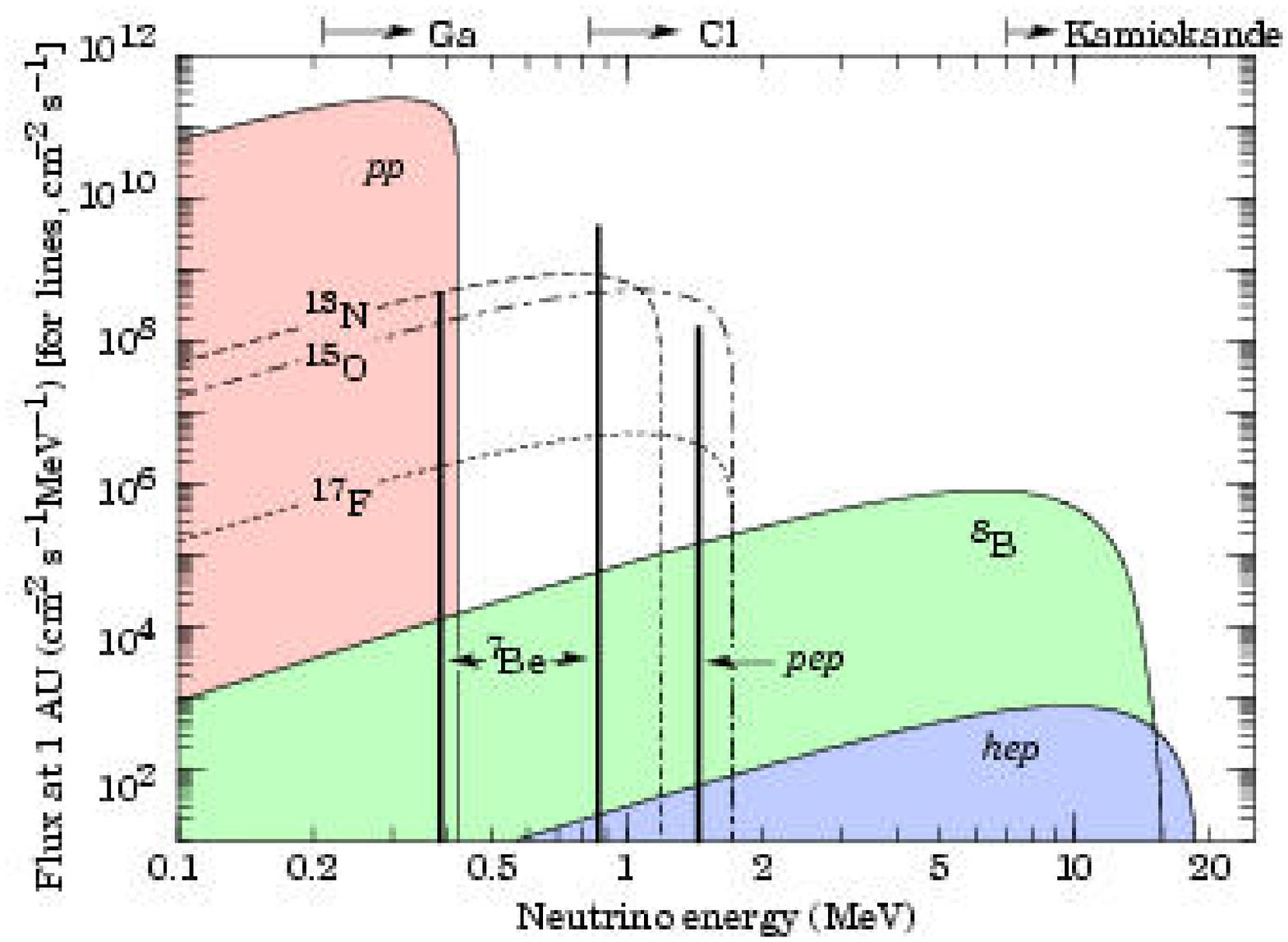}
         }
\epsfclipoff
\end{center}
\caption{(left) The chain of nuclear reactions and decays at the
center of the Sun; only $\nue$ are emitted. (right) The energy
spectrum of the solar neutrinos arriving on Earth; on the top are
indicated the energy thresholds of different $\nu$ detectors.}
\label{f:solarnu}
\end{figure}

\noindent \textit{SOLAR NEUTRINOS.}
According to the Solar Standard Model, all its energy is produced in a
series of thermonuclear reactions and decays at the center of the Sun;
this solar ``thermonuclear reactor'' is very small compared to the size
of the Sun. Neutrinos escape quickly from the Sun, while the emitted photons
suffer an enormous number of interactions (their mean free path is much
less than 1 cm) and reach the surface of the Sun in about one million years;
visible sunlight comes from a well defined surface, the photosphere.
An important fraction of the energy from the Sun is emitted in the form of
$\nue$ of energies from 0.1 to 14 MeV, see Fig. \ref{f:solarnu}.

Most of the emitted neutrinos come from the $p+p \rightarrow d+e^{+}+\nue$
reaction, which yields $\nue$'s with energies $0 < E_{\nue} < 0.42$ MeV;
they have interaction cross-sections of $\sim 10^{-45}$ cm$^{2}$. The highest
energy neutrinos, coming from $B^{8}$, have energies $0 < E_{\nue} < 14.06$ MeV
and cross-sections $\sim 3 \times 10^{-43}$ cm$^{2}$.
On Earth should arrive $\sim 7 \times 10^{10}$ $\nue$ cm$^{-2}$ s$^{-1}$.
The first experiment which detected solar neutrinos was the Homestake Chlorine radiochemical
experiment in the USA, via the reaction $\nue~+~^{37}Cl~\rightarrow~^{32}Ar~+~e^{-}$~;~the
produced gaseous $^{32}Ar$ atoms are flushed and counted; the detector threshold 
is 814 keV; it is thus sensitive to $B^{8}$ neutrinos only.

The Gallex (GNO) and Sage radiochemical experiments use Ga nuclei and have
a threshold of 233 keV; thus they are sensitive also to $pp$ neutrinos. They study 
the reaction ${\nu_{e}} + ^{71}Ga \rightarrow ^{71}Ge + e^{-}$; $^{71}Ge$ nuclei form 
gaseous Ge$\:$H$_{4}$ molecules. Every $\sim$ 20 days the tank is flushed with helium gas, 
which removes the few Ge$\:$H$_{4}$ molecules, which are chemically separated and then brought 
to a proportional counter, where one observes the decay 
$^{71}Ge \rightarrow ^{71}Ga + \gamma$ (1.2 keV, 10.4 keV), 
which has a decay half life of 11.4 days. The procedure was calibrated with an 800 k$\:$Ci
radioactive source of $^{51}Cr$, which yields $\nue$ of 430 keV (10\%) and 750 keV (90\%).
The water Cherenkov detectors (Kamiokande and SK) detect $\nue$'s
via their elastic interaction with electrons; the effective threshold
is $\simeq$7 MeV; thus they are sensitive only to $^{8}B$ neutrinos.

Here we shall recall the recent SNO results which imply neutrino 
oscillations and thus that the solar neutrinos reaching the Earth are not only 
$\nue$, but also $\num$, $\nut$. 
SNO is a 1000 t heavy water Cherenkov detector contained in a transparent acrylic 
spherical shell 12 m in diameter. Cherenkov photons generated in the heavy water are 
detected by 9456 photomultipliers (PMTs) mounted on a stainless steel sphere 17.8 m in diameter, 
immersed in 1500 t of light water, which provides shielding from ambient radioactivity.
SNO detects $^{8}B$ solar neutrinos through the reactions:
\vskip 0.2cm
\begin{tabular}{llll}
\it{Charged Current} (CC)    & $\nu_{e} + d     \rightarrow p       + p     + e^{-}   $ & ($\nue$) &(5.1)\\
\it{Neutral Current} (NC)    & $\nu_{x} + d     \rightarrow p       + n     + \nu_{x} 
$ & ($\nue+\nu_{\mu,\tau}$) &(5.2) \\
\it{Elastic Scattering} (ES) & $\nu_{x} + e^{-} \rightarrow \nu_{x} + e^{-}  $
 & ($\nue+0.154 \; \nu_{\mu,\tau}$) &(5.3) \\
\end{tabular}
\vskip 0.2cm
The charged current reaction (CC) is sensitive only to $\nue$, while the NC reaction 
is equally sensitive to all active neutrino flavours ($x = e, \mu, \tau$). 
The elastic scattering reaction (ES) is sensitive to all flavours, but with reduced sensitivity 
to $\nu_{\mu}$ and $\nu_{\tau}$. Sensitivity to the three reactions allows SNO to 
determine the electron and non-electron neutrino components of the solar flux.
$^{8}B$ solar neutrinos with energies $>$2.2 MeV (the NC reaction threshold) 
were measured with the NC signal (6.25 MeV $\gamma$ from the neutron capture 
$n + d \rightarrow t + \gamma$). 
The effective threshold, $E_{eff}\geq 5$ MeV, provides sensitivity 
to neutrons from the NC reaction, with an efficiency of $\sim$ 30\%.
The flux of $^{8}B$ neutrinos for $E_{eff}\geq 5$ MeV is (the 1st error is statistical, the 2nd systematical):
\begin{equation}
\phi_{CC} = 1.76^{+0.06+0.09}_{-0.05-0.09}, \ \ 
\phi_{ES} = 2.39^{+0.24+0.12}_{-0.23-0.12}, \ \ 
\phi_{NC} = 5.09^{+0.44+0.46}_{-0.43-0.43}
\end{equation}
The fluxes of electron neutrinos, $\phi_{e}$, and of $\num$+$\nut$, $\phi_{\mu\tau}$, are:
\begin{equation}
\phi_{e} = 1.76^{+0.05+0.09}_{-0.05-0.09}, \ \ \phi_{\mu\tau} = 3.41^{+0.45+0.48}_{-0.45-0.45}
\end{equation}
The total flux $\phi_{e} + \phi_{\mu\tau}$ is that expected from the Standard Solar Model.
Combining statistical and systematic uncertainties in quadrature,
$\phi_{\mu\tau}$ is $3.41^{+0.66}_{-0.64}$, which is 5.3$\sigma$ above zero, 
providing evidence for neutrino oscillations $\nu_{e} \rightarrow \nu_{\mu}, \nu_{\tau} \ $
with $\Delta m^{2} \simeq 5.0 \times 10^{-5}$ and tan$^{2} \theta \simeq 0.34$ \cite{sno2}.

\textit{NEUTRINOS FROM STELLAR GRAVITATIONAL COLLAPSES.}\\
Massive stars with masses larger than few solar masses ($m_{\odot}$), evolve gradually as increasingly
heavier nuclei are produced and then burnt at their centers in a chain of thermonuclear
processes ultimately leading to the formation of a core composed of iron and nickel.
Further burning in the shells surrounding the core may make the core mass exceed the
Chandrasekar limit, $m_{Ch} \simeq 1.4$ $m_{\odot}$.
Then the core implodes in a time only slightly longer than the freefall time
(a few ms) and leads to the formation of a neutron star (or of a black hole
if the star is very massive). In our galaxy one expects one such event about every 30 years.
The energy released during a stellar collapse is the gravitational binding energy
of the residual neutron star $\simeq 10^{53}$ ergs = 0.1 $m_{\odot}$, mostly in the form of $\nu$'s
with an average energy of $\sim$ 14 MeV; $4 \times 10^{57}$ $\nu$'s of each species are
emitted. If the collapse occurs at the center of our galaxy, the total neutrino flux at the
Earth would be $\sim 10^{12}$/cm$^{2}$ for each species, emitted in few seconds.

During the collapse, there are three main stages of $\nu$ emission:\\
\noindent i) \textit{Neutronization}: 
$e^{-} + p \rightarrow n + \nue$; only $\nue$ are emitted in a time of few ms.
A large number of multi-MeV photons is also generated; these, interacting in 
the surrounding matter, yield many $e^{+}e^{-}$ pairs, which lead to 
%$e^{+} + e^{-} \rightarrow \nue + \overline{\nue}$ and to 
$e^{+} + e^{-} \rightarrow 2\gamma$, 
followed by $\gamma$'s recreating $e^{+}e^{-}$ pairs.
%(these processes last for a time lightly longer than the freefall time).

\noindent ii) \textit{Matter accretion}:
All types of $\nu$'s are emitted in this phase and in the next one, e.g. 
$e^{+} + e^{-} \rightarrow \nue + \overline{\nue}$.

\noindent iii) \textit{Cooling}:
The neutron star is very hot and is ``cooled'' by thermal neutrino emission from the
``neutrinosphere'', which lasts $\sim$10 s. The bulk of the neutrino luminosity
is emitted during this phase, with $\langle E_{\overline{\nu}_{e}} \rangle
\simeq$ 15 MeV.

All SN neutrinos may be detected via their NC interactions with
electrons, $\nu_{x} + e^{-} \rightarrow \nu_{x} + e^{-}$; the cross-section for this 
process is small; the scattered electrons ``remember'' the direction of the incoming 
neutrinos. The dominant observed reaction is
$\overline{\nue} + p \rightarrow n + e^{+}$, which has $\sigma = 7.5 \times 10^{-44} E_{\nu}^{2}$
(MeV$^{2}$ cm$^{2}$). It is energetically possible on free protons, as in H$_{2}$O
and in C$_{n}$H$_{2n+2}$ detectors. The $e^{+}$ annihilates immediately,
$e^{+} + e^{-} \rightarrow 2\gamma$, while the neutron is captured after a mean time of
$\simeq 200 \mu$s ($n + p \rightarrow d + \gamma$, with $E_{\gamma} \simeq 2.2$ MeV).
D$_{2}$O detectors can also detect $\nue + n \rightarrow p + e^{-}$.
Because of the energy dependence of the neutrino cross-section on $E_{\nu}$, the average
positron energy is $\langle E_{e} \rangle \simeq$ 17 MeV.

The observation of Supernova 1987A in the Large Magellanic Cloud, 170.000 light years
away, in 1987 \cite{sn1987} has given impetus to the search for $\nu$'s from stellar
collapses and together with solar neutrinos opened up the field of \textit{Neutrino Astronomy}.
Bursts of neutrinos were observed by the Kamioka (12 events) and IMB (8 events) proton
decay detectors and by the Baksan (3 events) neutrino telescope; the Mont Blanc detector
(80 t of liquid scintillator) observed a probable signal of 5 neutrinos,
4 hours earlier. The $\nu$'s arrived hours earlier than visible light; they 
arrived all within 10 seconds and the flux was approximately the predicted one.

Many detectors are now capable of yielding information on neutrinos from supernovae
and they are linked in a supernova watch system \cite{watch}.

\textit{HIGH ENERGY MUON NEUTRINO ASTRONOMY.}
TeV $\num$'s are expected to come from several galactic and extragalactic sources, 
see Sect. 3. Neutrino production requires astrophysical accelerators of protons and
astrophysical beam dumps, where pions are produced and decay in $\num$.

A $\num$ interacts in or below a detector yielding the observed muon.
Large detectors are needed to obtain reasonable numbers of events. The detectors are
tracking detectors (MACRO) or water or ice Cherenkov detectors (SK, Baikal, AMANDA).
Future neutrino telescopes will be very large Cherenkov detectors either with 
water (ANTARES, km$^3$, NESTOR) or with ice (AMANDA, Ice Cube). The muon pointing accuracy
may be checked by measuring the location of the Moon and Sun shadows which lead to
a decrease of the number of downgoing muons in their direction \cite{moon}.

Atmospheric neutrinos are the main source of background for these searches. 
Tracking detectors have better angular resolution and use smaller angular
search cones (3$^\circ$), thus reducing the background from atmospheric $\nu$'s.
Larger search cones are used by water and ice detectors.

\begin{figure}[t]
\begin{center}
\epsfclipon
  \mbox{ \hspace{-1cm} \epsfysize=4.5cm
         \epsffile{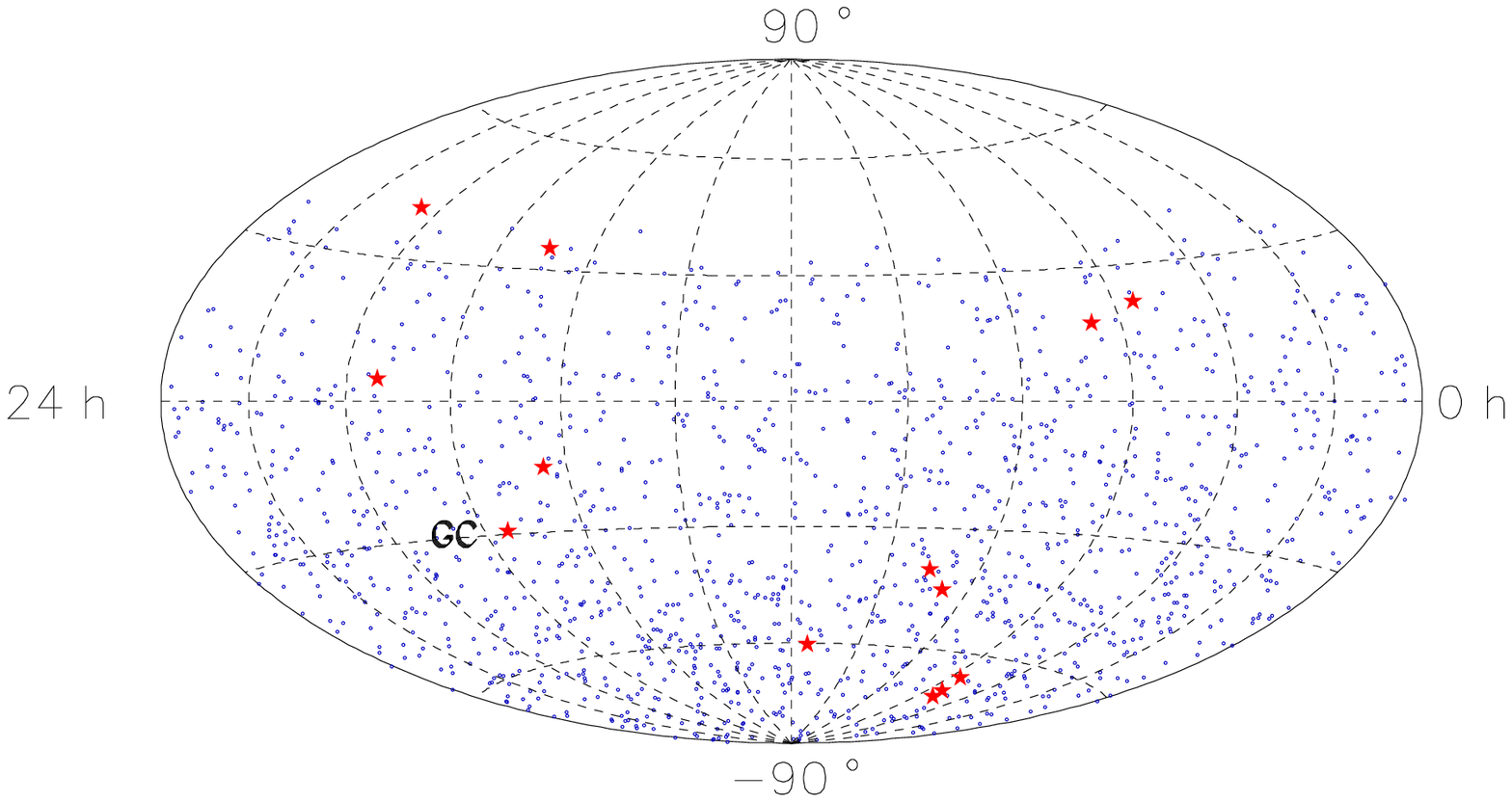} 
         \hspace{-1cm} \epsfysize=6.0cm
         \epsffile{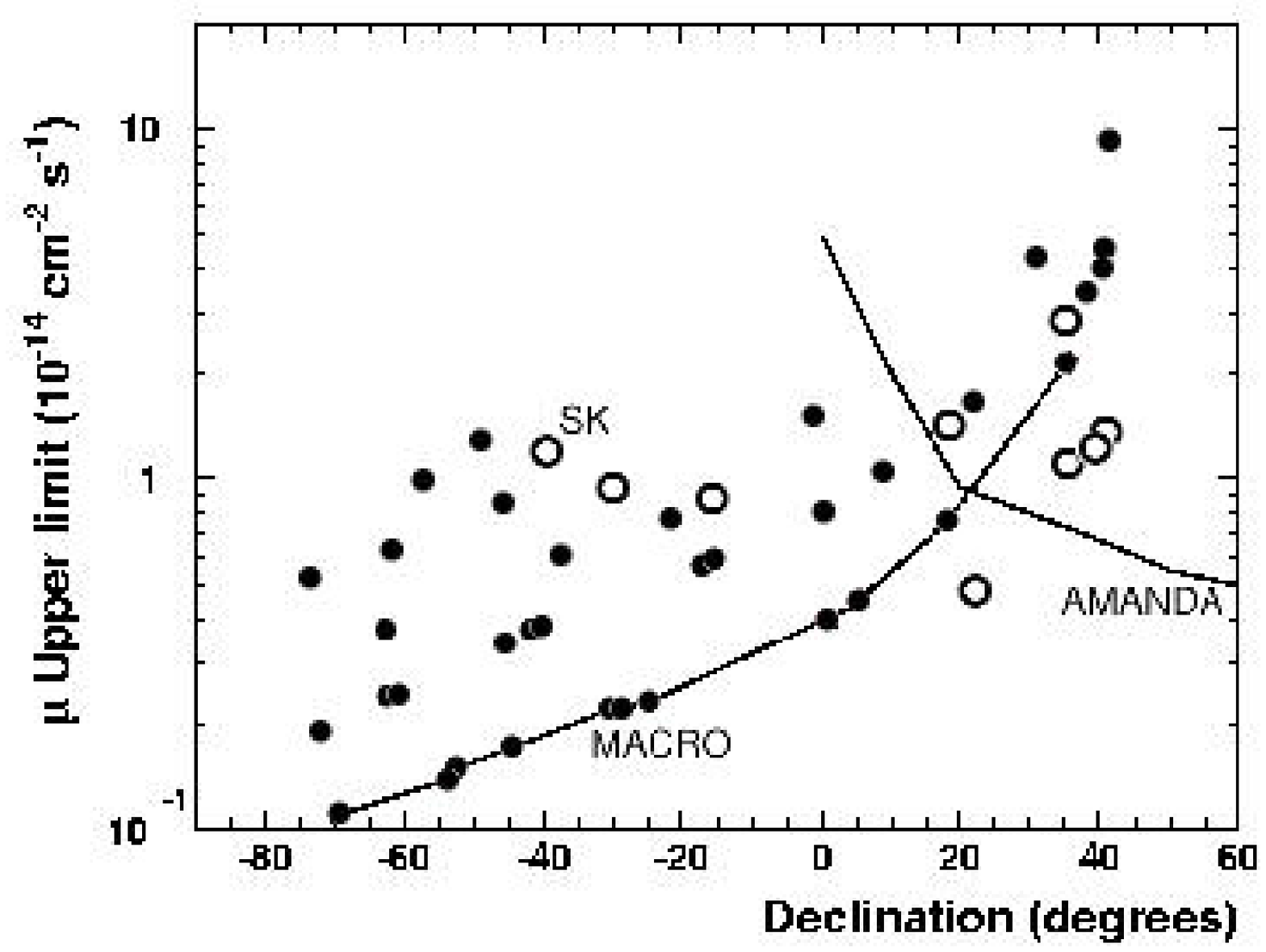} 
         }
{\small \hskip 5.0 truecm (a) \hskip 7.5 truecm (b)}
\epsfclipoff
\end{center}
\caption{HE $\num$ astronomy. (a) MACRO 1356 upgoing muons in equatorial coordinates.
(b) 90\% CL muon flux limits from MACRO (black points) as a function of the declination
for selected sources; the solid line refers to the limits obtained when the atmospheric
$\num$ background was zero or 1 event. The limits from SK (open circles) and
AMANDA (thin line) refer to higher energy $\num$'s.}
\label{f:nuastro}
\end{figure}

An excess of events is searched for around the positions of known sources. 
The 3$^\circ$ angular bins used by MACRO takes into account the angular
smearing produced by the muon multiple scattering in the rock below the detector
and by the energy--integrated angular distribution of the scattered muon, with
respect to the neutrino direction.
Fig. \ref{f:nuastro}a shows the 1356 MACRO events. The 90\% CL upper limits on the muon 
fluxes from specific celestial sources are shown in Fig. \ref{f:nuastro}b \cite{ar2001}. 
The solid line is the
MACRO sensitivity vs. declination. Notice that there are two cases, GX339-4
($\alpha$=255.71$^o$, $\delta$=-48.79$^o$) and Cig X-1 
($\alpha$=230.17$^o$, $\delta$=-57.17$^o$), with 7 events:
in Fig. \ref{f:nuastro}b they are
considered as background, therefore the upper flux limits are higher; but they
could also be indications of signals \cite{nuastro}.

%mettici referenze sulla caption.

MACRO searched for time coincidences of their upgoing muons with $\gamma$--ray bursts
as given in the BATSE 3B and 4B catalogues, for the period from April 91 to December 2000.
No statistically significant time correlation was found.
It also searched for a diffuse astrophysical neutrino flux for which they establish
a flux upper limit at the level of $1.5 \cdot 10^{-14}$ cm$^{-2}$s$^{-1}$ \cite{nuastro}.

\setcounter{figure}{0}\setcounter{table}{0}\setcounter{equation}{0}
\section{Proton decay}
\begin{figure}[t]
  \begin{center}
  \mbox{ \hspace{-0.5cm} \epsfysize=5.2cm
         \epsffile{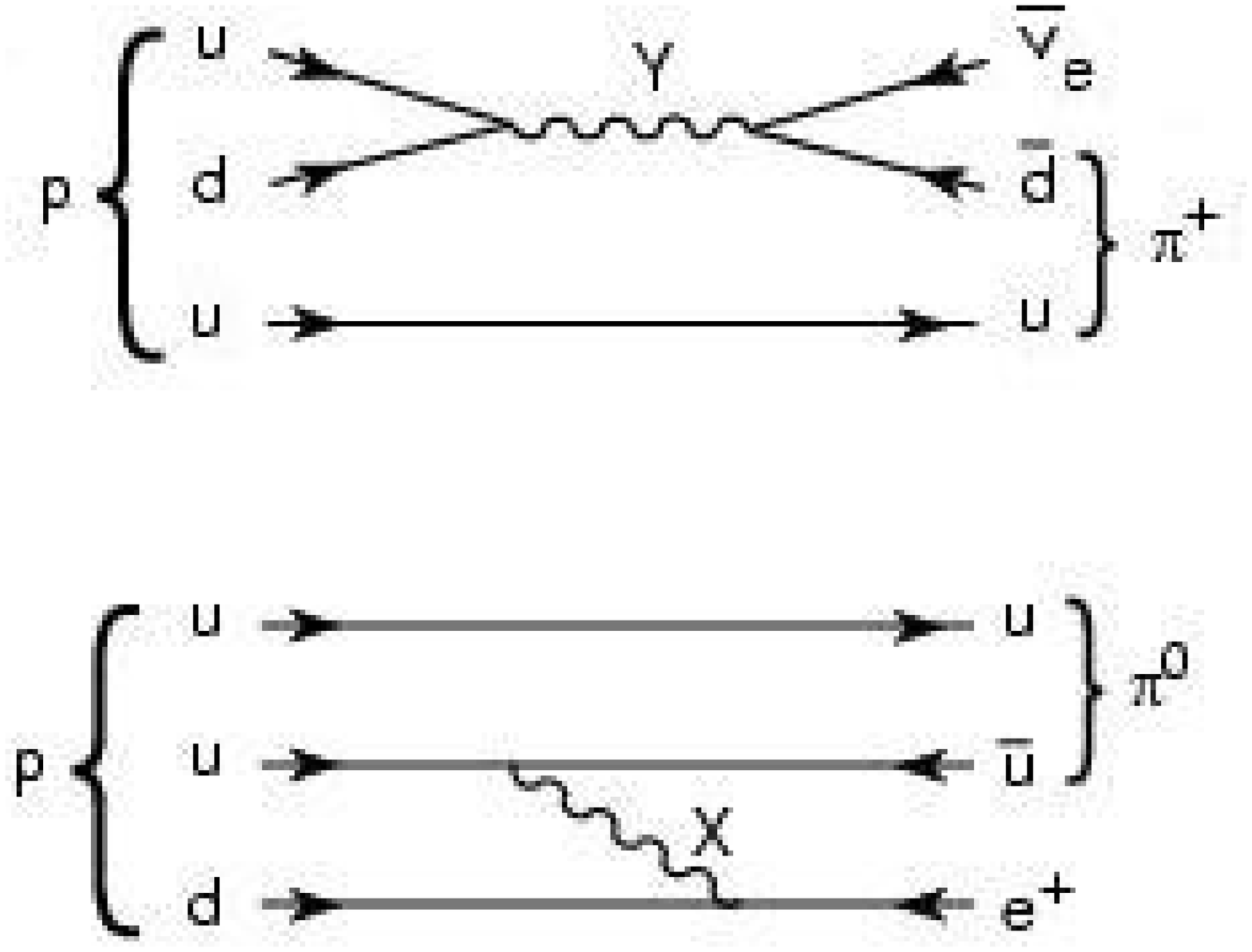} 
         \hspace{0.0cm}  \epsfysize=8.0cm
         \epsffile{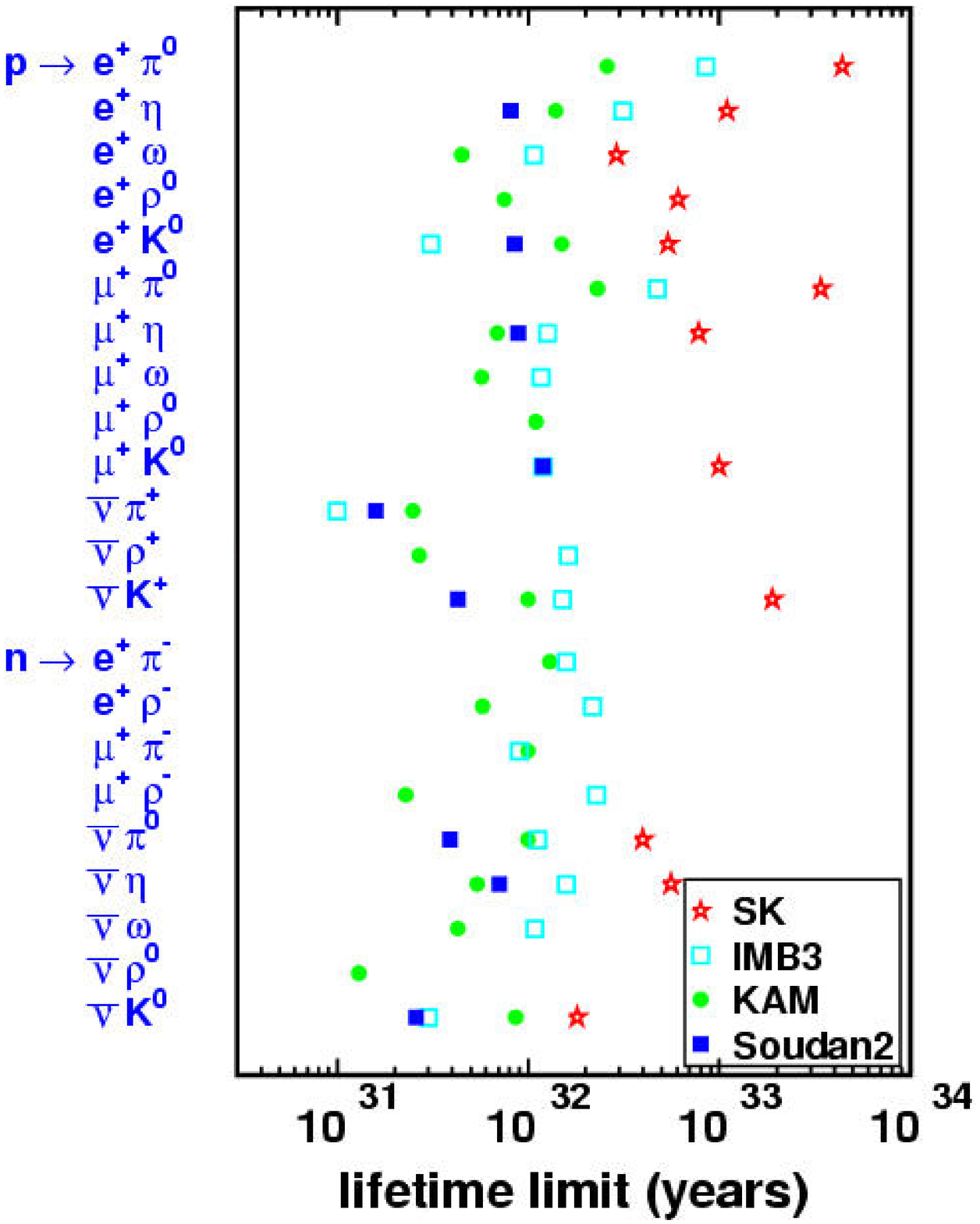} 
         }
{\small \hskip 5.0 truecm (a) \hskip 7.5 truecm (b)}
\end{center}
\caption{
(a) Proton decay, $p\rightarrow e^{+}\pi^{0}$ and $p\rightarrow \overline{\nue} \pi^{+}$, mediated
by the GUT X and Y bosons with electric charges 4/3 and 1/3, respectively.
(b) Present limits on selected proton decay modes.
For $p\rightarrow e^{+}\pi^{0}$, $\tau/BR > 4 \times 10^{33}$ y; 
for $p \rightarrow \bar{\nu}K^{+}$ (preferred in some supersymmetric models), 
$\tau/BR > 6.7 \times 10^{32}$ y [1.5, 6.3].}
\label{f:pdecay}
\end{figure}

In the SM of particle physics quarks and leptons are placed in separate doublets
and baryon number conservation forbids proton decay. However there is no known gauge symmetry which generates
baryon number conservation. Therefore the validity of baryon number
conservation must be considered an experimental question. 
GUT theories place quarks and leptons in the same multiplets; 
thus quark $\longleftrightarrow$ lepton transitions are possible: they are mediated by massive X,Y bosons
with electric charges of 4/3 and 1/3, respectively, see Fig. \ref{f:pdecay}a.
Sakharov suggested that in the early Universe at the end of the GUT phase transition the matter--antimatter
asimmetry was created with violations of CP and of baryon number \cite{sak}.

The mass scale of GUT theories depends on the extrapolation to very high energies of the inverse
of the 3 fundamental dimensionless coupling constants: $\alpha_{s}^{-1}$ for the strong interaction,
$\alpha_{U(1)}^{-1}$ and $\alpha_{SU(2)}^{-1}$ for the electroweak interaction. The simplest extrapolation
leads to an approximate common value at mass scales of $10^{13} \div 10^{15}$ GeV; 
the addition of a supersymmetry scale at $\sim$1 TeV leads to a better Grand Unification at 
$\sim 10^{16}$ GeV. But there are other possibilities which may yield lower mass scales \cite{lowmass}.
The simplest GUT model, SU(5), leads to a value of $\tau_{p}/B_{p\rightarrow e^{+}\pi^{0}} \sim 10^{30}$ years,
which is inconsistent with the measured limit 
$\tau_{p}/B_{p\rightarrow e^{+}\pi^{0}} > 4 \times 10^{33}$ y.
The SO(10) GUT leads to a considerably longer lifetime ($10^{32}\div 10^{33}$);
even larger values ($10^{32}\div 10^{39}$) are predicted by other GUT models.

The search for proton decay was the main reason for developing large underground detectors
(water Cherenkovs and tracking calorimeters).
%Water detectors have large masses, many free protons and yield the sense of the
%track direction. Tracking calorimeters have a higher spatial resolution and a better $\pi/\mu$ separation
%at energies of $\sim$200 MeV.
The main background in these detectors comes from low energy atmospheric neutrinos. Most of this background
can be eliminated selecting contained events and by topological and kinematical contraints.

\small
The expected number of events per year in a decay channel with branching fraction B is
$N = f \cdot N_{N} \cdot M_{F} \cdot T \cdot B/\tau \cdot \epsilon$
where $f$ is the nucleon fraction ($f=p/(p+n)$ or $f'=n/(p+n)$), $N_{N}$ is the number
of nucleons in 1 kt mass ($\sim 6\cdot 10^{23} \cdot 10^{9}$), $M_{F}$ is the fiducial mass (in kt), 
$T$ is the livetime (in years) and $\epsilon$ is the overall detection efficiency.
The decay mode $p\rightarrow e^{+}\pi^{0}$ is essentially background free: one has 
$\tau_{p}/B > 2.6 \cdot 10^{32} \cdot f \cdot M_{F} \cdot T \cdot \epsilon$. For 
decay modes with sizable background one has
$\tau_{p}/B > 4 \cdot 10^{32} \cdot f \cdot \epsilon \sqrt{M_{F}T/R_{b}}$,
where $R_{b}$ is the background rate. 
\normalsize
\vspace{0.2cm}
\par Fig \ref{f:pdecay}b shows present lower limits for many proton decay modes [1.5, 6.3].

\textit{$n\overline{n}$ OSCILLATIONS.}
If $\Delta{B}$=1 processes are allowed one should also expect $\Delta{B}$=2 processes,
like $n \rightarrow \overline{n}$ transitions, often called neutron--antineutron
oscillations. They have been searched for in a beam of thermal neutrons from a nuclear reactor
impinging in a large calorimeter, where a $\bar{n}$ would annihilate and could readily
be observed. Very cold neutrons were used and special care was taken to have zero magnetic fields. 
The probability $P$ of producing $\overline{n}$ in a $n$ beam, in vacuum and in the absence
of external fields, depends on the observation time $t$ as $P=(t/\tau_{n\overline{n}})^2$
where $\tau_{n\overline{n}}$ is the characteristic $n \rightarrow \overline{n}$ transition time.
The present limit for free neutrons is $\tau_{n\overline{n}} > 1.2 \times 10^{8}$ s [1.5, 6.4].
The Soudan 2 underground experiment estabilished a similar limit for neutrons bound in iron nuclei.
It is $\tau_{n\overline{n}} > 1.3 \times 10^{8}$ s \cite{chung}. The experiment is
background limited (from atmospheric neutrinos).

\setcounter{figure}{0}\setcounter{table}{0}\setcounter{equation}{0}
\section{Neutrinoless Double Beta decay}
\begin{figure}
\begin{center}
\vskip -2cm
  \epsfclipon
  \mbox{\epsfysize=9.0cm \epsffile{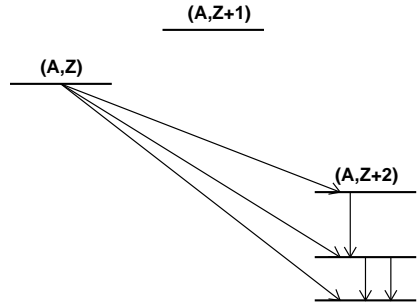}}
  \epsfclipoff
\end{center}
\vskip -3cm
\caption{Energy levels for nuclei considered for double beta decays.}
\label{f:levels}
\end{figure}

\begin{figure}[b!]
  \begin{center}
  \epsfclipon
  \mbox{ \hspace{-0.5cm} \epsfysize=6.0cm
         \epsffile{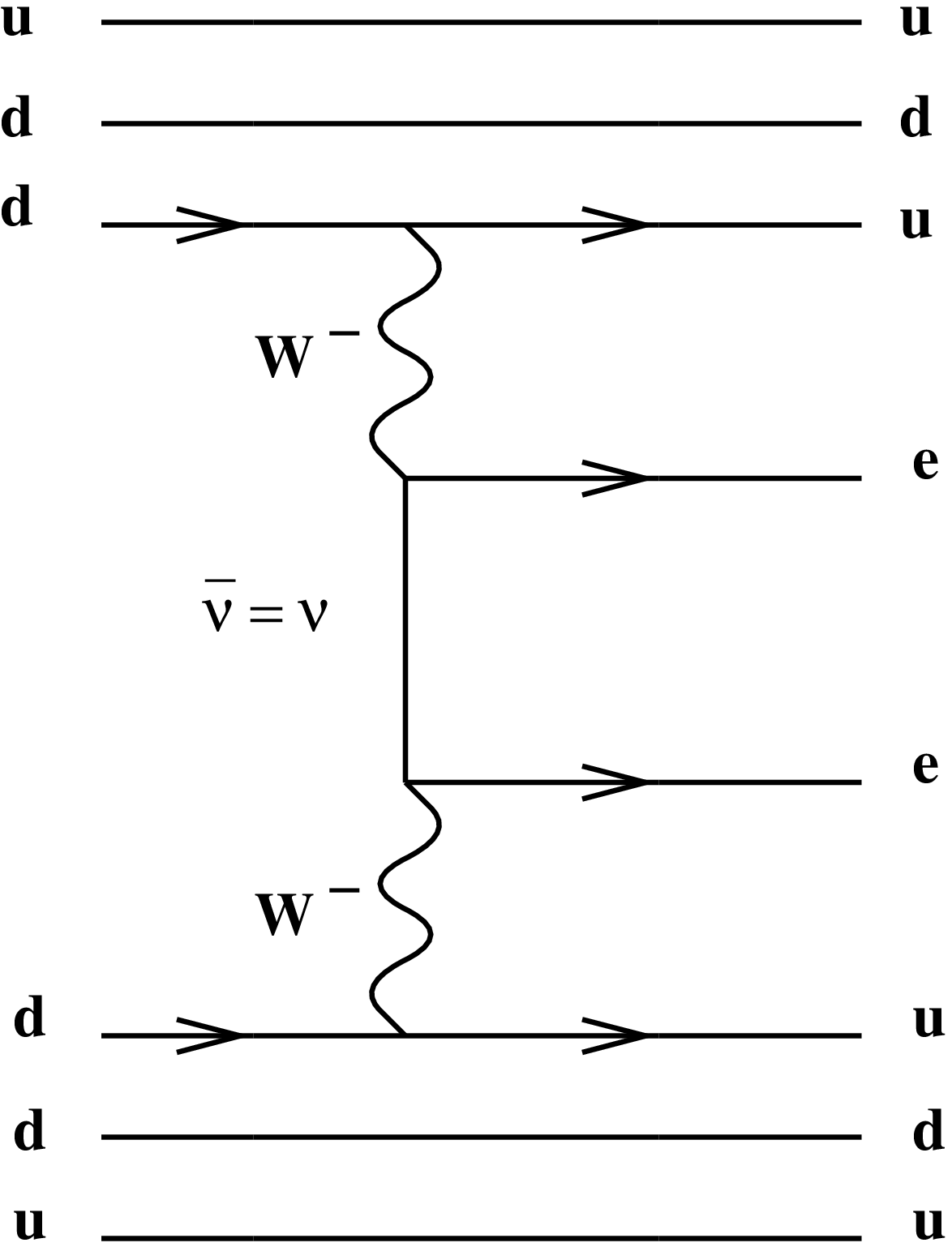} 
         \hspace{0.0cm}  \epsfysize=5.0cm
         \epsffile{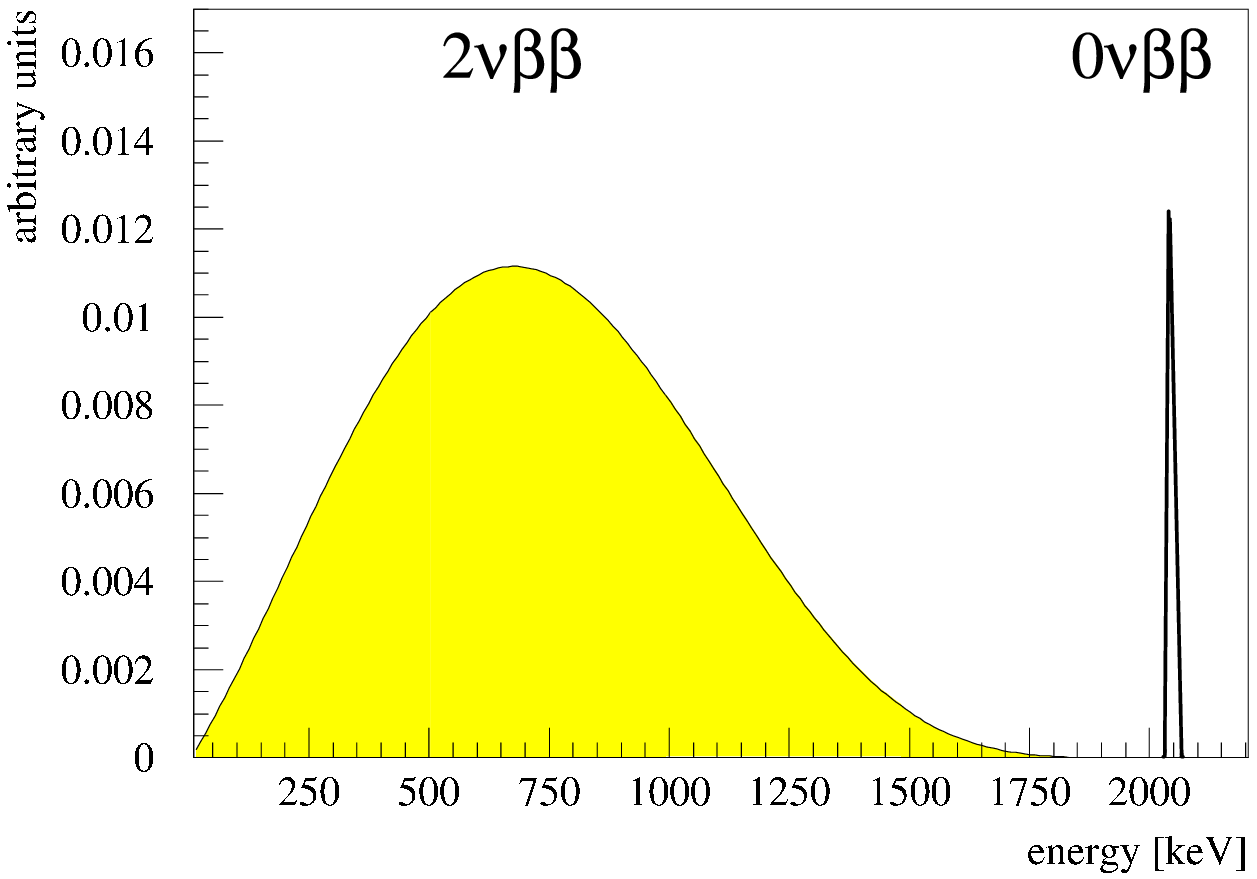}
         }
{\small \hskip 5.0 truecm (a) \hskip 7.5 truecm (b)}
  \epsfclipoff
\end{center}
\caption{
(a) Feynman diagram for neutrinoless double $\beta$ decay;
(b) Energy spectra for the sum of the two electron energies in double $\beta$ decay.}
\label{f:ndbd}
\end{figure}

For some even--even nuclei the decay chain 
\begin{eqnarray*}
        & (A,Z) \rightarrow & (A,Z+1) + e^{-} + \overline{\nu}_{e} \\
        & & \ \ \ \ \ \ \ \hookrightarrow (A,Z+2) + e^{-} + \overline{\nu}_{e}
\end{eqnarray*}
is forbidden by energy conservation (Fig. \ref{f:levels}). For some nuclei 
one could have $(A,Z) \rightarrow (A,Z+2)$ in a single step in 3
different ways:
\begin{eqnarray}
 & & (A,Z) \rightarrow (A,Z+2) + 2e^{-} + 2\overline{\nu}_{e}  \label{duenu}   \\
 & & (A,Z) \rightarrow (A,Z+2) + 2e^{-} + X^{0}                \label{majoron} \\
 & & (A,Z) \rightarrow (A,Z+2) + 2e^{-}                        \label{nuless}
\end{eqnarray}
The first mode \ref{duenu} is allowed but is very rare because it is a second
order CC weak interaction. The third decay (see Fig. \ref{f:ndbd}a), 
called \textit{neutrinoless double $\beta$ decay},
violates the conservation of the electron lepton number and would require 
$\nu_{e} = \overline{\nu}_{e}$ and non zero neutrino mass (\textit{Majorana neutrino}).
The second decay \ref{majoron} is highly hypothetical; it would require that the Majorana neutrino
would emit a new boson, called a \textit{Majoron}.

The energy spectra of the sum of the energies of the two emitted electrons, $E=E_1+E_2$, is different 
in the three cases (Fig. \ref{f:ndbd}b): it would be a line corresponding to the maximum energy for 
the decay (Eq. \ref{nuless}), a continuous spectrum peaked at low values of $E$ for (\ref{duenu})
and continuous spectra for the various Majoron--accompanied modes (Eq. \ref{majoron}), classified by 
their spectral index. Until now the experiments have found few events corresponding to the first case 
and none for the other two cases.

Double $\beta$ decay experiments are very refined, like those for direct DM searches (see Section 8).
They are placed underground and efforts are made to reduce all sources of environmental radioactivity 
(including neutrons) using extra shielding with radiopure elements. All experiments are
increasing their detector masses and their complexity.
Some groups use materials which are at the same time source and detector, for example $^{76}Ge$. 
The Heidelberg-Moscow experiment at Gran Sasso, using some kg of highly enriched $^{76}Ge$ 
($^{76}Ge \rightarrow ^{76}Se + 2e^{-}$), obtained $t_{1/2} > 10^{25}$y \cite{mancante1}.
The Milano group used a cryogenic $TeO_{2}$ 
detector obtaining $t_{1/2} > 5.6 \times 10^{32} y$ \cite{mibeta}.
For further discussions see references \cite{fiorini,aalseth} and other references in Section 8.

Future developments include the use of cryogenic detectors with the simultaneous
measurement of temperature variations and of $dE/dX$ (via ionization or excitation).

\setcounter{figure}{0}\setcounter{table}{0}\setcounter{equation}{0}
\section{Dark matter and dark energy}

\begin{figure}[t]
\begin{center}
\epsfclipon
\mbox{\epsfysize=6.0 cm \epsffile{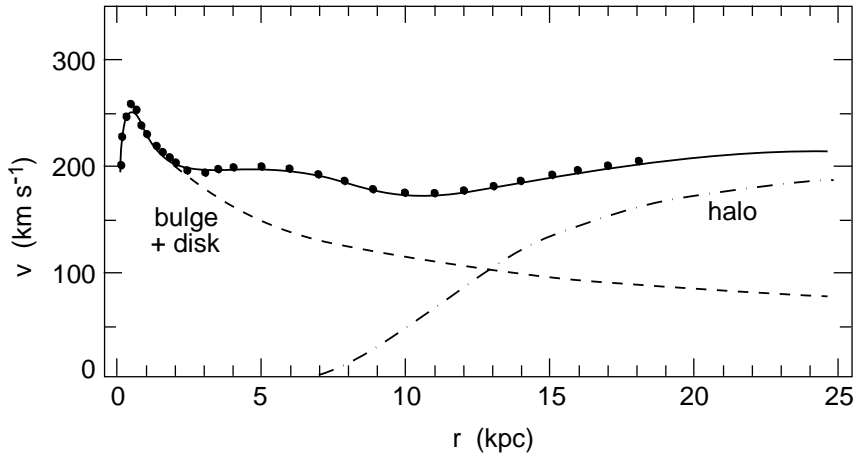}}
\epsfclipoff
\end{center}
\caption{DM in galactic halos. The black points are measured rotation velocities
of the stars in the spiral galaxy NGC3198, plotted vs the distance $r$ from the galactic
center. The dashed line is the visible contribution of the galactic bulge+disk; the dashed--dotted line
is the contribution of the DM halo. In our Galaxy the Sun is at $\sim$ 8 kpc from our galactic center: 
it has a velocity of 220 km/s; it would have a velocity of 160 km/s in the absence
of the DM halo.}
\label{f:ottouno}
\end{figure}

The matter content of galaxies may be estimated from the number of visible stars, assuming
an average star mass. The rotation curves of stars in spiral galaxies give evidence
for (non luminous) Dark Matter (DM), see Fig. \ref{f:ottouno}; 
the galactic DM should be more
than 10 times the visible matter, unless something is wrong with the gravitational
theory. The study of the dynamics of groups of galaxies indicate again the presence of DM,
about 60 times more abundant than the luminous matter. The study of $x$--ray emission
by clusters of galaxies indicates the presence of an ionized gas where fast electrons
give bremsstrahlung radiation ($x$--rays). The mass of the ionized gas is estimated
to be $>$5 times that of luminous matter. Astronomical informations yield
the visible, DM and dark energy given in Table \ref{t:dark}.

\begin{table}
        \begin{center}
                \begin{tabular}{ll}
                \hline
                Type  &  \hspace{4.2cm} $\Omega$ \\
                \hline
                Visible (luminous) baryonic matter    & \hspace{4.2cm}  $\Omega_{vis} \simeq 0.005$   \\
                Non luminous baryonic matter          & \hspace{4.2cm}  $\Omega_{b} \simeq 0.045$     \\
                Non baryonic Dark Matter              & \hspace{4.2cm}  $\Omega_{DM} \simeq 0.30$     \\
                (of which neutrinos                   & \hspace{4.2cm}  $\Omega_{\nu} \simeq 0.003 - 0.1$)  \\
                Quintessence (Dark Energy)            & \hspace{4.2cm}  $\Omega_{Q} \simeq 0.65$      \\
                \hline
                Total                                 & \hspace{4.2cm}  $\Omega_{tot} \simeq 1$       \\
                \hline
                \end{tabular}
        \end{center}
        \caption{Visible, baryonic and non-baryionic DM, and dark energy contributions 
        to the energy content of the Universe expressed in terms of $\Omega=\rho/\rho_{critical}$;
        $\rho_{critical} \simeq 3H^2/8\pi G_N \simeq 1.9 \times 10^{-29} h^0 (g cm^{-3})$ is
        the density of a flat euclidean universe. Massive neutrinos may contribute to the 
        non-baryonic DM ($\Omega_{\nu}$ at most 0.1) and may concentrate
        around large clusters of galaxies.}
  \label{t:dark}
\end{table}

The search for DM and the understanding of its nature is one of the
central problems of physics, astronomy and cosmology. Without DM it
is difficult to reconstruct the history of the Universe. 
Most of the DM should be in the form of particles in the halos of galaxies and 
of groups of many galaxies. Particle physics provides many candidates
for \textit{cold} DM:
they are globally called WIMPs (Weakly Interacting Massive Particles)
and would move with $\beta \sim 10^{-3}$.
The best example is the \textit{neutralino}, the lightest 
supersymmetric particle; it is a linear combination of the photino, the zino and
of two neutral higgsinos. 
But one also needs fast particles with small masses, like the neutrinos and the
axions; they are relativistic and constitute the \textit{hot DM} and would
be located mainly in the halos of clusters of galaxies \cite{bottino}.

\textit{MACHOS.} Candidates for non luminous baryonic matter are planets like
Jupiter, brown dwarfs (stars too cold to radiate), cold gas clouds, stellar
remnants (white dwarfs, neutron stars, stellar mass black holes), mid-mass and
very massive black holes, etc. Some of these objects (the \textit{MACHOs}), have been searched 
for using \textit{gravitational lensing} techniques \cite{machos}. Optical telescopes
have been looking night after night, at stars in the Large Magellanic Cloud, 
a small satellite galaxy of our Galaxy. 
The search was for intensity variations of the stars, in
different colours. If a MACHO in the halo of our galaxy would pass in front of the stars,
in the line of sight star-Earth, one should observe, by gravitational
lensing, an increase of light of these stars, in all colours, which would last
for few days. Several events were observed, after removing
the ``background'' due to variable stars; their number 
can only account for a fraction of the non visible baryonic matter.

\textit{DIRECT DM SEARCHES.} They are based on WIMP--nucleus 
collisions inside a detector and on the observation of the nuclear recoil in the
keV energy range via ionization or scintillation; cryogenic detectors can also measure
the temperature variation (via phonons); many experiments use thermal/scintillation
and thermal/ionizations detectors, thus reducing the background.
The following list tries to classify the experiments:\\
- Scintillation detectors           (Expts DAMA $\rightarrow$ LIBRA, NaI, NAIAD, ANAIS, SACLAY, 
Ca$F_{2}$, ELEGANTS, ...)\\
- Ge ionization detectors           (Expts IGEX, COSME, HM, HDMS, GENIUS, ...)\\
- Thermal detectors                 (Expts MIBETA, CUORICINO, CUORE, ...)\\
- Thermal + scintillation detectors (Expts ROSEBUD, CRESST, ...)\\
- Thermal + ionization detectors    (Expts EDELWEISS, CDMS, ...)

Many DM experiments were originally double $\beta$ decay experiments, see Section 7. 
Cryogenic technology should still have large improvements. 
The experiments are too numerous to summarize properly. 
Here we shall recall only a few and refer to recent summaries \cite{dmexp} and to the Appendix.

\begin{figure}
\begin{center}
\epsfclipon
\mbox{\epsfysize=6.1cm \epsffile{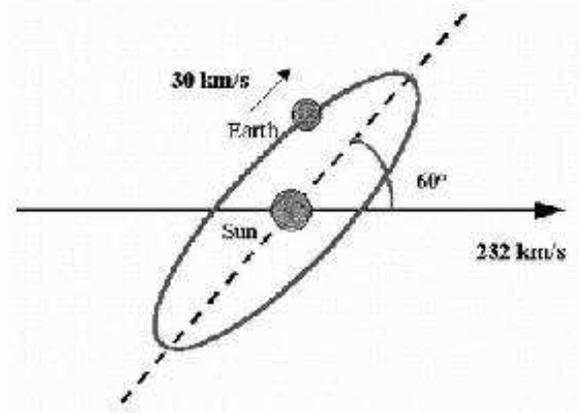}}
\epsfclipoff
\end{center}
\vspace{-0.5cm}
\caption{Sketch of the motion of the Earth around the Sun and of the solar
system in our galaxy in the WIMP ``wind''.}
\label{f:ottodue}
\end{figure}

\begin{figure}
\begin{center}
\epsfclipon
\mbox{\epsfysize=5.5cm \epsffile{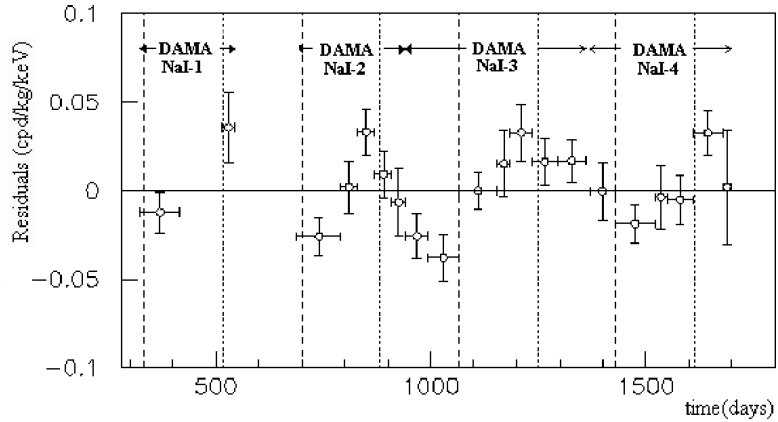}}
\epsfclipoff
\end{center}
\vspace{-0.5cm}
\caption{Residual counting rate vs time for the DAMA experiment \protect\cite{bernabei}.}
\label{f:ottotre}
\end{figure}

\begin{figure}
\begin{center}
\epsfclipon
\mbox{\epsfysize=8.0cm \epsffile{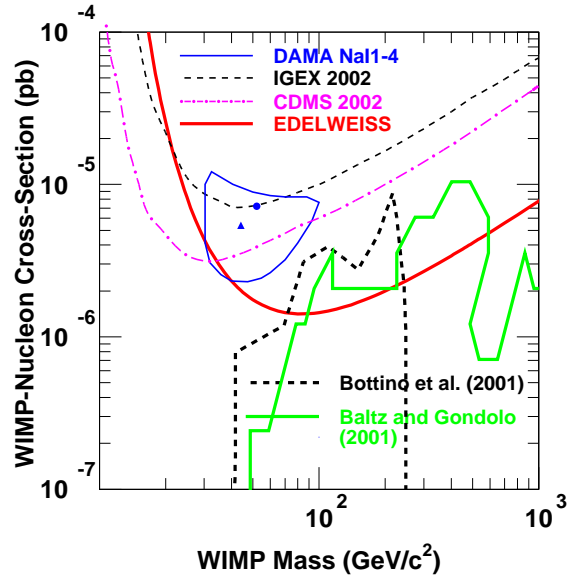}}
\epsfclipoff
\end{center}
\caption{Direct DM searches. Cross-section vs WIMPs mass. Allowed region by the DAMA experiment
and limits by the IGEX, CDMS and EDELWEISS experiments \protect\cite{igex,cdms,edelweiss}.
Also shown are the predictions of some supersymmetric models \protect\cite{bottino,baltz}.}
\label{f:ottoquattro}
\end{figure}

The DAMA experiment at Gran Sasso uses $\sim$100 kg of NaI(Tl) scintillators, each about
9.7 kg and each ``seen'' by two PMTs in coincidence \cite{bernabei}. 
It searches for WIMPs by the so called annual modulation signature.
The WIMPs should have a maxwellian velocity distribution in the galactic halo (with a
cut--off at the galactic escape velocity). Thus a WIMP ``wind'' would hit continuously
the Earth. Since the Earth rotates around the Sun, it should be invested by a larger
wind flux in june, when its rotational velocity adds up to the velocity of the
solar system in the galaxy; the flux should instead be smaller in december
when the two velocities are in opposite directions, see Fig. \ref{f:ottodue}
(there should also be a small daily effect caused by the rotation of the Earth).
DAMA observed a modulation for events with nuclear recoil energies
between 2 and 6 keV, Fig. \ref{f:ottotre}, and checked that it did not come from CR seasonal
variations \cite{ambrosio}.
If WIMPs are interpreted as neutralinos, DAMA favours a mass
$m_{\chi} \sim 59$ GeV and gives the 90 \% CL allowed region shown in Fig. \ref{f:ottoquattro}.

The CDMS experiment at Stanford uses ultra pure Si and Ge cristals 1 cm thick and 7.5 cm in diameter 
at a temperature of 20 mK. It measures phonon and ionization signals from the nuclear recoil. 
The energy resolution is 850 eV and 400 eV for the ionization and
phonon channels, respectively, the trigger threshold is at 3 keV \cite{cdms}.
Their upper limits vs WIMP mass are shown in Fig. \ref{f:ottoquattro}.

The HDMS ionization detector at Gran Sasso uses enriched $^{73}Ge$ \cite{hdms}.
The EDELWEISS experiment in the Frejus tunnel uses a cryogenic $Ge$ detector
in which heat and ionization are measured \cite{edelweiss}.
Their negative result is shown in Fig. \ref{f:ottoquattro}, which summarizes
the present situation for direct DM searches:
the experiments have sensitivities to cross-sections of $\sim 10^{-6}$ pb.
Future experiments should reach sensitivities about 10 times smaller.

%The Heidelberg--Moscow (HDMS) experiment at Gran Sasso uses 10.9 kg of highly enriched 
%$^{73}Ge$ detectors. A 2.3 kg prototype yielded the limits quoted in Fig. 8.4. 
%When fully operational it should be able to reach a sensitivity comparable to that of CDMS.

The CRESST experiment at Gran Sasso measures phonons and scintillation light \cite{cresst}.
The detectors consist of dielectric sapphire ($Al_{2}O_{3}$) cristals with superconducting 
tungsten film, which at a temperature of 15 mK functions as a sensitive thermometer: a small 
change in its temperature changes its resistance which is measured with a SQUID.
A separate detector measures the light emitted by the scintillating target. 
It has an energy threshold of 500 eV.

\textit{INDIRECT SEARCHES FOR WIMPS.} 
WIMPs could be intercepted by celestial bodies, slowed down and trapped in their centers, where
WIMPs and anti--WIMPs could annihilate yielding pions which decay into GeV-TeV neutrinos;
the $\nu$'s interact below or inside detectors yielding upgoing muons.
The MACRO experiment searched for upgoing $\num$ coming from the center of
the Earth, using various search cones
(10$^\circ$ --15$^\circ$) around the vertical. No signal was observed;
conservative upper limits were set at the level of $0.5 \times 10^{-14}$ cm$^{-2}$s$^{-1}$ \cite{wimp}.
If the WIMPs are identified with the smallest mass neutralinos, these limits may be used to constrain 
the stable neutralino mass, following the model of Bottino et al., and probably 
excluding $\sim$1/2 of the DAMA allowed region. A similar procedure was used to search for
$\num$ from the Sun [8.5, 2.5]. The AMANDA Coll. recently published the
results of $\num \rightarrow \mu$
from the center of the Earth \cite{amanda2}.

\textit{WARM DARK MATTER. THE AXION.}
The axions have been proposed by theories that try to explain why the weak interaction violates CP
while the strong interaction does not. Axions could be produced in the interior of the Sun, where
photons scatter from electrons and protons. If axions exist, the solar axions should be as common as solar
neutrinos, but they have energies much lower than those of solar neutrinos, and are thus difficult
to detect. The CAST experiment at CERN uses a strong magnetic field (9 Tesla) to convert the solar
axions into $x$-rays. It uses one 10 m long magnet planned for the Large Hadron Collider. Inside the
strong magnetic field there are two vacuum pipes, each of which will view the Sun with $x$-ray detectors
mounted at both ends. The ``telescope'' (the two pipes with detectors) 
will view the Sun through one end when it rises
and trough the other end when it sets. The experiment is presently being commissioned \cite{cast}
(see this reference for a list of previous experiments).

\begin{figure}
\begin{center}
\epsfclipon
\mbox{\epsfysize=10.5cm \epsffile{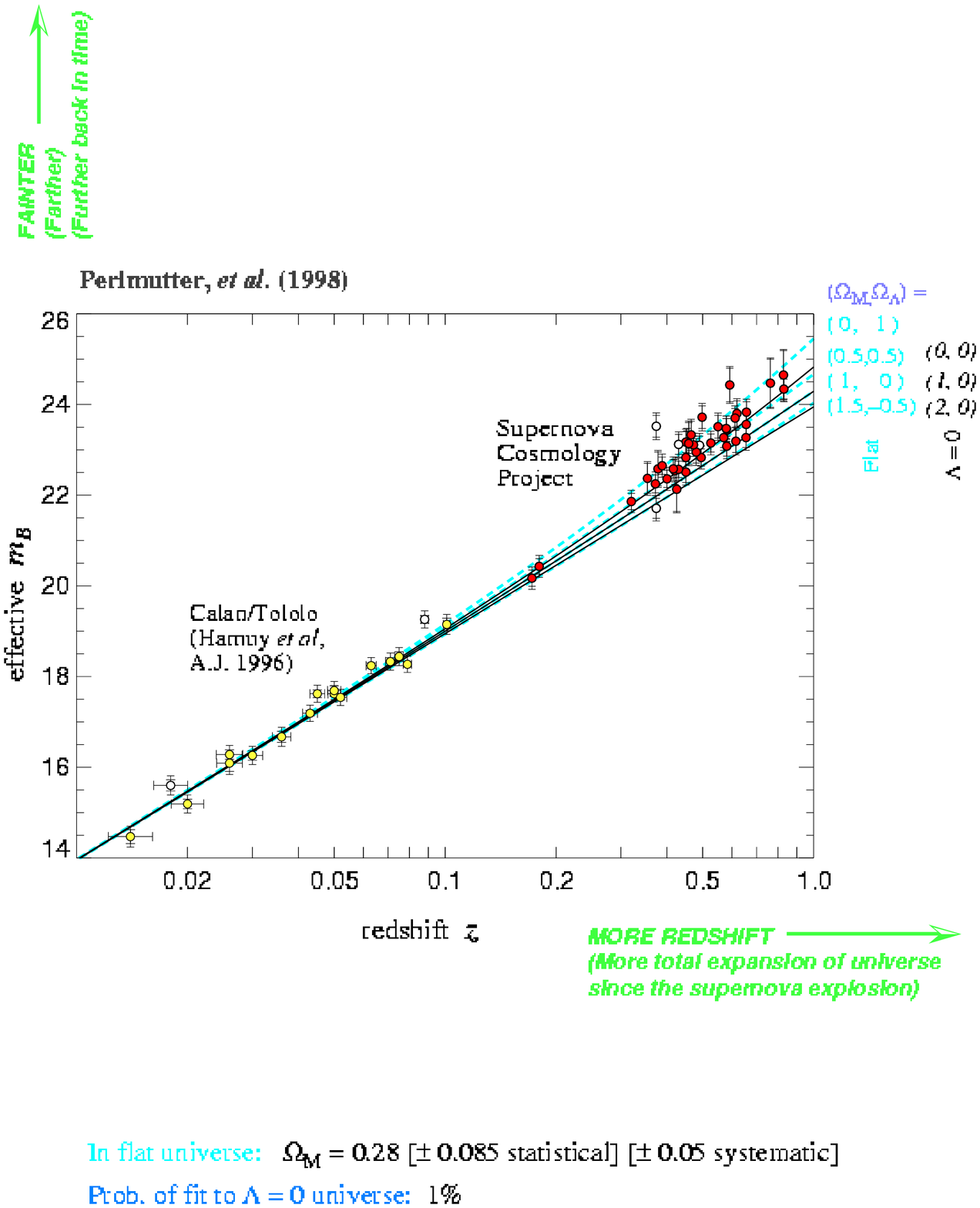}}
\epsfclipoff
\end{center}
\caption{Effective magnitude vs redshift for 42 High-Redshift Type 1a Supernovae.}
\label{f:ottocinque}
\end{figure}

\textit{DARK ENERGY. QUINTESSENCE.}
In 1998 the International Supernova Cosmology Project (ISCP) and the High-z Supernova Search Team (HSST) 
announced their results on the study of far away type 1a Supernovae (SN1a) \cite{perl}. 
These supernovae have for about one month luminosities comparable to that
of a whole galaxy: they are thus observable even if at large distances. Moreover their intrinsic
luminosity is the same for all SN1a. They are probably caused by a white dwarf 
star which increases its mass from the infall mass from a companion star till when the mass of the
white dwarf reaches 1.4 solar masses triggering a gravitational collapse. 
These supernovae can be used as standard candles of intrinsic intensity $I_0$. 
If they are at a distance $l$ from us the observed luminosity will be 
$I=I_0/l^2$, and one obtains the distance $l$ from a measurement of the observed luminosity $I$. 
The recession velocity can be obtained from a measurement of the Doppler shift towards the red 
(thus obtaining the $z$ parameter, defined as $z \equiv (\lambda_{obs} - \lambda_{0})/\lambda_{0}$,
where $\lambda_{obs}$ and $\lambda_{0}$ are the observed and emitted wavelengths respectively).
The measurements are difficult because SN1a are rare (a few per millenium in a Galaxy): one has
to plan a strategy for observation, which requires powerful telescopes on Earth and the Hubble 
telescope in space.
Several dozens SN1a have been measured at high $z$. The goal of these studies was to measure changes in the 
expansion rate of the Universe. One expected that, because of gravity attraction, the rate would
decrease with time, but the results indicate the opposite: the Universe expansion is accelerating.
Fig. \ref{f:ottocinque} shows the observed luminosity for these supernovae vs their red shift. 
Notice that at high $z$
the luminosity of the supernovae is above the expected value 
(they are dimmer, i.e. they have a magnitude $m_{B}$ larger, than would have been 
expected if the Universe expansion were slowing down under the influence of
gravity; they must thus be located farther
away than would be expected for a given redshift $z$; this can be explained if the expansion rate
of the Universe is accelerating.
\par A new project (SuperNova Acceleration Probe) would use a dedicated space telescope equipped with
a large CCD: it should allow the study of many more SN1a and also at larger distances.
\par Which is the force which accelerates the expansion? It may be a sort
of \textit{dark energy}, which is equivalent to a repulsive force. An estimate of the energy 
density of the dark energy is given in Table \ref{t:dark}.

Einstein introduced in 1916 the cosmological constant $\lambda$ in order that his general relativity
equations could allow a static Universe. Now $\lambda$ is useful to explain an accelerating Universe.
But the value of $\lambda$ is much smaller than what could be obtained from the Standard Model of
Particle Physics. Further measurements and further theoretical work are required to really understand
the situation.

\setcounter{figure}{0}\setcounter{table}{0}\setcounter{equation}{0}
\section{Cosmic microwave background radiation (CMB)}
\begin{figure}
\begin{center}
\epsfclipon
\mbox{\epsfysize=6.6 cm \epsffile{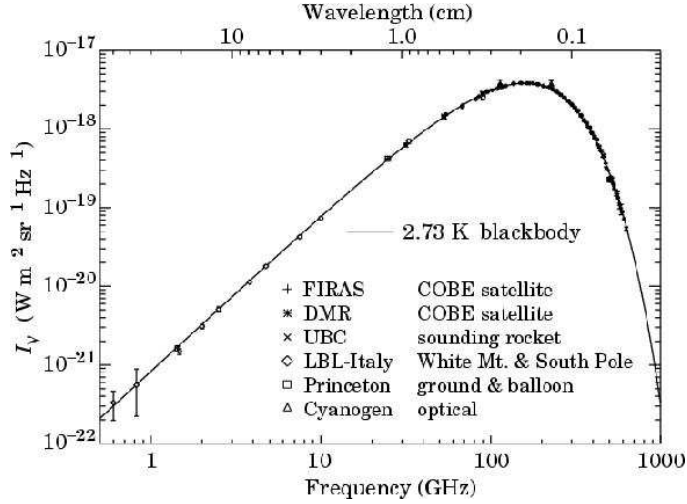}}
\epsfclipoff
\end{center}
\caption{Frequency spectrum of the CMB radiation for frequencies between 0.6 and 1000 GHz
(wavelengths $5 \div 0.05$ cm), measured by different experiments. The solid line is the prediction
of a black body spectrum with $T \simeq$ 2.73 K.}
\label{f:blackbody}
\end{figure}

%The inflation model of the Early Universe (at $t \sim 10^{-34}$ s) predicts peaks in the distribution
%of the Cosmic Microwave Background (CMB) power plotted versus the angular size of regions contributing to the CMB.
%The Boomerang experiment, installed in a balloon and exposed for about 10 days at 40 km heigth at the
%south pole, was the first experiment to reach angular resolutions of 1$^{o}$. 
%They observed the first peak at an angle at about 1 degree; it was confirmed by the MAXIMA Collaboration.
%Later Boomerang and the DASI (Degree Angular Scale Interferometer), using a detector on the roof of the
%South Pole Station, presented evidence for a second and a third peak.
%According to the inflation model the location of the first peak yields the total energy of the Universe,
%expressed as a fraction of the critical density; one has $\Omega_{total}$=1.03$\pm$0.06.
The CMB Radiation was discovered accidentally in 1964 by Penzias and
Arno at a Bell Telephone Lab. with a millimetric antenna used for satellite communications \cite{cmb}.
The CMB is a microwave e.m. radiation which arrives on Earth from every direction. Fig. \ref{f:blackbody}
shows its frequency spectrum measured by different experiments: the data have the characteristic
frequency distribution of a black body radiation, a Plank distribution, with $T_{CMB}=(2.726\pm 0.005)$ K.
The maximum of the distribution is at to $\nu_{m}=5.879 \cdot 10^{10}$ T(K) = 160.5 GHz
and thus $E_{\gamma m}=h\nu_{m}$=0.5 meV. Assuming that this black body radiation fills uniformely the
Universe one has $N_{\gamma} \simeq 411$ photons/cm$^{3}$, 
$\rho_{\gamma} = (\pi^{2}/15) T_{\gamma}^{4} \simeq 4.68 \cdot 10^{-34}$ g 
cm$^{-3}$ $\simeq$ 0.262 eV cm$^{-3}$.

The CMB radiation comes from the cosmic time $t \simeq$ 300000 y ($T \simeq
4000$ K), the time of atom formation, see Fig. 1.1.
Before this time, matter ($e^{-}$, $p$, nuclei) was totally ionised; matter and radiation interacted continuously
and were in thermodynamic equilibrium. After this time the photons did not have enough energy to ionize atoms:
neutral atoms were formed, the photons do not interact with atoms, and thus matter and radiation started
to have independent lives. The Universe continued its expansion, the temperature of matter and radiation decreased in temperature, and the photons eventually reached their present temperature of 2.73 K.

\begin{figure}
\begin{center}
\epsfclipon
\mbox{\epsfysize=7.5 cm \epsffile{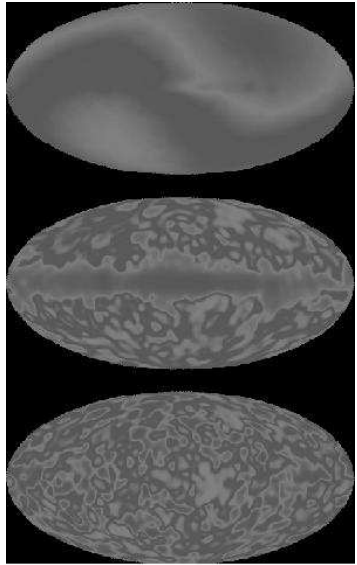}}
\epsfclipoff
\end{center}
\caption{Three false colour images of the sky in galactic coordinates as
seen at microwave frequencies by the COBE
experiment. The orientation of our galaxy runs horizontally across the center of the images.
The Cosmic Microwave Background radiation as measured by COBE is uniform to 1 part in 300.
At this level one observes a variation of temperature between the upper right band and the lower
band (top). After removing the dipole term they obtain the picture in the middle, where the fluctuations 
are at the level of 10$^{-4}$ and are dominated by the galactic
plane. After removing the galaxy signal one 
finds smaller fluctuations, at the level of 10$^{-5}$, all over the sky (bottom).}
\label{f:cmb}
\end{figure}

\begin{figure}
\begin{center}
\epsfclipon
\mbox{\epsfysize=6.6 cm \epsffile{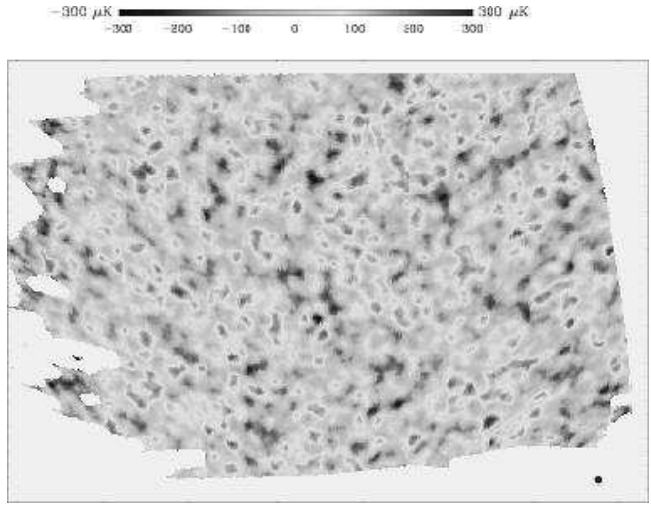}}
\epsfclipoff
\end{center}
\caption{Small angular scale variations of the CMB radiation observed by the Boomerang experiment.}
\label{f:boomerang}
\end{figure}

CMB precision measurements were performed by the COBE satellite \cite{cobe}, the Boomerang experiment in a
stratospheric balloon at the south pole \cite{boomerang} and the DASI detector at the south pole ground station
\cite{dasi}. The frequency spectrum of the CMB radiation, and thus of its temperature, is quite uniform in
every direction. However small temperature variations at the level of $10^{-2} \div 10^{-3}$ were soon found; 
they indicate the existence of a dipole term as shown in Fig. \ref{f:cmb} top. It arises from the motion of
our Solar System in the Galaxy and of our local group of galaxies with respect to the CMB radiation
(with $v \simeq 371$ km/s).
After removal of the dipole term COBE found temperature variations about 30 times smaller, which come from
sources in our Galaxy; they also found other more fundamental fluctuations,
Fig. \ref{f:cmb} center. After removal of the
galaxy effect, COBE observed small temperature variations over relatively large angular regions limited
by its 7$^{\circ}$ angular resolution, Fig. \ref{f:cmb} bottom. 
Later the experiment Boomerang \cite{boomerang}, which had an angular resolution of $\sim$ 15 arc minute, 
found smaller angular fluctuations over a large fraction of the sky, Fig. \ref{f:boomerang}.

\begin{figure}
\begin{center}
\epsfclipon
\mbox{\epsfysize=6.6 cm \epsffile{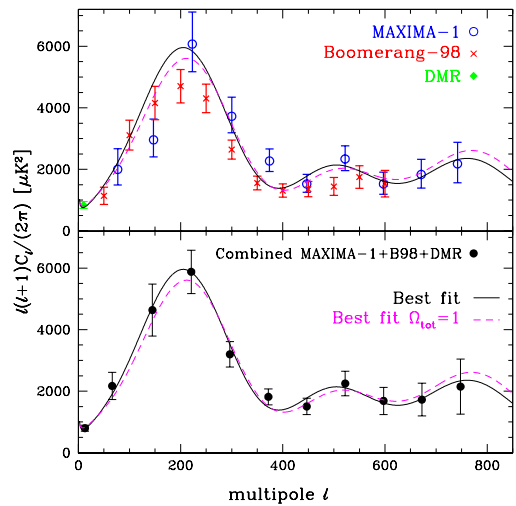}}
\epsfclipoff
\end{center}
\caption{CMB power spectra. 
Top: Data from MAXIMA-1, Boomerang and COBE-DMR. Bottom: maximum-likelihood fit to the combined data.
In both figures the curves show the best fit models.}
\label{f:power}
\end{figure}

The Boomerang and DASI experiments computed the angular power spectrum of the fluctuations 
from $l$=50 to 600, Fig. \ref{f:power}.
This spectrum is dominated by a peak at multipole $l \simeq 197 \pm 6$ and by two other peaks at $l \sim 500$
and 850. From these values one may obtain important cosmological parameters and a discrimination
between the inflationary model of the Early Universe (in which the primordial plasma undergoes ``acoustic''
oscillations) and the topological defect structure model.

The data favour the inflationary picture, from which one obtains fundamental parameters \cite{fisoggi}:
the position of the first peak yields $\Omega_{total} = \rho_{tot}/\rho_{c} = 1.03 \pm 0.06$ (flat universe) 
and the constant of the expansion of the Universe $h \simeq$ 0.7. The general structure of the data provides 
evidence for the existence of cold DM and Dark Energy, 
with $\Omega_{cDM} \simeq 1/3$ and $\Omega_{Q} (=\Omega_{DE}) \simeq 2/3$
(compare with the values in Table 8.1).

\setcounter{figure}{0}\setcounter{table}{0}\setcounter{equation}{0}
\section{Magnetic Monopoles}
\begin{figure}
\begin{center}
\mbox{\epsfig{file=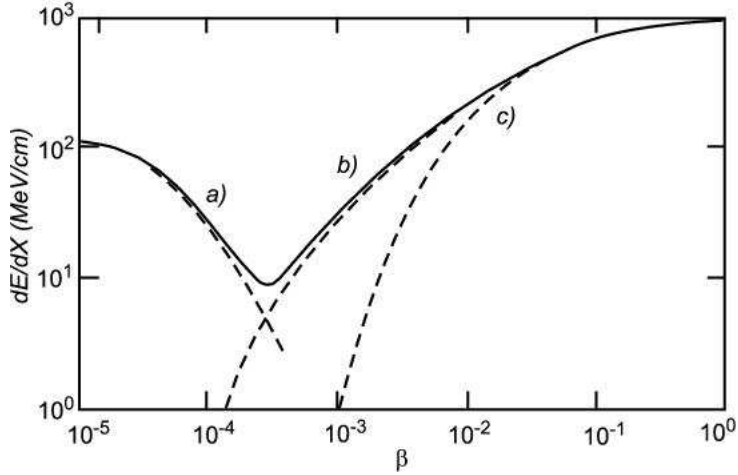,height=6.3cm}}
\end{center}
\caption{Energy losses, in MeV/cm, of $g=g_D$ MMs in liquid hydrogen vs ${\beta}$. 
a) corresponds to elastic monopole--hydrogen atom scattering; 
b) corresponds to interactions with level crossings; 
c) describes the ionization energy loss.}
\label{f:eneloss}
\end{figure}

In 1931 Dirac introduced the Magnetic Monopole (MM) in order to explain the
quantization of the electric charge, which follows from the existence of at 
least one free magnetic charge. He established the relationship between the 
elementary electric charge $e$ and the basic magnetic charge 
$g$:~$eg=n\hbar c/2$, where $n$ is an integer; 
$g_{D}=\hbar c/2e=137 e/2 \simeq 3.29 \cdot 10^{-8} cgs \simeq 68.5 e$ is called the unit
Dirac charge; because of the large magnetic charge,
a MM acquires a large energy in a magnetic field; a fast MM ($\beta > 10^{-2}$) behaves as an equivalent electric charge $(Ze)_{eq}=g_{D}\beta$, $\beta = v/c$. 
The MM energy loss in matter is sketched in Fig. \ref{f:eneloss}.
There was no prediction for the MM mass.
From 1931 searches for ''classical Dirac monopoles'' were carried out at every new
accelerator using relatively simple set--ups. 
Searches at the Fermilab $\bar{p}p$ collider established cross-section upper limits
$\sigma < 3\cdot 10^{-32}$ cm$^2$ up to 850 GeV mass for the reaction $p\bar{p} \rightarrow M\bar{M}$.
Experiments at LEP yielded 
$\sigma < 3 \cdot 10^{-37}$ cm$^2$ for $e^{+}e^{-} \rightarrow M\bar{M}$ for
masses up to $45$ GeV \cite{thooft}, and 
$\sigma < 5 \cdot 10^{-38}$ cm$^2$ for masses from 45 to 102 GeV 
\cite{michela}
\footnote{
Possible effects due to low mass MMs have been reported, see
f.e. Ref. \cite{russi}.
}

Electric charge is naturally quantized in GUT gauge theories of the basic interactions; such theories imply the existence of MMs, with  calculable properties. The MMs appear in the Early Universe at the phase transition corresponding to the spontaneous breaking of the unified group into subgroups, one of which is U(1) \cite{thooft}. The MM mass is related to the mass of the carriers X,Y of the unified interaction, $ m_{M}\ge m_{X}/G$, where G is the dimensionless unified coupling constant at E $\simeq m_{X}$. 
In GUTs with $m_{X}\simeq 10^{14}-10^{15}$ GeV and $G\simeq 0.025$, $m_{M} > 10^{16}-10^{17}$ GeV. 
This is an enormous mass: MMs cannot be produced at any man--made accelerator, 
existing or conceivable. They could only be produced in the first instants of 
our Universe.
Larger MM masses are expected if gravity is brought into the unification picture and in some 
SuperSymmetric models.

\begin{figure}
\begin{center}
\mbox{\epsfig{file=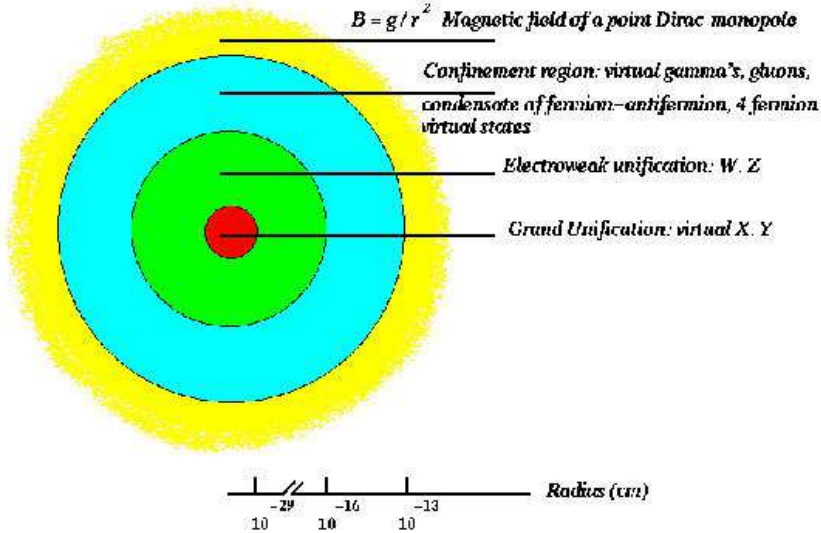,height=7.4cm}}
\end{center}
\caption{Structure of a GUT pole. The various regions correspond to: 
(i) Grand Unification ($r \sim 10^{-29}$ cm; inside this core one finds virtual $X$ and $Y$ bosons); 
(ii) electroweak unification ($r \sim 10^{-16}$ cm; inside this region one finds virtual $W^{\pm}$ and $Z^0$); 
(iii) confinement region ($r \sim 10^{-13}$ cm; inside one finds virtual $\gamma$, gluons and a condensate of fermion-antifermion pairs and 4-fermion virtual states); 
(iv) for $r>$ few fm one has the field of a point magnetic charge.}
\label{f:monostru}
\end{figure}

GUTs applied to the standard early universe scenario yield too many poles, while inflationary scenarios
lead to a very small number. Thus GUTs demand the existence of
MMs; however, the prediction of the MM mass is uncertain by 
orders of magnitude, the magnetic charge could be 1, 2, 3, ... Dirac units, and the expected flux 
could vary from a very small value to an observable one. The structure of a GUT pole is
sketched in Fig. \ref{f:monostru} \cite{thooft}.

Intermediate mass monopoles (IMMs) may have been produced in later phase transitions in the early universe, 
in which a semisimple gauge group yields a U(1) \cite{lazaride}. IMMs with masses $10^{5} \div 10^{12}$ GeV 
would have a structure similar to that of a GUT pole, but with a larger core, since $R\approx 1/m_{M}$.
%IMMs have been proposed as the sources of the highest energy cosmic rays ($E > 10^{20}$ eV). 
IMMs may be accelerated to relativistic velocities in the galactic magnetic field, and in several 
astrophysical sites.

The lowest mass MM is stable, since magnetic charge is conserved like electric charge. 
Therefore, the MMs produced in the early universe should still exist as cosmic relics, 
whose kinetic energy has been affected by their travel through galactic and 
intergalactic magnetic fields.

GUT poles in the CR should have low velocities and relatively large energy losses;
they are best searched for in the penetrating cosmic radiation. 
IMMs could be relativistic and may be searched for at high altitude
laboratories, in the downgoing CR and, if very energetic, also in the
upgoing CR.
\par

%\textit{COSMOLOGICAL AND ASTROPHYSICAL BOUNDS ON GUT POLES.}
Rough upper limits for a GUT monopole flux $F$ in the cosmic radiation were obtained on the basis of cosmological 
and astrophysical considerations. 

\ndt - {\it Limit from the mass density of the universe.} 
It is obtained requiring that the MM mass density be smaller than the critical density.
For $m_M \sim 10^{17}$ GeV the limit is $ F={n_Mc\over 4\pi}\beta<3\times 10^{-12}h^2_0\beta~(\mbox{cm}^{-2}\mbox{s}^{-1} \mbox{sr}^{-1}).$

\ndt - {\it Limit from the galactic magnetic field. The Parker limit.}
The $\sim 3\mu$G magnetic field in our Galaxy is stretched in the direction 
of the spiral arms; it is due to the non--uniform rotation of the Galaxy,
which generates a field with a time--scale approximately equal to the rotation period 
of the Galaxy $(\tau\sim 10^8$ yr). Since MMs are accelerated in magnetic fields, they 
gain energy, which is taken away from the stored magnetic energy. 
An upper bound for the MM flux is obtained by requiring that the kinetic energy gained 
per unit time by MMs be less than the magnetic energy generated by the dynamo effect. 
This yields the so--called Parker limit, $F<10^{-15}~\mbox{cm}^{-2}~\mbox{s}^{-1}$ sr$^{-1}$
\cite{parker}; in reality it is mass dependent \cite{parker}. 
An extended Parker bound, obtained by considering the survival of an early seed field, yields
$F\leq 1.2 \times 10^{-16}(m_M/10^{17}GeV)~\mbox{cm}^{-2}~\mbox{s}^{-1}~\mbox{sr}^{-1}$ \cite{parker}.
Other limits are obtained from the intergalactic field, from pulsars, etc.

\begin{figure}
\vspace{-3cm}
\begin{center}
\mbox{ \epsfysize=9.5cm
       \epsffile{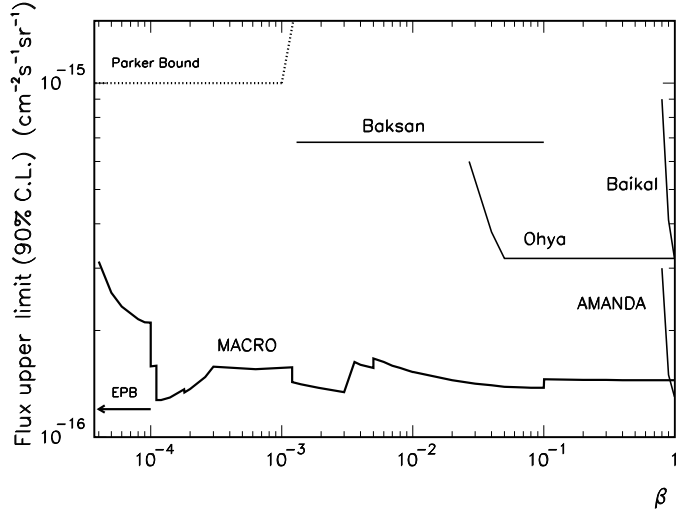}}
\end{center}
\caption{Magnetic monopole flux upper limits at the 90\% CL obtained by MACRO and
by other direct experiments. The limits apply to singly charged ($g=g_{D}$) MMs
assuming catalysis cross-sections smaller than a few mb.}
\label{f:flux}
\end{figure}

\textit{SEARCHES FOR SUPERMASSIVE GUT POLES.}
A flux of cosmic GUT poles may reach the Earth and may have done so for its whole life.
The velocity of these MMs could be in the range $4 \times 10^{-5} <\beta <0.1$, with possible peaks 
corresponding to the escape velocities from the Earth, the Sun and the Galaxy.
Searches for such MMs have been performed with superconducting induction devices, whose combined limit 
is at the level of $2 \times 10^{-14}~cm^{-2}~s^{-1}~sr^{-1}$, independent of $\beta$.
Direct searches were performed above ground and underground using scintillators, gaseous 
detectors and Nuclear Track Detectors (NTDs).
The most complete search was performed by MACRO, with three different types of subdetectors and with an 
acceptance of $\sim$ 10,000 m$^2$sr for an isotropic flux; no MMs were detected; the 90\% CL limits, 
shown in Fig. \ref{f:flux} vs $\beta$ \cite{mm_macro},
are at the level of $1 - 2\times 10^{-16}~cm^{-2}~s^{-1}~sr^{-1}$;
Fig. \ref{f:flux} shows also results from other experiments \cite{nakamura,amanda}.
Some indirect searches used ancient mica; if an incoming MM captures an 
aluminium nucleus and drags it through subterranean mica, it causes a trail of 
lattice defects which survive if the mica is not reheated.
The mica pieces analyzed ($13.5$ and $18$ cm$^2$) have been recording tracks for 
$4\div9\times 10^8$ years. The flux limits are 
$\sim 10^{-17} ~\mbox{cm}^{-2}~ \mbox{s}^{-1} $sr$^{-1}$ for $10^{-4}<\beta<10^{-3}$
\cite{price}.
There are many reasons why these indirect experiments might not be sensitive. 

\textit{INTERMEDIATE MASS MMs.}
IMMs could be present in the cosmic radiation. 
Detectors at the earth surface or underground/underwater would be capable
to detect MMs coming from above if they 
have intermediate masses \cite{slim}; lower mass MMs may be searched for with detectors 
located at mountain altitudes, in balloons and in satellites. 
Few experimental results are available \cite{nakamura}.
%the Baikal and AMANDA limits are shown in Fig.~10.3 \cite{amanda}. 
The SLIM experiment is searching for IMMs with 
NTDs at the Chacaltaya high altitude Lab \cite{slim}.

\textit{MONOPOLE CATALYSIS OF PROTON DECAY.}
A GUT pole may catalyze  proton decay, $ p+M\to M+e^+ +\pi^0$. 
The cross-section could be comparable with that of ordinary strong interactions, if the MM core is surrounded by a 
fermion-antifermion condensate, with some $\Delta B\ne 0$ terms extending up to the confinement region, Fig. \ref{f:monostru}.
%Thus MMs may capture a proton or a nucleus and lead to the catalysis reaction. 
For spin 1/2 nuclei, like aluminium, there should be an enhancement in the cross-section over that for free protons;
for spin-0 nuclei there should be a $\beta$--dependent suppression;
for oxygen the suppression could be $\sim 10^{-2}$ at $\beta=10^{-3}$, $\sim 10^{-5}$ at $\beta=10^{-4}$.
If the $\Delta B\ne 0$ cross-section for MM catalysis of proton decay 
were large, then a MM would trigger a chain of baryon ``decays'' along 
its passage through a large detector. MACRO made a thorough search for monopole catalysis of
proton decay, obtaining a MM flux upper limit at the level of 
$\sim 3 \cdot 10^{-16}$ cm$^{-2}$ s$^{-1}$ sr$^{-1}$ for $1.1 \cdot 10^{-4} \leq \beta \leq 5 \cdot 10^{-3}$
\cite{catalisi}.

\noindent - {\it Astrophysical limits from monopole catalysis of nucleon 
decay.} 
The number of MMs inside a star or a planet should  increase with
time, due to a constant capture rate and a  small pole--antipole
annihilation rate. The catalysis of nucleon decay by MMs could 
be a source of energy for these astrophysical bodies.
The catalysis argument, applied to the protons of our sun, leads to the 
possibility that the sun could emit high energy neutrinos. 
%The $\nu_e$'s could be detected via elastic scattering on electrons. 
Kamiokande gave the limit 
$F<8\times 10^{-10}\beta^2$ if the catalysis cross-section is $\sim$1 mb.
From such limit they placed a limit on the number of poles in the sun: less 
than 1 pole per $10^{12}$ g of solar material \cite{mm_macro}.\par
\par A speculative upper bound on the total number of MMs present 
inside the Earth can be made assuming that the energy released 
by MM catalysis of nucleon decay in the Earth does not exceed 
the surface heat flow.

\setcounter{figure}{0}\setcounter{table}{0}\setcounter{equation}{0}
\section{Searches for other exotica}
\begin{figure}[ht]
\begin{center}
\mbox{\hspace{-0.5cm}
\epsfig{file=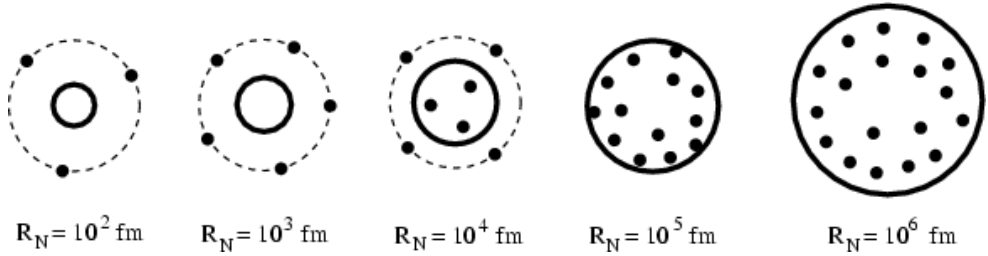,height=3.35cm}}
\vspace{-0.2cm}
\end{center}
\caption{Dimension of the quark bag (of radius $R_{N}$) and of the core+electrons of a nuclearite.
For $M_{N} < 10^{9}$ GeV the electron cloud is outside the
quark bag and the core+electrons system has a size
of $\sim 10^{5}$ fm = 1 \AA; for $10^{9} < M_N < 10^{15}$
GeV the electrons are partially inside the core; for $M_N > 10^{15}$
GeV all electrons are inside the core. The black dots are
electrons, the quark bag border is indicated by thick solid circles; the
border of the core+electron system is shown by dashed lines.}
\label{f:nuclearite}
\end{figure}

\begin{figure}[t]
\begin{center}
\mbox{\epsfig{file=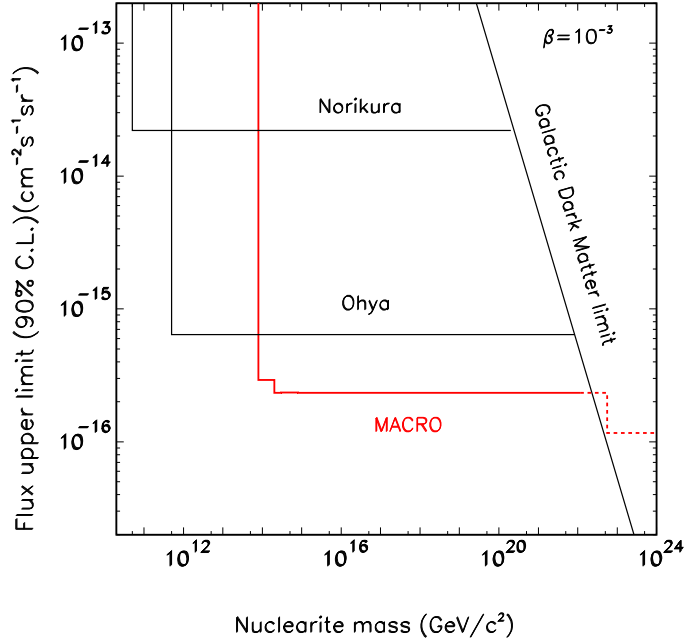,height=8.5cm}}
\vspace{-1.0cm}
\end{center}
\vspace{0.5cm}
\caption{90\% CL flux upper limits versus mass for nuclearites with $\beta = 10^{-3}$ at ground level.}
\label{f:nuclim}
\end{figure}

\textit{NUCLEARITES.}
{\it Strangelets, Strange Quark Matter (SQM)} consist of aggregates
of $u,~d$ and $s$ quarks; 
the SQM is a QCD colour singlet and has a positive integer electric charge \cite{witten}. 
The overall neutrality of SQM is ensured by an electron cloud which surrounds it, 
forming a sort of atom (the word \textit{nuclearite} denotes the core+electron system).
They could have been produced shortly after the Big Bang and may have survived as 
remnants; they could also appear in violent astrophysical processes, such as neutron star 
collisions. Nuclearites should have a density $\rho_N = M_N/V_N \simeq 3.5 \times 10^{14}$ g cm$^{-3}$,
larger than that of atomic nuclei, and they could be stable for all baryon numbers, from
$A \simeq$ few tens up to strange stars ($A \sim 10^{57}$) \cite{derujula2}. Nuclearites could 
be part of the cold dark matter.
[There have been many discussions on probable consequences of nuclearites: (i) the possible
observation of a strange quark star having the same mass of a neutron star, but of smaller size;
(ii) a nuclearite traversing the Earth could be a linear source of several earthquakes \cite{larousserie}; 
(iii) some disasters like the Tunguska one in Siberia in early 1990's could be due to the arrival of a large 
nuclearite].
The relation between mass and size of nuclearites is illustrated in Fig. \ref{f:nuclearite}.
The main energy loss mechanism for low velocity nuclearites in matter is that
of atomic collisions: a nuclearite should displace the matter in its path by elastic or quasi-elastic collisions with the ambient atoms.
The energy loss rate is large and nuclearites should be easily 
seen by scintillators and NTDs.
Nuclearites are expected to have galactic velocities, $\beta \sim 10^{-3}$,
and traverse the Earth if $M_{N} >$ 0.1 g.
Most nuclearite searches were made as byproduct of MM searches. 
The best {\it direct} flux limits for nuclearites, Fig. \ref{f:nuclim}, come from three large area experiments: 
two used CR39 NTDs at mountain altitude \cite{nakamura} and in the Ohya mine;
the MACRO experiment used liquid scintillators and NTDs \cite{macro2,ggichep}.

\textit{Q-BALLS.}
They should be aggregates of squarks $\tilde q$, sleptons $~\tilde l$ and 
Higgs fields \cite{coleman,kusenko1,kusenko2}. The scalar condensate inside a Q-ball core
has a global baryon number $Q$ (and may be also a lepton number). 
Protons, neutrons and may be electrons
could be absorbed in the condensate.
There could exist neutral (SENS) and charged Q-balls (SECS).
SENS do not have a net electric charge, are massive, may 
catalyse proton decay and may capture a proton yielding 
SECS, charged objects of lower masses; for SECS
the Coulomb barrier could prevent the capture of nuclei. SECS have integer
charges because they are colour singlets.
Q-balls with sleptons in the condensate can also absorb electrons. 
A SENS which enters the earth atmosphere could absorb a
nucleus of nitrogen which would give it the charge $Z=+7$.
Other nuclear absorptions are prevented by Coulomb repulsion. If the  
Q-ball can absorb electrons at the same rate as protons, the positive charge
of the absorbed nucleus may be neutralized by the charge of absorbed $e^{-}$.
If the absorption of $e^{-}$ is slow or impossible, the Q-ball carries a positive 
electric charge after the capture of the first nucleus.

Q-balls could be cold DM candidates. 
SECS with $\beta \simeq 10^{-3}$ and $M_Q < 10^{13}$ GeV/c$^2$ could reach 
an underground detector from above, SENS also from below 
\cite{kusenko2}. SENS may be detected by their almost continuous 
emission of charged pions ($dE/dx \sim $100 GeV g$^{-1}$cm$^{2}$); 
SECS may be detected via their large energy losses in scintillators and in NTDs.

\textit{FRACTIONALLY CHARGED PARTICLES.}
They are expected in GUTs as deconfined quarks; their charges range from 
$Q=e/5$ to $2/3e$. When traversing a medium they release a fraction $(Q/e)^{2}$ 
of the energy deposited by a muon. Fast Lightly Ionising Particles 
(LIPs) have been searched for by
Kamiokande and MACRO. The 90\% CL flux upper limits for LIPs with charges $2e/3$, $e/3$ and
$e/5$ are at the level of $10^{-15}cm^{-2}s^{-1}sr^{-1}$ \cite{lips1}.
Limits on nuclei with fractional charge were obtained at accelerators \cite{lips2}.
New bulk matter searches for fractional charge particles yielded new limits \cite{lee}.

\setcounter{figure}{0}\setcounter{table}{0}\setcounter{equation}{0}
\section{Gravitational waves}
The Earth should be continously bombarded by gravitational waves (GW) produced by distant celestial bodies subject to ``strong'' gravitational effects.
GWs are emitted when the quadrupole moment of an object of large mass is subject to large and fast variations. Only large celestial bodies should produce gravitational radiation measurable on Earth. These bodies may be binary systems of close--by stars (in particular when a neutron star is about to fall on the other); they should yield a periodic emission of GWs, with frequencies from few hundred Hz to 1 MHz. The amplitude of the emitted wave should increase as the stars approach each other and should become very large when one star is about to fall on the other. Asymmetric stellar gravitational collapses may give bursts of GW, with frequencies of the order of 1 kHz and durations of few ms. Also vibrating black holes, star accretions, galaxy formation, and the Big Bang (at the end of the epoch of unification of the gravitational force with the others) may produce or have produced GWs \cite{amaldi,pizzella,barish}.

\begin{figure}
\begin{center}
\epsfclipon
\mbox{\epsfysize=7.8cm \epsffile{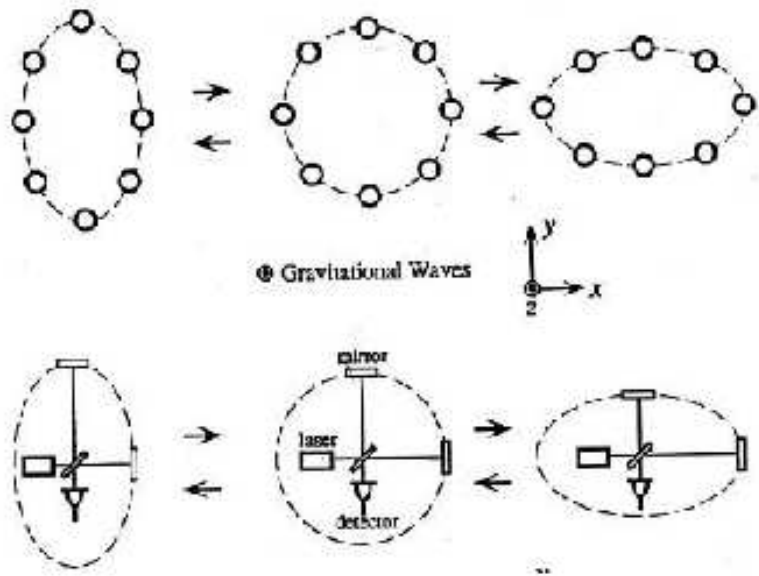}}
\epsfclipoff
\end{center}
\caption{Top figure: the effect of gravitational waves is shown 
on a ring of free particles. The circle alternately elongates vertically while
squashing horizontally and vice versa with the frequency of the gravitational wave.
The interferometry detection technique can measure this difference
in length as indicated in the lower figure. An interferometer measures the difference
in distance in two perpendicular directions.}
\label{f:dodiciuno}
\end{figure}

\begin{figure}
\begin{center}
\epsfclipon
\mbox{\epsfysize=10.0cm \epsffile{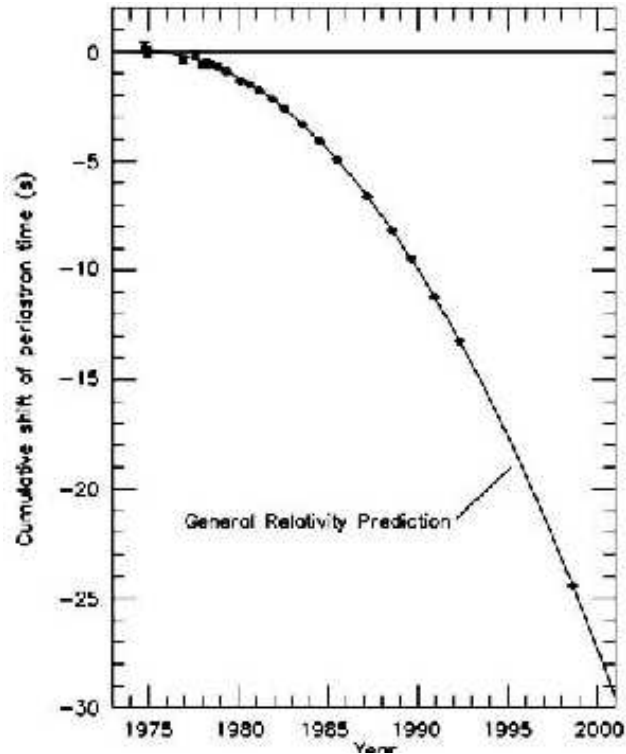}}
\epsfclipoff
\end{center}
\caption{The binary system PSR1916+13, containing two neutron stars, exhibits a speedup
of the orbital period. Over 25 years the total shift recorded is about 25 s. 
The black points are the data; the solid line is the prediction (not a fit to the data) 
from the parameters of the system. The agreement is impressive and this experiment provides 
strong evidence for the existence of gravitational waves.}
\label{f:dodicidue}
\end{figure}

\begin{figure}
\begin{center}
\epsfclipon
\mbox{\epsfysize=6.65cm \epsffile{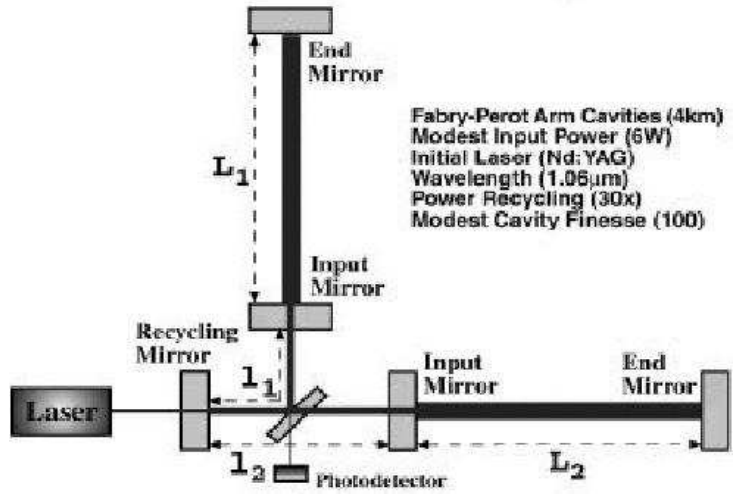}}
\epsfclipoff
\end{center}
\caption{The layout of the LIGO Michelson interferometer with Fabry--Perot cavities.}
\label{f:dodicitre}
\end{figure}

\begin{figure}
\begin{center}
\epsfclipon
\mbox{\epsfysize=7.65 cm \epsffile{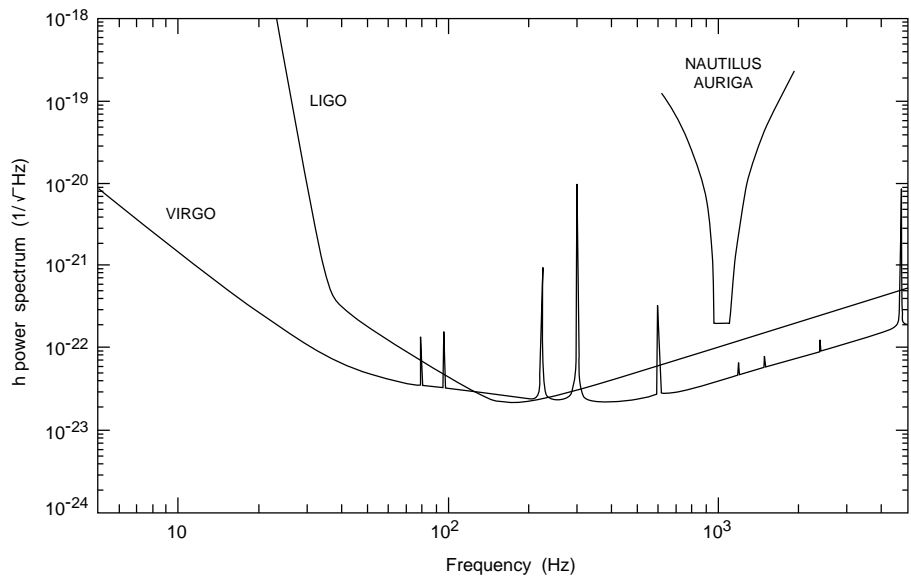}}
\epsfclipoff
\end{center}
\caption{Sensitivities of gravitational wave detectors: cryogenic detectors
(AURIGA and NAUTILUS) and appended interferometers (LIGO and VIRGO).}
\label{f:dodiciquattro}
\end{figure}

A GW is a transverse wave which travels at the speed of light. A gravitational wave should modify the distances between objects in the plane perpendicular to the direction of propagation of the wave, Fig. \ref{f:dodiciuno}. These deformations are expected to be extremely small. It has been estimated that a star collapse at the center of our galaxy may produce a variation of the order of $h \sim 10^{-18}$ metre per metre of separation of two objects on Earth. The Supernova 1987A in the large Magellanic Cloud could probably have produced a distortion 10 times smaller. A collapse in the Virgo cluster (at $\sim$ 17 MPc), should yield relative variations of $10^{-21}$. Two astronomers, R. Hulse and J. Taylor, studied the neutron star binary system PSR 1913+16. Both neutron stars have 1.4 solar masses; they are at a distance of about 2 million km and the revolution time is 8 hours. Hulse and Taylor demonstrated that the motion of the pulsar around its companion could not be understood unless the dissipative force associated with GW emission were included. 
The two neutron stars spiral in toward one another speeding up the motion (see Fig. \ref{f:dodicidue}). The inspiral is only 3 mm per orbit; so it will take more than one million years before they coalesce. 
The expected rate of coalescing binary neutron star systems is about a few per year if the detectors are sensitive up to 200 Mpc.

Very sensitive instruments are needed to directly observe GWs. The two lines developed until now are resonating metallic bars at very low temperatures and two-arms laser interferometric systems. A major program is underway for both types of detectors, with the purpose to detect gravitational star collapses up to the Virgo cluster.
The major difficulties arise from every type of noise, each of which should be carefully minimized.

There are at present five cryogenic resonant bars in operation: ALLEGRO, AURIGA, EXPLORER, NAUTILUS and NIOBE. They have roughly the same experimental sensitivity, $h \sim 10^{-22}$ at their resonant frequency. NIOBE, made with niobium, has a resonance frequency of 700 Hz, the others, of aluminium, have resonance frequencies of 900 Hz. 
The 5 groups search for coincidences among their detectors \cite{amaldi}. 
Some possible signals in coincidence may have been detected \cite{ronga}. 
NAUTILUS, in Frascati, detected CR showers in coincidence with a counter array.
The GRAVITON project plans to study and implement spherical cryogenic detectors, which would be omnidirectional.
In the first step three equal spherical detectors of 0.65 m diameter, made of a copper alloy
(94\% Cn, 6\% Al) are beeing planned, SCHENBERG in Sao Paulo, Brasil, MINIGRAIL in Holland and SFERA in Frascati, Italy.
The SCHENBERG antenna should have a sensitivity $h \sim 10^{-21}$ Hz$^{-1/2}$ at 3.2 kHz at temperatures of 15--20 mK \cite{aguiar}.

Table 12.1 gives the main characteristics and the sensitivities of the interferometers (see also Fig. \ref{f:dodiciquattro}) [12.6-12.9].
LIGO consists of a pair of interferometers (Fig. \ref{f:dodicitre}) located in the US at a large distance one from the other. The VIRGO detector is being completed near Pisa by a French--Italian collaboration.
TAMA is run by a Japanese collaboration and GEO by a UK--Germany collaboration. The interferometers should be starting to take data soon. Also in this case coincidences will be arranged.

The project LISA (Laser Interferometer Space Antenna) aims to detect GWs in the frequency range $10^{-3}\div 10^{-4}$ produced from binary systems and stochastic GW radiation. LISA will consist of an array of 4 satellites that are the ends of two interferometer arms with lengths of 5 million km; they will be located in almost circular orbits around the sun.

Stochastic signals from GW emitted in the first instants of the Universe, at the Plank time, $10^{-43}$ s, could be possibly detected through correlation of the background signals from several detectors. 

The detection of gravitational waves would have far reaching consequences. It would probe the general theory of relativity and it would open up a new observational window related to violent astrophysical phenomena.

\begin{table}
        \begin{center}
                \begin{tabular}{cccccc}
                \hline
                detector & arm length & bandwidth   & $\widetilde{h}$   & h                 & h                 \\
             &            & [Hz]        & $1/\sqrt{Hz}$     & pulse             & continuous        \\
    \hline
                        LIGO   & 4 km       & $\sim 300$  & $       10^{-24}$ & $10^{-22}$        & $3\cdot 10^{-28}$ \\
                 VIRGO   & 3 km       & $\sim 1000$ & $4\cdot 10^{-23}$ & $10^{-21}$        & $3\cdot 10^{-26}$ \\
                GEO600   & 600 m      & $\sim 400$  & $8\cdot 10^{-23}$ & $4\cdot 10^{-21}$ & $2\cdot 10^{-26}$ \\
                        TAMA   & 300 m      & $\sim 300$  & $8\cdot 10^{-23}$ & $5\cdot 10^{-21}$ & $8\cdot 10^{-26}$ \\
                \hline
                \end{tabular}
        \end{center}
        \caption{Main features and target sensitivity for the gravitational
                interferometers. $\widetilde{h}$ is a quantity related to
                the frequency (and energy) spread of the signal.}
\end{table}

\setcounter{figure}{0}\setcounter{table}{0}\setcounter{equation}{0}
\section{Conclusions}
Non-accelerator astroparticle physics has become an interesting and lively
interdisciplinary field of research at the frontier of particle physics, astrophysics
and cosmology, and thus involving researchers from different fields. The number of
experiments in this field has been growing considerably (see list in Appendix)
and so have theoretical and phenomenological papers (see \cite{ultima} and the
large number of references in this lecture notes and in those of P. Lipari).

Non-accelerator astroparticle physics is performed in underground halls,
under-water, under-ice, at the surface, with balloons and in satellites.

In cosmic rays we are interested in particles at extreme energies
and in the search for rare particles and rare phenomena.
There now are good indications that some HE cosmic rays come from 
expanding supernovae shells. CRs and the Universe seem to be made 
only of matter.

The satellites looking for nuclear tests in the atmosphere lead to the discovery 
of $\gamma$--ray bursts; they occur at the level of 1 per day and they are followed 
by afterglows. Many experiments were performed on this topic and more will follow.

There are now very strong indications for atmospheric and solar neutrino oscillations and thus for tiny
neutrino masses, which represent clear departures from the Standard Model of
particle physics. A variety of more precise experiments and of long baseline experiments
are in preparation. One expect to see soon the detection of a $\nut$ in a
$\num$ beam, and some oscillation patterns. Considerable theoretical work is
being performed on neutrino oscillations and on their implications.
The neutrino masses seem to be too small to provide an appreciable contribution to the DM.

The field of neutrino astrophysics concerns solar neutrinos, neutrinos from stellar gravitational 
collapses and HE muon neutrinos. Large neutrino telescopes should open up this last field.

Further searches for proton decay will involve much larger detectors. More refined cryogenic detectors 
and more massive detectors are required for new improved searches for neutrinoless double beta decays.

Many direct searches for cold DM have been performed, with some controversial
indication for WIMPs. New, more massive and more refined detectors will allow direct
searches at a higher level of sensitivity. Also indirect searches will be
carried out with improved sensitivities.

The searches for magnetic monopoles and other exotic particles in the cosmic radiation
has reached new levels of sensitivity. Improved searches will require very large
surfaces covered with detectors.

The experimental study of the CMB radiation at new precision levels has led to the
observation of the first three ``peaks'' in the spectral analysis. From their locations
several important parameters of the Universe have been computed.

The study of type 1a supernovae has given strong indications for the accelerated expansion 
of the Universe.

The Universe seems to be flat and to have the critical density.
But baryons constitute only 5\% of it, and visible baryons only 0.5\%.
Non-baryonic dark matter is approximately 30\%, whereas ``dark energy'' is
about 65\%: why our Universe has these properties?

There are plenty of exotic astrophysical objects: neutron stars, solar mass black
holes, supermassive black holes and may be also quark stars made of $u$, $d$, $s$ quarks.

A large effort is presently made on very sensitive cryogenic and interferometer gravitational 
wave detectors and one may be close to the birth of gravitational wave astronomy.

Astroparticle physics is an interesting field of research for the young generation in
general and for young researchers from developing countries in particular.

We conclude recalling that the 2002 Nobel Prize in Physics was given to
R. Davis and M. Koshiba ``for pioneering contributions to
astrophysics, in particular for the detection of cosmic neutrinos", and
to R. Giacconi for ``for pioneering contributions to astrophysics, which 
have led to the discovery of cosmic $x$-ray sources''.

\section*{Acknowledgements}
We thank our MACRO colleagues, especially those in Bologna, for their cooperation.
We acknowledge fruitful discussions with 
E. Bellotti, S. Cecchini, M. Cozzi, M. Giorgini, P. Lipari, G. Palumbo, 
N. Paver, Q. Shafi and many others.

%\newpage
%C1-Intro.tex:      \cite{ictp6,amaldi} --------------------> c1, c12
%C1-Intro.tex:      \cite{ictp6,oujda}
%C2-Detectors.tex:  \cite{ictp6,listaexp,mauri} ------------> c1, c2, c2
%C3-HECR.tex:       \cite{enomoto,aharonian}
%C6-Proton.tex:     \cite{pdg,hayato}
%C6-Proton.tex:     \cite{pdg,hayato}
%C6-Proton.tex:     \cite{pdg,baldo}
%C8-Dark.tex:       \cite{igex,cdms,edelweiss}
%C8-Dark.tex:       \cite{bottino,baltz}
%C8-Dark.tex:       \cite{wimp,ar2001} ---------------------> c8, c2
%C10-Monopoles.tex: \cite{nakamura,amanda}
%C11-Exotica.tex:   \cite{macro2,ggichep}
%C11-Exotica.tex:   \cite{coleman,kusenko1,kusenko2}

\newpage
\section*{Appendix A1}
A list of astroparticle physics experiments follows. Though the list is extensive, it may not
contain all experiments: we apologize for this and hope to be informed of possible omissions.
General information is found in:\\
\noindent \footnotesize \ttfamily http://www.mpi-hd.mpg.de/hfm/CosmicRay/CosmicRaySites.html\normalfont\normalsize,\\ 
\footnotesize \ttfamily http://www.slac.stanford.edu/spires/experiments/online\_exp.html\#N 
\normalfont \normalsize and\\
\footnotesize \ttfamily http://server11.infn.it/comm2/schede/index.htm \normalfont \normalsize

\vspace{0.5cm}
\small
\noindent \textbf{Space experiments}

\begin{list}{--}{\itemindent -13pt}
  \item \textit{ACCESS} [Advanced Cosmic ray Composition Experiment for the Space Station] to start in 2005
  \item \textit{AGILE} [Astro-rivelatore Gamma a Immagini LEggero] To be launched in 2004
  \item \textit{AirWatch} Proposed experiment for the International Space Station (ISS) to study HE CRs from space. R\&D status; 2nd phase of EUSO
  \item \textit{AMS} [Alpha Magnetic Spectrometer] Antimatter detector for the ISS, to start in 2004
  \item \textit{ASCA} [Advanced Satellite for Cosmology and Astrophysics] Ended in 2000
  \item \textit{BeppoSAX} $x$-ray astronomy satellite, ended in 2002
  \item \textit{BLAST} [Burst Locations with an Arc Second Telescope] $\gamma$-ray bursts
  \item \textit{Chandra} Advanced $x$-ray astrophysics facility
  \item \textit{CGRO} [Compton Gamma Ray Observatory] Ended in 2000; it had 4 experiments:
     \textit{BATSE} [Burst And Transient Source Experiment];
     \textit{OSSE} [Oriented Scintillation Spectrometer Experiment];
     \textit{COMPTEL} [imaging COMPton TELescope];
     \textit{EGRET} [Energetic Gamma Ray Experiment Telescope]
  \item \textit{COBE} See Cosmic Microwave Background
  \item \textit{EUSO} [Extreme Universe Space Observatory] ESA, Europe, in 2007
  \item \textit{GLAST} [Gamma ray Large Area Space Telescope] HE $\gamma$-ray astronomy
  \item \textit{HETE} [High-Energy Transient Experiment], HETE-2, launched in 2000
%  \item \textit{IMP-8} (Interplanetary Monitoring Platform)
  \item \textit{INTEGRAL} [INTErnational Gamma-Ray Astrophysics Laboratory] Being launched
  \item \textit{NINA} [New Instrument for Nuclear Analysis] Mission for low energy CRs
  \item \textit{OWL} [Orbiting Wide-angle Collector] To study HE CRs, NASA
  \item \textit{PAMELA} Magnetic spectrometer, to be launched in 2002
  \item \textit{PLANCK} Cosmic Background Radiation anisotropies, to be launched in 2007
  \item \textit{Rosat} $x$-ray satellite, ended in 1999
  \item \textit{RXTE} [Rossi X-ray Timing Explorer] Launched in 1995
  \item \textit{SWIFT} $\gamma$-ray burst mission, to be launched in 2003
  \item \textit{VELA} Military satellites, which discovered $\gamma$-ray bursts in the 1970's
%  \item \textit{WIND} Explores solar wind and plasma processes near the Earth as well as gamma-ray bursts
  \item \textit{XMM-Newton} [X-ray Multi-mirror Mission]
\end{list}

\noindent \textbf{Balloon experiments}

\begin{list}{--}{\itemindent -13pt}
%  \item \textit{ATIC} [Advanced Thin Ionization Calorimeter] 
  \item \textit{BESS} [Balloon Expt. with a superconducting Solenoid Spectrometer] 
  \item \textit{BETS} [Balloon borne Electron Telescope with Scintillating fibers] 
  \item \textit{CAPRICE} [Cosmic AntiParticle Ring Imaging Cherenkov Experiment] 
  \item \textit{CAKE} [Cosmic Abundance around Knee Energies] Flight from Sicily to Spain
  \item \textit{GRATIS} [Gamma-Ray Arcminute Telescope Imaging System] 
  \item \textit{GRIP} [Gamma-Ray Imaging Payload] 
  \item \textit{GRIS} [Gamma-Ray Imaging Spectrometer] 
  \item \textit{HEAT} [High Energy Antimatter Telescope]
  \item \textit{HIREGS} [HIgh REsolution Gamma-ray and hard $x$-ray Spectrometer]
  \item \textit{IMAX} [Isotope Matter Antimatter eXperiment]
  \item \textit{ISOMAX} [ISOtope MAgnet eXperiment] 
  \item \textit{JACEE} [Japanese-American Collaborative Emulsion Experiment] Completed
  \item \textit{MASS} [Matter Antimatter Superconducting Spectrometer]
%  \item \textit{MAXIMA} (see Cosmic Microwave Background)
  \item \textit{RUNJOB} [RUssian-Nippon JOint Balloon experiment] 
  \item \textit{SMILI} [Superconducting Magnet Instrument for Light Isotopes] 
  \item \textit{TIGRE} [Tracking and Imaging Gamma Ray Experiment]
  \item \textit{TIGER} [Trans Iron Galactic Element Recorder] 
  \item \textit{TRACER} [Transition Radiation Array for Cosmic Energetic Radiation] 
\end{list}

\noindent \textbf{Atmospheric Cherenkov experiments}

\noindent \textit{Telescopes and telescope systems}
\begin{list}{--}{\itemindent -13pt}
  \item \textit{CANGAROO} [Collaboration between Australia and Nippon for a GAmma Ray Observatory in the Outback] At Woomera, Australia 
  \item \textit{CAT}  [Cherenkov Array at Themis] Imaging telescope, HE $\gamma$-rays
  \item \textit{CLUE} [Cherenkov Light Ultraviolet Experiment] At HEGRA site, La Palma
  \item \textit{HEGRA} [High Energy Gamma Ray Astronomy] Cherenkov Telescopes on La Palma, Canary Islands, started in 1987, ended in 2000
  \item \textit{HESS} [High Energy Stereoscopic System] Under construction in Namibia
  \item \textit{MAGIC} [Major Atmospheric Gamma Imaging Cherenkov] 17 m telescope, in construction in Canary Islands
  \item \textit{PACT} [Pachmarhi Array of Cherenkov telescopes] At the High Energy Gamma Ray Observatory at Pachmarhi, India 
  \item \textit{VERITAS} [Very Energetic Radiation Imaging Telescope Array System] 
  \item \textit{Whipple} $\gamma$-ray telescope on Mt. Hopkins, Arizona 
\end{list}
\noindent \textit{Solar power facilities as light collectors:}
\begin{list}{--}{\itemindent -13pt}
  \item \textit{CELESTE} [CErenkov Low Energy Sampling and Timing Experiment] At Themis, France
  \item \textit{GRAAL} [Gamma-Ray Astronomy at ALmeria] Near Almeria, Spain 
  \item \textit{STACEE} [Solar Tower Air Cherenkov Experiment] At Sandia Labs, N.M.
\end{list}
\noindent \textit{Cherenkov counter arrays}
\begin{list}{--}{\itemindent -13pt}
  \item \textit{AIROBICC} Non-imaging counters in the HEGRA array
  \item \textit{BLANCA} [Broad LAteral Non-imaging Cherenkov Array] At CASA
  \item \textit{TUNKA-13} array of non-imaging counters near Lake Baikal
\end{list}

\noindent \textbf{Air shower experiments}

\begin{list}{--}{\itemindent -13pt}
  \item \textit{AGASA} [Akeno Giant Air Shower Array] In Japan
  \item \textit{AirWatch, EUSO and OWL} (fluorescence, see space experiments)
  \item \textit{ARGO-YBJ} Resistive Plate Chamber detector of 6500 m$^2$ in Tibet
  \item \textit{CASA-MIA} [Chicago Air Shower Array] Ended in 1998
  \item \textit{EAS-TOP} Above the Gran Sasso massif, Italy, ended in 2000
  \item \textit{Haverah Park} Ended in 1993
  \item \textit{GRAND} [Gamma Ray Astrophysics at Notre Dame] Tracking detector array
  \item \textit{KASCADE} [KArlsruhe Shower Core and Array DEtector]
  \item \textit{MILAGRO} Water Cherenkov $\gamma$-ray experiment Los Alamos; MILAGRITO.
  \item \textit{Norikura Observatory} In Gifu, Japan
  \item \textit{Pierre Auger Project} Giant Airshower Detector Project, in Argentina and USA, with counters and
  fluorescence detectors
  \item \textit{SPASE 2} [South Pole Air Shower Experiment]
  \item \textit{TA} [Telescope Array project] CR beyond the GZK cutoff (fluorescence)
  \item \textit{Tian-Shan Mountain Cosmic Ray Station} Lebedev Institute
  \item \textit{Tibet AS-gamma experiment} Air shower array, in Yangbajing, Tibet
\end{list}

\noindent \textbf{Other ground-based cosmic-ray experiments}

\begin{list}{--}{\itemindent -13pt}
  \item \textit{ALTA} [Alberta Large area Time coincidence Array]
  \item \textit{NALTA} [North American Large area Time coincidence Arrays]
  \item \textit{Pamir} Emulsion chamber experiment
  \item \textit{SLIM} [Search for LIght Monopoles] Nuclear Track Detectors at Chacaltaya Lab
%  \item \textit{University of Adelaide Cosmic Ray Muon Monitor}
\end{list}

\noindent \textbf{Cosmic Microwave Background experiments}
\begin{list}{--}{\itemindent -13pt}
  \item \textit{COBE} [COsmic Background Explorer] Satellite with 3 instruments (FIRAS, DMR, DIRBE)
  CMB radiation, lunched in 1989, completed
  \item \textit{DASI} [Degree Angular Scale Interferometer] CMB anisotropies, at South Pole
  \item \textit{CBI} [Cosmic Background Imager] In Chile
  \item \textit{BOOMERANG} [Balloon Observatory Of Millimetric Extragalactic Radiation ANd Geophysics]
  \item \textit{MAXIMA} [Millimiter Anisotropy eXperiment IMaging Array]
\end{list}

\noindent \textbf{Long-baseline neutrino experiments}

\begin{list}{--}{\itemindent -13pt}
  \item \textit{KARMEN} [KArlsruhe Rutherford Mittel-Energy Neutrino]
  \item \textit{MINOS} [Main Injector Neutrino Oscillation Search] from FNAL to Soudan mine
  \item \textit{OPERA} [Oscillation Project with Emulsion-tRacking Apparatus] in the CERN to Gran Sasso beam
\end{list}

\noindent \textbf{Neutrino experiments at reactors}

\begin{list}{--}{\itemindent -13pt}
  \item \textit{CHOOZ} In Ardennes, France, completed
  \item \textit{KamLAND} In the Kamioka mine, Japan ($>$ 100 km effective baseline)
  \item \textit{MUNU} Neutrino-electron scattering at the Bugey reactor, France
  \item \textit{Palo Verde} Reactor neutrino oscillation experiment
\end{list}

\noindent \textbf{Underground experiments}

\begin{list}{--}{\itemindent -13pt}
  \item \textit{BOREXINO} Liquid scintillator solar neutrino experiment at Gran Sasso
  \item \textit{BUST} [Baksan Underground Scintillator Telescope] Baksan, Russia
  \item \textit{L3-Cosmics/CosmoLEP}, CERN underground muon experiment with CRs
  \item \textit{Frejus} Ionization detectors, iron calorimeters in Frejus tunnel (1984-1988)
  \item \textit{GNO} [Gallium Neutrino Observatory] At Gran Sasso; GALLEX successor
  \item \textit{Homestake-Chlorine} Experiment for solar neutrinos
  \item \textit{Homestake-Iodine} Experiment for solar neutrinos
  \item \textit{ICARUS} [Imaging Cosmic And Rare Underground Signal] 
Liquid argon TPC detector at Gran Sasso; it is also a long-baseline $\nu$ expt.
  \item \textit{IMB} [Irvine Michigan Brookhaven] in a USA mine (1980-1991)
        \item \textit{Kamiokande} 5000 t water detector in the Kamioka Mine, Japan (1983-1995)
  \item \textit{KGF} [Kolar Gold Field] Calorimeter detector with scintillators, flash tubes and iron, in KGF mine, South Africa (completed)
  \item \textit{LSD} [Liquid Scintillator Detector] 90 t, in Mt Blanc tunnel (completed)
  \item \textit{LVD} [Large Volume Detector] 1000 t liquid scintillator at Gran Sasso
  \item \textit{MACRO} [Monopole, Astrophysics and Cosmic Rays Observatory] At Gran Sasso (1989-2000)
  \item \textit{NUSEX} [NUcleon Stability EXperiment] In the Mt Blanc tunnel (completed)
  \item \textit{OMNIS} [Observatory for Multiflavor NeutrInos from Supernovae] 
  \item \textit{SAGE} [Soviet-American Gallium Experiment] Baksan, Russia
  \item \textit{SNO} [Sudbury Neutrino Observatory] 1000 t D$_{2}$O detector, Creighton mine, Sudbury, Canada
  \item \textit{SOUDAN-2} Soudan Iron Mine, Minnesota, USA (1989-1993-ongoing)
  \item \textit{Super Kamiokande} 50000 t water Cherenkov detector, Kamioka Mine, Japan
\end{list}

\noindent \textbf{Underwater experiments}

\begin{list}{--}{\itemindent -13pt}
  \item \textit{ANTARES} [Astronomy with a Neutrino Telescope and Abyss environmental RESearch] Under construction in the Mediterrean sea near Toulon, France (\ttfamily http://antares.in2p3.fr \normalfont)
  \item \textit{Baikal} Underwater neutrino experiment in Lake Baikal, Russia
  \item \textit{NEMO} [NEutrino subMarine Observatory] Planned km$^3$-scale water detector
        \item \textit{NESTOR} [Neutrino Extended Submarine Telescope with
          Oceanographic Research] 
In the Mediterrean sea near Pylos, Greece, \ttfamily http://www.nestor.org.gr \normalfont
\end{list}

\noindent \textbf{Experiments in Antarctic ice (at the South Pole)}

\begin{list}{--}{\itemindent -13pt}
  \item \textit{AMANDA} [Antarctic Muon And Neutrino Detector Array]
  \item \textit{RAND} [Radio Array Neutrino Detector] 
  \item \textit{RICE} [Radio Ice Cerenkov Experiment] 
  \item \textit{ICECUBE} (a planned kilometer-scale ice neutrino observatory) 
\end{list}

\noindent \textbf{Other cosmic neutrino experiments}

\begin{list}{--}{\itemindent -13pt}
  \item \textit{Goldstone} Radio signals from UHE neutrino interactions in the moon
\end{list}

\noindent \textbf{Neutrino mass, double $\beta$ decay and direct dark matter searches}

\begin{list}{--}{\itemindent -13pt}
  \item \textit{CAST} Axion search with a strong magnetic field and $x$-ray detectors at CERN
  \item \textit{CDMS} [Cryogenic Dark Matter Search] at Stanford and in the Soudan mine
  \item \textit{CRESST} [Cryogenic Rare Event Search with Superconducting Thermometers] $Al_{2}O_{3}$ and $CaWO_{4}$ scintillating bolometers, at Gran Sasso
  \item \textit{CUORE} [Cryogenic Underground Observatory for Rare Events] at Gran Sasso
  \item \textit{CUORICINO} Test of CUORE
  \item \textit{DAMA} [particle DArk MAtter searches] $NaI(Tl)$ scintillators, at Gran Sasso
        \item \textit{DBA} Search for double $\beta$ decay of $^{100}$Mo at Gran Sasso
  \item \textit{EDELWEISS} [Experience pour DEtecter Les WIMPs En SIte Souterrain] Ge thermal and ionization cryogenic detector in the Frejus tunnel
  \item \textit{ELEGANTS} NaI and CaF2 scintillators in OTO Lab, Japan
  \item \textit{EROS} [Experience de Recherche d'Objets Sombres]
  \item \textit{GENIUS-TF} Test of Germanium detectors, DM and solar neutrinos, Gran Sasso
  \item \textit{HDMS} [Heidelberg Dark Matter Search] 
  \item \textit{IGEX} Ge ionization detectors in Baksan, Russia
  \item \textit{MACHO} [search for MAssive Compact Halo Objects]
  \item \textit{MAINZ} Neutrino mass experiment ($\nue$)
  \item \textit{MANU2} Cryogenic micro-calorimeters, Genova, Italy ($\nue$)
  \item \textit{MIBETA} Development of microbolometers for $x$-ray and $\beta$ decay studies
  \item \textit{ORPHEUS} Superconductiong detectors at Bern, CH
  \item \textit{ROSEBUD} [Rare Objects SEarch with Bolometers UndergrounD] 
Canfranc, underground lab., Spain; see also expts. COSME, IGEX, ANAIS
  \item \textit{TROITSK} Neutrino mass experiment ($\nue$)
  \item \textit{UK-DMC} [UK Dark Matter Collaboration] 
Ge detectors, in Boulby Mine; see also expts. NaI, NAIAD, ZEPLIN
  \item \textit{WARP} Test of a liquid argon scintillator detector, Legnaro and Gran Sasso
  \item \textit{ZEPLIN} Xe scintillation/ionization detectors, in Boulby Mine, UK
\end{list}

\noindent \textbf{Dark Energy experiments}

\begin{list}{--}{\itemindent -13pt}
  \item \textit{ISCP} [International Supernova Cosmological Project]
  \item \textit{HSST} [High-z Supernova Search Team]
  \item \textit{SNAP} [SuperNova Acceleration Probe]
\end{list}

\noindent \textbf{Gravitational wave experiments}

\begin{list}{--}{\itemindent -13pt}
        \item \textit{ACIGA} [Australian Consortium for Interferometric Gravitational Astronomy]
  \item \textit{AURIGA} Gravitational cryogenic resonant antenna in Legnaro, Italy
  \item \textit{GEO 600} British-German 600 m interferometer project
  \item \textit{GRAVITON project} cryogenic spherical resonant antennae: 
  SCHENBERG in S. Paulo, Brasil; MINIGRAIL in Holland; SFERA in Frascati, Italy
  \item \textit{LIGO} [Laser Interferometer Gravitational-Wave Observatory] 
  One detector at Hanford, Washington, and one at Liningstone, Louisiana
  \item \textit{LISA} [Laser Interferometer Space Antenna] (proposed);
    SMART2: test of LISA
  \item \textit{LSU Gravitational Wave Experiment} (Louisiana State University) 
  \item \textit{NAUTILUS} Gravitational cryogenic resonant antenna in Frascati, Italy
        \item \textit{NIOBE} Resonant-Bar gravitational detection project, western Australia
  \item \textit{ROG} [Ricerca Onde Gravitazionali] with gravitational 
  cryogenic resonant antennae in Frascati (NAUTILUS) and at CERN (EXPLORER)
  \item \textit{TAMA 300} [Tokyo Advanced Medium-scale Antenna] 300 m length, Japan
  \item \textit{VIRGO} Laser interferometric gravitational wave detector, near Pisa, Italy 
\end{list}

\normalsize

\end{document}